\documentclass[11pt]{amsart}\usepackage[]{graphicx}\usepackage[]{xcolor}
\makeatletter
\def\maxwidth{ %
  \ifdim\Gin@nat@width>\linewidth
    \linewidth
  \else
    \Gin@nat@width
  \fi
}
\makeatother

\definecolor{fgcolor}{rgb}{0.345, 0.345, 0.345}

\usepackage{framed}
\makeatletter
 {\par\unskip\endMakeFramed%
 \at@end@of@kframe}
\makeatother

\definecolor{shadecolor}{rgb}{.97, .97, .97}
\definecolor{messagecolor}{rgb}{0, 0, 0}
\definecolor{warningcolor}{rgb}{1, 0, 1}
\definecolor{errorcolor}{rgb}{1, 0, 0}
\newenvironment{knitrout}{}{} 

\usepackage{alltt}

\usepackage{accents,setspace,graphicx,srcltx,enumitem}
\usepackage[authoryear,longnamesfirst]{natbib}

\usepackage{amsmath}
\usepackage{amssymb}
\usepackage{amsfonts}
\usepackage{amsthm}
\usepackage{amsxtra}
\usepackage{amstext}
\usepackage{subfigure}
\usepackage{bbm}
\usepackage{amsaddr}
\usepackage[english]{babel}
\usepackage{hyperref}

 \usepackage{MnSymbol}  
 \usepackage{adjustbox}
 \usepackage{caption} 
 \newtheorem{theorem}{Theorem}[section]
 \newtheorem{lemma}[theorem]{Lemma}
 \newtheorem{corollary}[theorem]{Corollary}
 \newtheorem{assumption}{Assumption}[section]
 \newtheorem{proposition}{Proposition}[section]
 
 \theoremstyle{definition}
 \newtheorem{definition}{Definition}[section]
 
 \theoremstyle{remark}
 \newtheorem{remark}[]{Remark}
 
 \theoremstyle{remark}
 \newtheorem{example}[]{Example}
 
 \DeclareMathOperator*{\argmin}{arg\,min}

 \allowdisplaybreaks

\newcommand{\IR}{\mathbb{R}}

\newcommand{\vdagger}{\varphi_{\dagger}}
\newcommand{\un}{\hbox{\it 1\hskip-3pt I}}
\newcommand{\dps}{\displaystyle}

\IfFileExists{upquote.sty}{\usepackage{upquote}}{}
\begin{document}
\linespread{1.25}

\title[Iterative Nonparametric IV with Discrete Instruments]{Iterative Estimation of Nonparametric Regressions with Continuous Endogenous Variables and Discrete Instruments\vspace{-4ex}}

\author{Samuele Centorrino$^\ast$}
\address[S.\ Centorrino]{$(^\ast)$ Corresponding Author. International Monetary Fund and Economics Department, Stony Brook University, USA. \\\textit{Email address}: \texttt{\textup{scentorrino@imf.org}}.}
\author{Fr\'ed\'erique F\`eve}
\address[F. F\`eve]{Toulouse School of Economics, University of Toulouse Capitole, Toulouse, France. \\\textit{Email address}: \texttt{\textup{frederique.feve@tse-fr.eu}}.}

\author{Jean-Pierre Florens}
\address[J.\ P.\ Florens]{Toulouse School of Economics, University of Toulouse Capitole, Toulouse, France. \\\textit{Email address}: \texttt{\textup{jean-pierre.florens@tse-fr.eu}}.}

\thanks{A previous version of this paper has been circulated under the title: ``Nonparametric instrumental regressions with (potentially discrete) instruments independent of the error term". The authors would like to thank the editor, Xiaohong Chen, the associate editor, two anonymous referees, Yingyao Hu and Ben Williams for their comments and remarks, and Enno Mammen for helpful preliminary conversations on this topic. Samuele Centorrino gratefully acknowledges support from the Stony Brook University Initiative for the Arts, Humanities, and Lettered Social Sciences. Fr\'ed\'erique F\`eve and Jean-Pierre Florens acknowledge funding from the French National Research Agency (ANR) under the Investments for the Future program (Investissements d'Avenir, grant ANR-17-EURE-0010).}

\date{This version: \today; First version: December 20, 2017.}

\begin{abstract}
We consider a nonparametric regression model with continuous endogenous independent variables when only discrete instruments are available that are independent of the error term. Although this framework is very relevant for applied research, its implementation is challenging, as the regression function becomes the solution to a nonlinear integral equation. We propose a simple iterative procedure to estimate such models and showcase some of its asymptotic properties. In a simulation experiment, we detail its implementation in the case when the instrumental variable is binary. We conclude with an empirical application to returns to education.
\bigskip

\medskip

\end{abstract}

\maketitle

\setstretch{1.5}

\section{Introduction}

Instrumental variables are a workhorse of applied research in economics. The instrumental variable approach is based on two main assumptions that instruments need to satisfy: exogeneity, i.e., instruments are uncorrelated with or otherwise independent of the structural error term, and relevance, i.e., instruments are correlated with or otherwise dependent on the included endogenous regressor. The ability to identify the main causal effect of interest depends on the complexity of the model we wish to estimate and on the strength of these two requirements.

In this work, we consider the nonparametric regression model
\begin{equation} \label{eq:mainmodintro}
Y=\vdagger (X) + U, \quad E(U) = 0,
\end{equation}
where the explanatory variable $X$ is endogenous. This model is usually identified and estimated by taking the instrumental variable, $W$, as mean-independent of the error term. That is, $E(U\vert W)=0$, an exogeneity restriction on $W$. 

Under this relatively weak hypothesis, the regression function $\vdagger$ is the solution of the integral equation
\begin{equation} \label{eq:linillposed}
E(Y\vert W)= E(\varphi (X)\vert W).
\end{equation}
An estimator of $\vdagger$ is obtained by solving a regularized version of this functional equation \citep[see][among others]{chen2012,darolles2011,florens2003,hall2005,horowitz2011,newey2003}. 

Uniqueness of the solution to equation \eqref{eq:linillposed} is usually obtained under the so-called \textit{completeness condition}, or strong identification, which is the relevance assumption in this model (see \citet*{florens1990} and more recently \citet*{andrews2011}, \citet*{canay2013}, \citet*{xavier2011} and \citet*{freyberger2015}, among others). This condition dictates that
\begin{equation} \label{eq:npivcomplete}
E(\vdagger (X) \vert W)\overset{a.s.}{=}0 \; \Rightarrow \; \vdagger \overset{a.s.}{=} 0,
\end{equation}
which implies a strong dependence between the endogenous variable, $X$, and the instrument, $W$. In particular, whenever $X$ is a continuous variable, completeness requires the instrument $W$ to be itself continuous.\footnote{If $W\in \{1,...,L\}$ the set of equations $\int \varphi(x) f(x \vert W=l) dx=0 \; \forall l$ cannot characterize a function $\varphi$. This is true also when $L = \infty$, as we show in Example \ref{ex:infsuppins} in Appendix \ref{annexA4}.}

This condition is strong and may pose some challenges to the empirical application of fully nonparametric estimators of model \eqref{eq:mainmodintro}. Many instruments employed in empirical work are only binary or discrete. This is true, for instance, for randomized experiments with partial compliance, where the intent-to-treat provides an obvious source of exogenous variation \citep[see][]{krueger1999,torgovitsky2015}. 

In many of these cases, the endogenous variable is continuous, while the instrument is binary and, therefore, cannot satisfy the completeness condition stated above. In general, under standard assumptions of uncorrelated or mean-independent instruments, it is not possible to identify any, albeit known, function of $X$ with more than two parameters when only a binary instrument is available \citep[see][]{lochner2015,loken2012}.

Discrete instruments can be accommodated in this setting either by considering a pseudo-solution to equation \eqref{eq:linillposed} \citep{babii2017a}; or by strengthening the exogeneity condition and, as a consequence, weakening the relevance condition. The latter route is the one we undertake in this paper. 

We consider the estimation of the model in equation \eqref{eq:mainmodintro} where the instrument, $W$, is taken to be fully independent of the error term $U$. That is, $U\upmodels W$. This strong exogeneity condition characterizes $\vdagger$ as a solution to a nonlinear integral equation of the first kind \citep{dunker2014,dunker2018,kalten2008}.

We detail the implementation of an iterative Landweber-Fridman (LF) algorithm that can be used to estimate the function $\vdagger$ under the restriction of independence, and we derive some of its asymptotic properties.

In a previous paper, \citet{dunker2014} have considered the estimation and asymptotic properties of nonparametric estimators for nonlinear integral equations. The model analyzed in our paper is presented there as an example \citep[see also][]{dunker2018}. The approach of \citet{dunker2014} is extremely general and oriented to a class of regularization schemes less familiar to econometricians. We believe, however, that the case of discrete instruments is sufficiently important for applied research in econometrics to be separately and carefully developed. \citet{loh2022} takes a first step in this direction by considering nonparametric identification and estimation in a model with discrete instruments and endogenous variables. Our paper focuses on the case where $X$ is continuous \citep[see also][]{loh2023}. 

We contribute to the existing literature in several directions. First, the LF technique that we propose is both computationally efficient and easy to implement. Its properties in this class of statistical problems are new to the best of our knowledge. Finally, while our approach is not as general as those provided by \citet{dunker2014} and \citet{dunker2018} and is limited to the statistical model at hand, our convergence rates clearly spell out the sources of the estimation error and allow us to provide better guidance for implementation in this particular example. Ultimately, our paper has both theoretical and empirical value when compared to existing literature.

The independence assumption is often used in the following nonseparable triangular model 
\begin{align*}
Y =& g_1 (X,\varepsilon)\\
X =& g_2 (W,V),
\end{align*}
with $(\varepsilon,V) \upmodels W$, $\varepsilon$ and $V$ are standard uniform random variables, and $g_1(x,\cdot)$ and $g_2(w,\cdot)$ are monotone in their second argument \citep[see][]{fevrier2015,torgovitsky2015}. Identification and estimation of this model is usually based on a control function assumption \textit{\`a la} \citet{imbens2009} (see also \citealp{chernozhukov2007} and \citealp{horowitzlee2007}). The relative advantage of instrumental variables over control functions is that we do not need to specify a triangular structure, i.e., we do not need to specify a first-stage equation that relates the endogenous variable to the instrument. More generally, we allow for a simultaneous dependence between the endogenous regressor, $X$, and the dependent variable, $Y$, that is excluded by a control function approach. However, our model assumes that the function $g (X, \varepsilon)$ has a separable form, with $U = F^{-1}_U(\varepsilon)$, and $F_U$ the cumulative distribution function of $U$. In this respect, our approach is less general than control functions that can account for unobserved multivariate heterogeneity.

The paper is organized as follows. In Section \ref{sec:identification}, we briefly discuss some local identification conditions in the independence case in relation to the usual completeness condition. In Section \ref{sec:estimation}, we present the practical implementation of our estimator, whose properties are detailed in Section \ref{sec:asymprop}. We discuss the specific case of a binary instrument in Section \ref{sec:mcsim}. We conclude our work with an empirical application estimating the returns to education, using data from \citet{card1995}.

\subsection*{Notations} In the following, we let $L^2$ be the space of square-integrable functions with respect to the Lebesgue measure. For a real-valued function $\varphi \in L^2$, we let $\Vert \varphi \Vert$ be the $L^2$-norm. For an operator $T$, we let $\Vert T \Vert$ be the operator norm induced by the $L^2$ norm. For two random variables $X_1$ and $X_2$, we denote $X_1 \upmodels X_2$, if $X_1$ and $X_2$ are independent. If $a$ and $b$ are scalar, $a \vee b = \max\{ a,b\}$, and $a \wedge b = \min\{a,b\}$. For two sequences $a_n$ and $b_n$, we use the notation $a_n \asymp b_n$ to signify that the ratio $a_n/b_n$ is bounded away from zero and infinity. Finally, for an even integer $a \geq 2$, we let $E \Vert \cdot \Vert^a = E \left( \Vert \cdot \Vert^a \right)$.

\section{Local identification under independence} \label{sec:identification}

We consider a random element $(Y,X,W) \in \IR \times \IR^p\times\IR^q$. We analyze the model
\begin{equation}
Y=\vdagger (X) + U, \quad \varphi \in L^2_X,
\end{equation}
where the random vector $X$ only contains endogenous regressors, i.e., $X$ and $W$ do not have elements in common.

This model has been considerably studied under the following mean independence condition
\begin{equation} 
E(U \vert W) =0. \tag{$MI$}
\label{eq:mindependence}
\end{equation}
This condition characterizes $\vdagger$ as the solution of the linear equation
\begin{equation} \label{eq:varphisolmeanind}
E(\varphi(X) \vert W) = E(Y \vert W).
\end{equation}

We replace Assumption \eqref{eq:mindependence} with a stronger Assumption of full independence.
\begin{equation}
U\upmodels W \text{ and } E(U)=0. \tag{$FI$} 
\label{eq:findependence}
\end{equation}

The condition $E(U) = 0$, which implies $E(Y) = E (\vdagger(X))$, is a normalization condition. It may be thought to be implicit in the definition of the error term $U$, and if it is not satisfied, $\varphi$ may be identified up to location only. In the following, and without loss of generality, we assume that $E(Y) = E (\vdagger(X)) = 0$, as the mean of $Y$ can always be identified (provided it exists) and estimated at a parametric rate. Therefore, we assume that $\vdagger \in \mathcal{E}$, where $\mathcal{E} = \lbrace \varphi \in L^2_X : E(\varphi) = 0 \rbrace$

The \eqref{eq:findependence} condition implies \eqref{eq:mindependence}. Thus, if the completeness condition between $X$ and $W$ is verified, $\vdagger$ is identified under \eqref{eq:findependence}, while the reverse cannot be true. 

We, therefore, explore weaker local identification conditions under \eqref{eq:findependence}, which may be satisfied in particular in the case where $X$ is continuous and $W$ discrete.\footnote{Global identification conditions could also be analyzed, but they are less interpretable and difficult to verify in practice \citep[see, e.g.][]{chernozhukov2005,feve2015,beyhum2024}. \citet{loh2023} studies global identification conditions in a similar setting, but he focuses on the space of essentially bounded functions.}

We recast the independence condition in \eqref{eq:findependence} in the following way. We use the notations
\begin{equation}
F(y,x\vert w) = \frac{\partial}{\partial x_1}\cdots \frac{\partial}{\partial x_p} P (Y\leq y, X\leq x\vert W=w),
\end{equation}
and
\begin{equation}
F(y,x) = \frac{\partial}{\partial x_1}\cdots \frac{\partial }{\partial x_p} P (Y\leq y, X\leq x).
\end{equation}
Roughly speaking $F$ is a cdf as a function of $y$, and a density as a function of $x$.

The condition in \eqref{eq:findependence} is equivalent to
\begin{equation}
P(U\leq u\vert W=w)= P(U\leq u) \quad \forall u, w,
\end{equation}
or
\begin{equation} \label{eq:nlinversepr}
\int F(\vdagger (x) + u , x\vert w) dx = \int F(\vdagger (x) +u,x) dx.
\end{equation}

Equation \eqref{eq:nlinversepr} may be written as
\begin{equation} \label{eq:nlmomeq}
T(\vdagger)=\int \left[ F(\vdagger (x) + u , x\vert w)- F(\vdagger (x) +u,x) \right] dx=0, 
\end{equation}
and is a nonlinear integral equation, where $\vdagger$ denotes its solution, and $T:\mathcal{E} \rightarrow L^2_{U \times W}$ is a nonlinear operator. 

We consider the following definition of local identification. 
\begin{definition} \label{prop:onetoone}
$\vdagger$ is locally identified on $\mathcal{B}_{\dagger} \subseteq \mathcal{E}$ if $\Vert T(\varphi) \Vert > 0$ for all $\varphi \in \mathcal{B}_{\dagger}$, with $\varphi \neq \vdagger$. 
\end{definition}

Let
\[
\mathcal{B}_R =\lbrace \varphi \in \mathcal{E}: \Vert \varphi - \vdagger \Vert < R \rbrace,
\]
an open ball of radius $R$ around $\vdagger$. Local identification of $\vdagger$ is given in the following Proposition which recasts a more general result given by \citet{chenlee2013}.

\begin{proposition} \label{prop:onetoone}
Let the following hold.
\begin{itemize}
\item[(i)] The operator $T$ is Fr\'echet differentiable for all $\varphi \in \mathcal{E}$, with $T^\prime_{\varphi}$ its Fr\'echet derivative. 
\item[(ii)] $T^\prime_{\vdagger}$ is a one-to-one linear operator.
\item[(iii)] For all $\varphi \in \mathcal{B}_{R}$, there exists a constant $M > 0$, such that
\[
\Vert T(\varphi) - T(\vdagger) - T^\prime_{\vdagger} (\varphi - \vdagger) \Vert \leq M \Vert \varphi - \vdagger \Vert^2. 
\]
\item[(iv)] For a constant $M > 0$, $\mathcal{B}_{M} = \lbrace \varphi \in \mathcal{E}: \Vert T^\prime_{\vdagger} (\varphi - \vdagger) \Vert > M \Vert \varphi - \vdagger \Vert^2 \rbrace$.
\end{itemize}
Then, the separable NPIV model is locally identified on $\mathcal{B}_\dagger = \mathcal{B}_R \cap \mathcal{B}_M$ under Assumption $FI$.
\end{proposition}

Condition (i) is about the Fr\'echet differentiability of the operator $T$ \citep[see][Assumption 2, p. 789]{chenlee2013}, which helps control the behavior of our nonlinear problem in the vicinity of the solution. Moreover, the Fr\'echet derivative is taken to be injective, which is a rank condition on $T^\prime_{\vdagger}$ \citep[see][Assumption 1, p. 788]{chenlee2013}. Condition $(iii)$ restricts the amount of nonlinearity that is allowed for the ill-posed inverse problem at hand by requiring the continuity of the Fr\'echet derivative \citep[see][Assumption 2, p. 789]{chenlee2013}. Finally, the restriction on the set $\mathcal{B}_M$ is a source condition, as explained by \citet{chenlee2013}. In Hilbert spaces, a sufficient condition for $(iv)$ to hold is to restrict the identification set to an ellipsoid or a hyper-rectangle whose elements are sufficiently smooth in the sense that we specify below.

We now provide more explicit sufficient conditions on the primitives of our model to obtain local identification according to Proposition \ref{prop:onetoone}.

\begin{assumption} \label{ass:differentiability}
$F(y,x\vert w)$ and $F(y,x)$ are differentiable with respect to $y$ with continuous first partial derivatives for all $(x,w)$. Their first partial derivatives with respect to $y$ are the conditional density and the marginal density of $(Y,X)$ denoted $f_{Y,X\vert W}(y,x\vert w)$ and $f_{Y,X}(y,x)$, respectively. 
\end{assumption}

Under Assumption \ref{ass:differentiability}, condition (i) of Proposition \ref{prop:onetoone} is satisfied and
\begin{equation} \label{eq:momcond}
T^\prime_\varphi\tilde{\varphi} = \int \left(f_{Y,X \vert W}(\varphi (x) + u ,x\vert w) - f_{Y,X}(\varphi (x) + u,x)\right)\tilde{\varphi} (x) dx,
\end{equation}
where $T^\prime_\varphi: \mathcal{E} \rightarrow L^2_{U \times W}$ denotes the Fr\'echet derivative of $T$ computed in $\varphi$ as a linear function of $\tilde{\varphi}$.

\begin{assumption}[Conditional completeness] \label{ass:condcomp}
Let $\varphi \in \mathcal{E}$. Then
\[
E(\varphi (X)\vert U=u,W=w)\overset{a.s.}{=}E(\varphi (X)\vert U=u) \Rightarrow \varphi \overset{a.s.}{=} 0.
\]
\end{assumption}

Assumption \ref{ass:condcomp} states that the projection of $\varphi$ under $L^2_{U\times W}$ differs from the projection of $\varphi$ under $L^2_U$ except if $\varphi$ is constant. Moreover, this constant equals $0$ under the additional condition $E(U) = 0$. In other words, this condition requires that the dependence structure between $X$ and $U$ changes with the value of the instrument $W$. This Assumption implies condition $(ii)$ in Proposition \ref{prop:onetoone}. 

\begin{remark}
In settings where the IV is independent of $U$, and the completeness condition in \eqref{eq:npivcomplete} holds, Assumption \ref{ass:condcomp} is a weaker condition than \eqref{eq:npivcomplete}. In particular, under independence, $E[\varphi(X) \vert W = w;U = u] \overset{a.s.}{=} E[\varphi(X) \vert U =u]$ implies $E[\varphi(X) \vert W = w]  \overset{a.s.}{=} 0$, which in turn implies $\varphi \overset{a.s.}{=} 0$, our Assumption \ref{ass:condcomp}.\footnote{We thank an anonymous referee for highlighting this point.} However, when $X$ is continuous, condition \eqref{eq:npivcomplete} is satisfied only when $W$ is continuous. Thus, more generally, Assumption \ref{ass:condcomp} neither implies nor is implied by \eqref{eq:npivcomplete}.
\end{remark}

\begin{remark}
There could be cases when $U \upmodels X$ but $U$ and $X$ are dependent conditional on $W$ (see Examples \ref{ex:example1} and \ref{ex:example2} below). In these cases, $X$ is exogenous and Assumption \ref{ass:condcomp} still holds and can be used for point identification of $\vdagger$. However, when $U \upmodels (X,W)$, Assumption \ref{ass:condcomp} goes back to the usual completeness condition and cannot be satisfied with purely discrete instruments. In that case, $\vdagger$ is only partially identified by condition \eqref{eq:findependence}.
\end{remark}

\begin{assumption}[Uniform bounds]~ \label{ass:scalab}
\begin{enumerate}
\item[(i)] The conditional density of $Y$ given $(X,W)$, $f_{Y \vert X,W} (y \vert x,w)$, is uniformly bounded above by a positive constant $M_1$.
\item[(ii)] The first partial derivative of $f_{Y \vert X,W} (y \vert x,w)$ exists and is continuous almost everywhere. Moreover, for all $(x,w)$, 
\begin{align*}
\left\vert \frac{\partial}{\partial y} f_{Y \vert X,W} (y \vert x,w) \right\vert \leq M_2,\\
f_{X \vert W}(x \vert w) \leq M_3 f_X(x).
\end{align*}
\end{enumerate}
\end{assumption}
Assumption \ref{ass:scalab} implies condition $(iii)$ in Proposition \ref{prop:onetoone} for some radius $R>0$ \citep{chernozhukov2007}, and, more generally, it implies that $T^\prime_{\varphi}$ is a Hilbert-Schmidt operator, for all $\varphi \in \mathcal{E}$, and thus compact \citep[see][]{carrasco2007h}. The last condition also implies the local ill-posedness of the inverse problem as the eigenvalues of $T^\prime_{\vdagger}$ have zero as an accumulation point. Proofs of these statements are given in Section \ref{annexA2} of the Appendix.

Finally, we restrict the set $\mathcal{B}_M$ in a way that is amenable to imposing a \textit{source condition}. That is, we impose a restriction on the smoothness properties of local deviations from $\vdagger$ as a function of the local ill-posedness of the inverse problem, as determined by the speed of convergence of the eigenvalues of $T^\prime_{\vdagger}$ to $0$. 

\begin{definition}[\protect{\citealp{carrasco2007h}, \citealp[][Corollary 5, p. 794]{chenlee2013}}] \label{ass:sourcecondID}
Let $\lbrace \chi_1,\chi_2,\dots \rbrace$ be an orthonormal basis in $\mathcal{E}$, and $\lbrace t_1,t_2,\dots \rbrace$ be a bounded positive sequence, such that $\lbrace (t^2_j,\chi_j),j = 1,2,\dots \rbrace$ are the eigenvalues and the associated eigenvectors of the operator $T^{\prime \ast}_{\vdagger} T^\prime_{\vdagger}$. Then
\[
\mathcal{B}^\ast_{M} = \left\lbrace \varphi = \vdagger + \sum_{j = 1}^\infty b_j \chi_j : \sum_{j} \frac{b_j^2}{t_j^2} < \frac{1}{M^2} \right\rbrace,
\]
where $M = M_2 M_3$.
\end{definition}

Definition \ref{ass:sourcecondID} imposes that the set of functions in the ellipsoid $\mathcal{B}^\ast_{M}$ around $\vdagger$ be sufficiently smooth to guarantee that its Fourier coefficients go to zero faster than the eigenvalues of $T^\prime_{\vdagger}$. We have that $\mathcal{B}^\ast_{M} \subseteq \mathcal{B}_{M}$, so that this condition is sufficient to satisfy the requirement of Proposition 2.1(iv).

Let $s$ and $a$ be positive constants. When $b_j \asymp \exp( - s j)$, then this condition is satisfied whenever the eigenvalues of $T^\prime_{\vdagger}$ decay polynomially ($t_j \asymp j^{-a}$, mildly ill-posed inverse problem) or exponentially ($t_j \asymp \exp( - a j)$, severely ill-posed inverse problem) as long as $s > a$. If the $b_j$'s have polynomial decay, that is $b_j \asymp j^{-s}$, then Definition \ref{ass:sourcecondID} cannot be satisfied if the inverse problem is severely ill-posed. If the inverse problem is mildly ill-posed with $t_j\asymp j^{-a}$, then a sufficient condition to characterize the ellipsoid $\mathcal{B}^\ast_M$ is to impose that $s > a + 0.5$. In particular, if we interpret $s$ as the smoothness of $\varphi - \vdagger$ with respect to the Hilbert scale generated by the operator $\left(T^{\prime \ast}_{\vdagger} T^\prime_{\vdagger} \right)^{-1/2}$, then $s$ is determined by the least smooth element of this difference.\footnote{Results exist about the relationship between the number of continuous derivative of a function and the decay of its Fourier coefficients \citep{grafakos2014,katznelson2004}. For instance, a function whose Fourier coefficients decay as $j^{-s}$ has $s-1$ continuous derivatives, and its $s$-th derivative is integrable (that is, having a continuous $s$-th derivative is sufficient but not necessary for the function's integrability). Similarly, analytic functions have exponentially decaying Fourier coefficients. For a general Fourier decomposition wrt the eigenvectors of $T^{\prime \ast}_{\vdagger} T^\prime_{\vdagger}$, one can define a differential operator $D =\left(T^{\prime \ast}_{\vdagger} T^\prime_{\vdagger} \right)^{-1/2}$, i.e., the inverse of our integral operator, so that one could interpret smoothness in a standard sense.}

We thus revisit the conditions in Proposition \ref{prop:onetoone} to obtain the following.

\begin{proposition} \label{prop:condcomp}
Let Condition \eqref{eq:findependence} and Assumptions \ref{ass:differentiability}-\ref{ass:scalab} hold. The separable NPIV model is locally identified on $\mathcal{B}_{\dagger} = \mathcal{B}_{R} \cap \mathcal{B}^\ast_{M}$ under Assumption $FI$.
\end{proposition}

\begin{example} \label{ex:example1}
Let us assume that $\left(\begin{array}{l} U\\
X\end{array}\right) \vert W \sim N\left( \left(\begin{array}{l} 0\\0\end{array}\right), \left(\begin{array}{ll}
1&\rho(W)\\
\rho(W)&1
\end{array}\right)\right)$ and $W \in\{0,1\}$. Assumption \ref{ass:differentiability} is trivially satisfied. Also, if $\vert \rho (0) \vert \neq \vert \rho (1) \vert$, Assumption \ref{ass:condcomp} is verified (see \citealp{dunker2014}, for a formal proof). The conditional density of $Y$ given $(X,W)$ and its derivative are given by
\begin{align*}
f_{Y\vert X,W}(y \vert x,w) =& \frac{1}{\sqrt{1 - \rho^2(w)}}\phi \left( \frac{y - \rho(w)x}{\sqrt{1 - \rho^2(w)}}\right) \\
\frac{\partial}{\partial y}f_{Y\vert X,W}(y \vert x,w) =& \frac{y - \rho(w) x}{(1 - \rho^2(w))^{3/2}}\phi \left( \frac{y - \rho(w)x}{\sqrt{1 - \rho^2(w)}}\right),
\end{align*}
where $\phi$ is the pdf of a standard normal distribution, and the marginal and conditional density of $X$ given $W$ are equal and equal to a standard normal. Then, the uniform bounds in Assumption \ref{ass:scalab} are satisfied. Finally, let $H_j$ denote the $j$-th Hermite polynomial, with $\varphi(x) = \sum_{j = 1}^\infty H_j(x)b_j$. From \citet{dunker2014}, we have that $$E \left[ \varphi(X) \vert U,W = w \right] = \sum_{j = 1}^\infty E\left[ H_j(X) \vert U, W = w \right] b_j = \sum_{j = 1}^\infty H_j(U) \rho(w)^j b_j.$$ Hence, $T^\prime_{\vdagger} \varphi = \sum_{j = 1}^\infty H_j(U) (\rho(1)^j - \rho(0)^j) b_j$. Let us assume that $\vert \rho(1) \vert > \vert \rho(0) \vert$. As
\begin{equation*}
\sum_{j = 1}^\infty H_j(U) (\rho(1)^j - \rho(0)^j) = \sum_{j = 1}^\infty H_j(U) (\rho(1) - \rho(0)) \sum_{k = 0}^{j-1} \rho(1)^{j-1-k} \rho(0)^k,
\end{equation*}
we can bound
\begin{align*}
\left\vert \sum_{k = 0}^{j-1} \rho(1)^{j-1-k} \rho(0)^k\right\vert \leq & \vert \rho(1) \vert^{j-1} \sum_{k = 0}^{j-1} \left( \frac{\vert \rho(0)\vert}{\vert \rho(1)\vert} \right)^k =  \vert \rho(1) \vert^{j-1} \frac{\left( 1 - \left( \frac{\vert \rho(0)\vert}{\vert \rho(1)\vert} \right)^{j - 1} \right)}{1 - \left( \frac{\vert \rho(0)\vert}{\vert \rho(1)\vert} \right)}.
\end{align*}
Since the last term is $O(1)$, because $\vert \rho(1) \vert > \vert \rho(0) \vert$, then the $j$-th singular value of $T^\prime_{\vdagger}$ is bounded above by $\vert \rho(1) \vert^{j-1}$. A similar line of proof can be used when $\vert \rho(0) \vert > \vert \rho(1) \vert$.
This shows that the eigenvalues of $T^\prime_{\vdagger}$ have exponential decay, and the inverse problem is severely ill-posed. Thus, the model $Y=\vdagger (X) + U$, with $U\upmodels W$ and $E(U)=0$, is locally identified on $\mathcal{B}_\dagger$, as defined in \ref{ass:sourcecondID}, if $b_j \asymp \exp( - s j)$ for $s > -log(\vert \rho(0) \vert \vee \vert \rho(1) \vert)$. In this example, the dependence structure between $X$ and $U$ is solely captured by the correlation coefficient, which must change with the instrument $W$ (see also \citealp{hoderlein2011}).
\end{example}

\begin{example}[Conditional completeness for multivariate $X$] \label{ex:example2}
Let us generalize the above example to a multivariate case. Let $X \in \IR^p$ and $W$ be discrete with $L \geq 2$ mass points. We take a single discrete instrument with multiple mass points wlog. For instance, two binary instruments can be considered as one instrumental variable with four mass points. For $l = 1,\dots,L$, let $\tau_l = P(W = l) > 0$, i.e., a category with zero probability of occurrence can be removed, and $\sum_l \tau_l = 1$. We let 
\[
\left(\begin{array}{l} U\\
X\end{array}\right) \vert W \sim N\left( \left(\begin{array}{l} 0\\0\end{array}\right), \left(\begin{array}{ll}
1 &\Sigma_{UX}^\prime(W)\\
\Sigma_{UX}(W) & \Sigma_X
\end{array}\right)\right),
\]
where $\Sigma_X$ is a positive definite symmetric matrix, and $\Sigma_{UX}(W)$ is a $p \times 1$ vector of conditional covariances between $U$ and all the elements of $X$. The conditional distribution of $X$ given $(U = u,W = w)$ is normal with mean $\Sigma_{UX}(w) u$ and variance $\Sigma_X - \Sigma_{UX}(w) \Sigma^\prime_{UX}(w)$. While because of the independence of $W$ and $U$, the marginal distribution of $X$ given $U$ is a mixture of $L$ normal distributions, whose mixture weights are given by $\tau_l$. The marginal distributions of $X$ and $U$ are normal distributions with variances $\Sigma_X$ and $1$, respectively. Therefore, the probability measures of $X$ and $U$ are square integrable with respect to the normal measure. Hermite polynomials form a complete orthogonal system with respect to the normal measure, so we use them as eigenfunctions of the operator $T^\prime_{\vdagger}$ \citep[][p. 120]{letac1995}. All other assumptions for identification being satisfied as in the example above, we focus our discussion on Assumption \ref{ass:condcomp}. The function $\vdagger$ can be written as
\[
\vdagger (x) = \sum_{j = 1}^\infty b_j^\prime \mathbf{H}_j(x),
\]
where $\mathbf{H}_j$ is a vector valued Hermite polynomial \citep[as defined in][]{holmquist1996}, and $b_j$ a $jp$-vector of Fourier coefficients with $\frac{(j + p-1)!}{j!(p-1)!}$ unique values.\footnote{For instance, when $j = p = 2$, the vector of $b_j$ is $4 \times 1$ with $3$ unique entries.} For a matrix $A$, we let $A^{\langle j \rangle} = \bigotimes_{i = 1}^j A$, the $j$-th order Kronecker product. We have that
\begin{align*}
E \left[ \vdagger (X) \vert U = u, W = w \right] =& \sum_{j = 1}^\infty b_j^\prime \Sigma^{\langle j \rangle}_{UX}(w) H_j(u) \\
E \left[ \vdagger (X) \vert U = u \right] =& \sum_{j = 1}^\infty b_j^\prime \sum_{l= 1}^L \Sigma^{\langle j \rangle}_{UX}(w_l) \tau_l H_j(u),
\end{align*}
where the last result follows from the law of iterated expectations. Therefore, conditional completeness can be seen to be equivalent to 
\[
\sum_{j = 1}^\infty b_j^\prime \left( \Sigma^{\langle j \rangle}_{UX}(w) - \sum_{l= 1}^L \Sigma^{\langle j \rangle}_{UX}(w_l) \tau_l \right) H_j(u) \overset{a.s.}{=} 0 \quad \Rightarrow \quad \sum_{j = 1}^\infty b_j^\prime \mathbf{H}_j(x) \overset{a.s.}{=} 0,
\]
which, given the orthogonality between the Hermite polynomials, also implies
\[
b_j^\prime \left( \Sigma^{\langle j \rangle}_{UX}(w_k) -  \sum_{l= 1 }^L \tau_l\Sigma^{\langle j \rangle}_{UX}(w_l) \right) \left( \Sigma^{\langle j \rangle}_{UX}(w_k) -  \sum_{l= 1 }^L \tau_l\Sigma^{\langle j \rangle}_{UX}(w_l) \right)^\prime  b_j \overset{a.s.}{=} 0 \quad \Rightarrow \quad b_j \overset{a.s.}{=} 0,
\]
for all $j = 1,2,\dots$. A necessary and sufficient condition for this to hold is that $Var \left( \Sigma^{\langle j \rangle}_{UX}(W) \right)$ is non-singular. This implies that this matrix needs to have rank equal $\frac{(j + p-1)!}{j!(p-1)!}$ (the number of unique elements of the $j$-th time Kronecker product). Assuming $\lbrace \Sigma^{\langle j \rangle}_{UX}(w_k), k = 1,\dots,L-1 \rbrace$ are linearly independent, the rank of this covariance matrix is at most $L(L- 1)/2$. This is because
\begin{align*}
Var \left( \Sigma^{\langle j \rangle}_{UX}(W) \right)  =& \sum_{k = 1}^L  \tau_k \left( \Sigma^{\langle j \rangle}_{UX}(w_k) - \sum_{l= 1 }^L \tau_l\Sigma^{\langle j \rangle}_{UX}(w_l) \right) \left( \Sigma^{\langle j \rangle}_{UX}(w_k) -  \sum_{l= 1 }^L \tau_l\Sigma^{\langle j \rangle}_{UX}(w_l) \right)^\prime\\
=& \sum_{k = 1}^L  \sum_{l = 1,l\neq k}^L   \sum_{l^\prime = 1,l^\prime \neq k}^L \tau_k\tau_l \tau_{l^\prime} \left( \Sigma^{\langle j \rangle}_{UX}(w_k) - \Sigma^{\langle j \rangle}_{UX}(w_l) \right) \left(  \Sigma^{\langle j \rangle}_{UX}(w_k) - \Sigma^{\langle j \rangle}_{UX}(w_{l^\prime}) \right)^\prime\\
=& \sum_{k = 1}^{L-1}  \sum_{l > k}   \tau_k\tau_l \left( \Sigma^{\langle j \rangle}_{UX}(w_k) - \Sigma^{\langle j \rangle}_{UX}(w_l) \right) \left(    \sum_{l^\prime >k} \tau_{l^\prime} \left( \Sigma^{\langle j \rangle}_{UX}(w_k) - \Sigma^{\langle j \rangle}_{UX}(w_{l^\prime}) \right) \right)^\prime
\end{align*}
The summation with respect to $l$ is over $L-k$ linearly independent terms (because the $\Sigma^{\langle j \rangle}_{UX}(w_k)$'s are linearly independent), and therefore the rank of that matrix is $L-k$. Hence $\sum_{k = 1}^{L-1} (L-k) = L(L-1)/2$. Therefore, for any $p > 1$, conditional completeness requires a discrete instrument with infinite support or a continuous instrument.

Identification using discrete instruments with finite support can be restored if other conditions are imposed. For instance, under additivity, i.e., $\vdagger(x) = \sum_{k=1}^p \varphi_{\dagger,k}(x_k)$, we only need $L(L-1)/2 \geq p$, for all $j = 1,2,\dots$. For instance, if $p=2$, this requires $L \geq 3$ for conditional completeness to hold.

Similarly, if we wish to truncate the order of the basis approximation at $J^\ast \geq 1$ for $p$ given, we need to have
\[
\frac{L(L-1)}{2} \geq\frac{(J^\ast + p-1)!}{J^\ast !(p-1)!},
\]
which, for $p = J^\ast = 2$, again implies we must have $L \geq 3$. In the same scenario, and if we were to use only the restriction $Cov(U,W) = 0$, we would have four parameters and thus need $L \geq 4$.
\end{example}

\section{Estimation} \label{sec:estimation}

Recall that the model is written as $Y=\vdagger(X)+U$, where $X$ potentially depends on $U$, and $\vdagger$ is the true value of the function of interest. We assume there exists an instrumental variable, $W$, such that $U\upmodels W$ with $E(U) =0$. 

We observe an IID sample $\lbrace (Y_i, X_i, W_i), i=1,\dots,n \rbrace$ from the joint distribution of the random vector $(Y,X,W)$, and we take the support of $X$ to be compact and restricted to be the interval $[0,1]^p$ without loss of generality. 

The estimation procedure is based on equation \eqref{eq:nlmomeq}, where $F$ is replaced by a nonparametric estimator. Equation \eqref{eq:nlmomeq} is a nonlinear integral equation that defines an ill-posed inverse problem. 

In the case of nonlinear equations, we may consider both concepts of global and local ill-posedness. Global ill-posedness implies that the equation does not have a unique solution that depends continuously on $F$. This property has been studied by \citet{gagliardini2012} in a more general case. Our approach is based on local ill-posedness. This property is characterized by the linear approximation at the true value, $T^\prime_{\vdagger}$, which exists, and it is a compact operator under Assumptions \ref{ass:differentiability} and \ref{ass:scalab}. Then $T^\prime_{\vdagger}$ does not have a continuous inverse, and the problem is locally ill-posed.

The local ill-posedness of equation \eqref{eq:nlmomeq} requires using a regularization technique to obtain consistent estimators. Many solutions exist, but it is commonly accepted in the mathematical literature that iterative methods are the easiest to implement. We limit ourselves to the LF method, which is the simplest iterative method. More sophisticated methods can be used that do not require some of the smoothness restrictions imposed in our work. However they seem less suited for econometric problems, in which both the operator $T$ and the function of interest $\vdagger$ can be taken to be very regular. The main issue with LF regularization for nonlinear inverse problems is that it cannot take full advantage of the smoothness of the regression function (i.e., it \textit{saturates} at a certain level of regularity). In particular, Definition \ref{ass:sourcecondID} requires that $s$, the decay of the generalized Fourier coefficients of $\vdagger$, is larger than $a$, the decay of the eigenvalues of $T^\prime_{\vdagger}$ to zero. However in this context and under the additional conditions imposed below, LF regularization saturates at $s = a$ for both the mildly and the severely ill-posed inverse problem and prevents achieving a faster convergence rate. Therefore, to better accommodate the smoothness imposed by our identification conditions, we consider the estimation of the first partial derivatives of $\vdagger$, rather than of the function itself. As we explain in more detail below, this approach allows us to exploit the additional regularity of the class of continuously differentiable functions. Once the first derivative is estimated, we integrate with respect to the corresponding endogenous variable to recover the function of interest. Moreover, the first derivative has an important meaning in empirical analysis, as it measures \textit{marginal effects}.

Let us denote by $\vdagger^\prime = D\vdagger$ the first partial derivative of $\vdagger$ with respect to any of its arguments. $D$ is a differential operator, with $D^{-1}$ being an integral operator. Without loss of generality, we assume that the partial derivative is always taken with respect to the first component of $X$, $X_1$, and we denote as $X_{-1}$ the remaining components. That is,
\[
(D^{-1} \vdagger^\prime)(x) = \int^{x_1}_{-\infty} \vdagger^\prime(\xi,x_{-1})d\xi.
\]
We let $\mathcal{E}^\prime \subseteq \mathcal{E}$ be the space of centered, one-time differentiable functions of $X$, whose derivative is square-integrable. $\mathcal{E}^\prime$ is dense in $\mathcal{E}$. $D^{-1}$ is a linear operator, and the Fr\'echet derivative of the operator $T$ exists under the same conditions as above. $D$ is playing the role of Hilbert scale, and the ill-posedness of the inverse problem and the smoothness of $\vdagger$ are going to be linked to the properties of $D$. Finally, we define $A_{\vdagger} = T^\prime_{\vdagger} D^{-1}$, as the Fr\'echet derivative of $T$ with respect to $\vdagger^\prime$. 

We first describe the numerical algorithm and its implementation. We defer the study of the properties of this estimator to Section \ref{sec:asymprop}. The LF algorithm is based on the following recursive definition
\begin{equation} \label{eq:lfalg}
\varphi^\prime_{j+1} = \varphi^\prime_j - c A^{\ast}_{\varphi_j} T(\varphi_j), \text{ and } \varphi_{j} = D^{-1}\varphi^\prime_j,
\end{equation}
where $j = 0,1,2,\dots,N(c)-1$ is an integer and $N(c) >0$ is the total number of iterations, our regularization parameter, and $c > 0$ determines the size of the step between consecutive iterations. 

The algorithm in \eqref{eq:lfalg} requires a closed-form expression for $A^{\ast}_{\varphi}$, the adjoint of $A_\varphi$. $A_\varphi$ is a linear operator from $L^2_{X}$ into $L^2_{U\times W}$. Therefore, $A^{\ast}_{\varphi}$ is a linear operator from $L^2_{U\times W}$ into $L^2_{X}$ which ought to satisfy the following relation

\begin{align*}
\dps\int \int & \left[A_{\varphi} \tilde{\varphi}^\prime\right](u,w)\; \psi (u,w) f_{U,W} (u,w) du dw\\
=& \dps\int \tilde{\varphi}^\prime (x) \left[A^{\ast}_{\varphi} \psi\right] (x) f_X (x) dx \quad \forall \tilde{\varphi}^\prime \in L^2_{X}, \quad \psi \in L^2_{U\times W},
\end{align*}
with
\begin{align*}
& \dps\int \int \left[A_{\varphi} \tilde{\varphi}^\prime\right](u,w)\; \psi (u,w) f_{U,W} (u,w) du dw \notag \\
=& \dps\int \int_{-\infty}^{x_1} \tilde{\varphi}^\prime (\xi,x_{-1}) d\xi \psi (u,w) \left(f_{Y,X\vert W} (\varphi (x) + u, x\vert w) - f_{Y,X} ( \varphi (x) + u, x)\right) f_U (u) f_W (w) dx dw du\\
=& \dps\int \tilde{\varphi}^\prime (\xi,x_{-1})   f_X(\xi,x_{-1} ) \left\lbrace \frac{1}{f_X(\xi,x_{-1} )} \int \un(x_1\geq \xi )  \right. \times \\
& \qquad \left. \left[ \int \int \psi (u,w) \left(f_{Y,X\vert W} (\varphi (x) + u, x\vert w) - f_{Y,X} ( \varphi (x) + u, x)\right) f_U (u) f_W (w) dw du \right] dx \right\rbrace d\xi .
\end{align*}

Some algebra leads to
\begin{equation*}
\begin{aligned}
\left(A^{\ast}_{\varphi} \psi\right) (\xi ) =& \frac{1}{f_X(\xi,x_{-1} )} \int \un(x_1 \geq \xi ) \times \\
& \quad \left[ \int \int \psi (u,w) \left[f_{Y,X,W} (\varphi (x) + u, x,w) - f_{Y,X} (\varphi (x) + u, x) f_W(w) \right] f_U (u) du dw\right] dx_1,
\end{aligned}
\end{equation*}
which reduces to
\begin{equation}
\left(A^{\ast}_{\varphi} \psi\right) (x) = \frac{1}{f_X(x)}  \int \un(\xi \geq x_1)E\left[ \left( \psi (u,w) - E_W\left[ \psi (u,W) \right] \right) f_U(u) \vert X=(\xi,x_{-1})\right]f_{X_1}(\xi)d\xi,
\end{equation}
where $E_W$ denotes the expectation taken with respect to the marginal distribution of $W$, and we inverted the role of $\xi$ and $x_1$ to simplify notations. The LF estimator of $\vdagger^\prime$ is therefore given by
\begin{equation}
\hat{\varphi}^\prime_{j+1} = \hat{\varphi}^\prime_j - c \hat{A}^\ast_{\hat{\varphi}^\prime_j} \hat{T} (\hat\varphi_j),
\end{equation}
with $\hat{T}(\hat\varphi_j)$ an estimator of $T(\varphi)$ computed at the point $\hat\varphi_j$, and $\hat{A}^\ast_{\hat{\varphi}^\prime_j}$ an estimator of the adjoint of $A_{\hat\varphi_j}$. These objects are constructed as follows.
\begin{enumerate}
\item Select an initial value, $\hat{\varphi}^\prime_0$, to start our iterative procedure. The choice of the initial condition is crucial for the convergence of the algorithm, and it is thus advisable to experiment with several possible choices of it. One could take $\hat{\varphi}^\prime_0$ equal to the nonparametric estimation of the first derivative of $E\left[ Y | X \right]$. Another possible choice is to use a control function $V$, such that $U \upmodels X \vert V$, and estimate the first derivative with respect to $X$ of $E\left[ Y | X,V \right]$, an additive function of $X$ and $V$ in our model specification. Finally, it is possible to use flexible parametric estimation that embeds the independence assumption. We introduce this parametric estimator in Section \ref{sec:mcsim}. The uninformative initial condition, $\hat\varphi^\prime_0 =\hat\varphi_0= 0$, may be a bad choice, as at that point, the Fr\'echet derivative is always equal to $0$, and the algorithm may get stuck in a local solution.
\item At each iteration $j\geq 0$, we compute the centered residuals
\begin{equation}
\hat{U}_{ji} = Y_i - \hat\varphi_j (X_i) - \left(\frac{1}{n} \sum \left(Y_i - \hat\varphi_j (X_i)\right)\right).
\end{equation}
$T (\hat\varphi_j)$ is the difference between the cdf of $\hat{U}_{ji}$ conditional on $W$ and the unconditional cdf. In this paper, we use a nonparametric kernel estimator
\begin{equation} \label{eq:that}
\hat{T} (\hat\varphi_j) (u,w) = \frac{\dps\sum^n_{i=1} \bar{C}_{h_u} \left(u - \hat{U}_{ji}\right) L_{h_w} \left(w-W_i\right)}{\dps\sum^n_{i=1} L_{h_w} \dps\left(w-W_i\right)} - \dps\frac{1}{n} \dps\sum^n_{i=1} \bar{C}_{h_u} \left(u - \hat{U}_{ji}\right),
\end{equation}
where $C$ is a nonparametric kernel (e.g., Gaussian, Epanechnikov), $\bar{C}$ its distribution function, $L$ is a discrete kernel defined in Assumption \ref{assapp3b}, and $h_u$ and $h_w$ are bandwidths \citep[see, e.g.,][]{li2008}. When $W$ is binary, $\hat{T} (\hat\varphi_j)$ may also be computed by sorting with respect to the different values of $W$ (see Section \ref{sec:mcsim}).
\item The density $f_U(u)$ is estimated by the usual kernel density method, using the fitted residuals, $\hat{U}_{ji}$. Finally, $\hat{A}^\ast_{\hat\varphi_j} \hat{T} (\hat\varphi_j) $ is given by
\begin{equation}
(\hat{A}^\ast_{\hat\varphi_j} \hat{T} (\hat\varphi_j))(x)  =\dps \frac{\frac{1}{n}\dps\sum^n_{i=1} \bar{\hat{T}} (\hat{U}_{ji}, W_i) \hat{f}_{\hat{U}_j} (\hat{U}_{ji}) \left[ 1 - \bar{C}_{h_x}\left(x_1-X_{1,i}\right)\right]C_{h_x}\left(x_{-1}-X_{-1,i}\right) }{\frac{1}{nh_x}\dps\sum^n_{i=1} C_{h_x}\left(x-X_i\right)},
\end{equation}
with tuning parameter $h_x$, where
\[
\bar{\hat{T}} (\hat{U}_{ji}, W_i) = \hat{T} (\hat\varphi_j) (\hat{U}_{ji}, W_i) - \frac{1}{n} \sum_{i^\prime = 1}^n\hat{T} (\hat\varphi_j) (\hat{U}_{ji}, W_{i^\prime}).
\]
\item The constant $c > 0$ needs to be chosen in such a way that $c\|A^\ast_{\varphi}A_{\varphi}\|<1$, for $\varphi^\prime \in L^2_X$. $T^{\prime}_{\varphi}$ and $T^{\prime \ast}_{\varphi}$ are defined as products of a conditional expectation operator with $f_U(u)$. Thus $\|T^{\prime\ast}_{\varphi}T^{\prime}_{\varphi}\|\leq M^2_1$, where $M_1$ is defined in Assumption \ref{ass:scalab}(i)\footnote{Let us denote $r(u,w) = E\left[\tilde{\varphi}(X)  \vert U = u, W= w\right]$. Using the independence between $U$ and $W$, we can write $$\Vert T^\prime_{\varphi} \tilde{\varphi} \Vert^2 = E_{(U,W)} \left\lbrace f_U(U) \left[ r(U,W) - E_W \left( r(U,W) \right) \right] \right\rbrace^2,$$ where $E_{U,W}$ and $E_W$ denote expectations with respect to the distribution of $(U,W)$ and $W$, respectively. Therefore,
\begin{align*}
E_{(U,W)} & \left\lbrace f_U(U) \left[ r(U,W) - E_W \left( r(U,W) \right) \right] \right\rbrace^2 = E_U \left\lbrace f^2_U(U) E_W \left[ r(U,W) - E_W \left( r(U,W) \right) \right]^2 \right\rbrace\\
& \qquad  = E_U \left\lbrace f^2_U(U)  Var_W \left( r(U,W) \right) \right\rbrace \leq E_{(U,W)} \left[f^2_U(U) r^2(U,W) \right] \leq M_1^2 E\left[ \tilde{\varphi}^2(X) \right].
\end{align*} 
This implies $\Vert T^\prime_\varphi \Vert \leq M_1$. As $T^{\prime \ast}_\varphi$ is the adjoint of $T^\prime_\varphi$, we directly have that $\Vert T^\prime_\varphi \Vert^2 = \Vert T^{\prime \ast}_\varphi \Vert^2 = \Vert T^{\prime \ast}_\varphi T^\prime_\varphi  \Vert \leq M_1^2$}, and $\| D^{-1} \| \leq 0.5\vert \mathcal{X}_1 \vert$, where $\vert \mathcal{X}_1 \vert$ is the length of the support of $X_1$ \citep{florens2012}. Therefore, any $c<\frac{2}{\vert \mathcal{X}_1\vert M_1^2}$ guarantees convergence of our iteration scheme, i.e., it guarantees that $c A^\ast_{\varphi}A_{\varphi}$ is a contraction. Values of $c$ closer to the upper bound imply a bigger step size and a smaller number of iterations, although the approximation may be rather imprecise. While values of $c$ closer to $0$ imply a smaller step size, a larger number of iterations to achieve convergence, and better precision. Besides this restriction, the choice of $c$ \emph{is not essential} for the consistency of the LF estimator. To balance precision and computational speed, we take $c=0.5$. This choice should be conservative enough when $M_1 \leq 1$. For $M_1 > 1$, smaller values of $c$ may be necessary to achieve convergence.
\item Bandwidths can be chosen by Silverman's rule of thumb \citep[e.g., for $X \in \IR^p$, $h_j = (4/n(p+2))^{\frac{1}{p+4}}\sqrt{Var(X_j)}$, for $j = 1,\dots,p$; see][]{silverman1986} or by leave-one-out cross-validation. The bandwidth $h_u$ needs to be updated at each iteration as we update $\hat{U}_{ji}$. Cross-validation may lead to better finite sample properties as bandwidths adapt to the smoothness of the unknown population objects. However it can be computationally intensive, and users may prefer to use rule-of-thumb smoothing parameters to speed up computations. Proving the optimality of these choices is beyond the scope of the present paper. 
\item The last point is the choice of the stopping rule. This choice is crucial, as the stopping rule provides the regularization of the ill-posed inverse problem. It is common in the mathematical literature to adopt the so-called Morozov's discrepancy principle \citep[see][]{blanchard2017,kalten2008,morozov1967}. This principle leads to iterating up to $N_0(c) > 0$, such that $\|\hat{T} (\hat\varphi_{N_0(c)})\|^2$ is smaller than or equal to a pre-specified tolerance level. In practice, it is hard to determine this tolerance level in economic applications. Therefore, we choose the number of iterations by simple minimization of the error norm, $\Vert \hat{T} (\hat\varphi_{j}) \Vert^2$. \citet{florens2012} have successfully applied a similar approach to linear ill-posed inverse problems with LF regularization. However, to avoid over-regularizing (i.e., choosing an excessive number of iterations), we also suggest fixing a ceiling, $N_{max}$, which is based on the asymptotic theory derived below. We then proceed as follows: at each iteration $j = 1,2,\dots,N_{max}$, we compute $\Vert \hat{T} (\hat\varphi_{j}) \Vert^2$, and take $N_0(c)=\argmin_{j \in [1,N_{max}]}\Vert \hat{T} (\hat\varphi_{j}) \Vert^2$. We assess the finite sample properties of this criterion in Section \ref{sec:mcsim}.
\item At each iteration, we can compute $\hat\varphi_j$ as 
\[
\hat\varphi_j(x) = D^{-1}\hat\varphi^\prime_j(x) + \frac{1}{n} \sum_{i = 1}^n \left( Y_i - D^{-1}\hat\varphi^\prime_j(X_i) \right).
\]
The integral operator can be approximated using, for instance, the trapezoidal rule. The numerical approximation error in computing the integral operator is negligible compared to the nonparametric estimation error, so we take the operator $D^{-1}$ as known.
\end{enumerate}

\section{Asymptotic Properties} \label{sec:asymprop}

For nonlinear inverse problems, iteration methods like the one used here would generally not converge globally. We can prove local convergence by appropriately restricting the initial condition and controlling the behavior of the Fr\'echet derivative of the operator $\hat{T}$. In the following, the number of iterations, $N$, should be taken as a function of $c$ and the sample size, $n$, although we shall not make this dependence explicit. We also let $\varphi_0$ and $\varphi^\prime_0$ denote the probability limit of the (random) initial conditions $\hat\varphi_0$, and  $\hat\varphi^\prime_0$. We assume the following.

\begin{assumption}\label{ass:identification}
$\vdagger \in \mathcal{E}^\prime$ and is locally identified on $\mathcal{B}_\dagger$.
\end{assumption}

\begin{assumption}\label{ass:initcond}
$\varphi_0 \in \mathcal{B}_{\dagger}$.
\end{assumption}

Assumption \ref{ass:identification} imposes smoothness restrictions and local identification of the parameter of interest (see Section \ref{sec:identification}). Assumption \ref{ass:initcond} is a sufficient condition to guarantee local convergence of the iterative procedure. Under these conditions, we prove the following result. 

\begin{proposition} \label{prop:convlf}
Let Assumptions \ref{ass:identification} and \ref{ass:initcond} hold and 
\begin{equation} \label{ass:lfconv} 
\Vert \left( T^{\prime\ast}_{\vdagger} T^{\prime}_{\vdagger} \right)  \left(T^{\prime\ast}_{\varphi_j} T^{\prime}_{\varphi_j} \right)^{-1} \Vert \leq \frac{1}{M_4},
\end{equation}
with $M_4 \geq 1$, uniformly on $\mathcal{B}_{\vdagger}$, for all $j = 0,1,2,\dots$. Then, $\varphi_j \in \mathcal{B}_\dagger$ for all $j = 1,2,\dots$, and the LF algorithm in \eqref{eq:lfalg} converges to $\vdagger$.
\end{proposition}
This Proposition states that, at each iteration, the LF algorithm approaches $\vdagger$. Under Assumption \ref{ass:initcond}, this also implies that the LF algorithm stays in $\mathcal{B}_\dagger$. The proof is provided in Section \ref{annexA3} of the Appendix. The condition in \eqref{ass:lfconv} implies that the eigenvalues of $T^{\prime\ast}_{\varphi_j} T^{\prime}_{\varphi_j}$ go to zero at a speed that is at most the one of the eigenvalues of $ T^{\prime\ast}_{\vdagger} T^{\prime}_{\vdagger}$. As the decay of the singular values of $T^{\prime}_{\vdagger}$ determines the ill-posedness of the inverse problem, we essentially require that the degree of ill-posedness is no larger than the true one at each step of the LF algorithm. We further impose the following.

\begin{assumption}[Source Condition]\label{ass:sourcecond}
Let Assumption \ref{ass:initcond} hold. There exists $\beta > 0$, such that
\[
\varphi^\prime_0 - \vdagger^\prime = \left( A^{\ast}_{\vdagger} A_{\vdagger}\right)^{\frac{\beta}{2}} v,
\]
with $v \in L^2_X$, and $\Vert v \Vert =1$. 
\end{assumption}

\begin{assumption}[Link condition]\label{ass:linkcond}
For $a > 1$, $\left(T^{\prime \ast}_{\vdagger} T^{\prime}_{\vdagger} \right)^{\frac{1}{2a}} \sim D^{-1}$.
\end{assumption}

Assumption \ref{ass:sourcecond} is a H\"older \textit{source condition} \citep{engl2000}. This condition is widely used in linear inverse problems to relate the properties of the function of interest with the properties of the conditional expectation operator \citep[see][among others]{carrasco2007h}. In the nonlinear inverse problem literature, a similar condition is imposed \citep{dunker2014,dunker2018,kalten2008}. However this condition does not apply directly to the smoothness properties of $\vdagger^\prime$, but it is rather a condition on the local properties of our initial guess. With $a$ measuring the rate of decay of the eigenvalues of $T^{\prime}_{\vdagger}$, Assumption \ref{ass:linkcond} links the smoothness of the conditional expectation operator to the one of the integral operator, $D^{-1}$ (\citealp{engl2000}, Section 8.1, and \citealp{chen2011}). The notation $\sim$ is used here to indicate that the eigenvalues of $D^{-a}$ decay as fast as the singular values of $T^{\prime}_{\vdagger}$, i.e. the operator $T^{\prime}_{\vdagger}$ smooths a function at least as much as $D^{-a}$. If $s$ is the speed of decay of the generalized Fourier coefficients of $\varphi_0 - \vdagger$, $\beta = \frac{s-1}{a+1}$, then Assumption \ref{ass:sourcecond} is equivalent to $\varphi^\prime_0 - \vdagger^\prime = \left( T^{\prime \ast}_{\vdagger} T^{\prime}_{\vdagger} \right)^{\frac{(s-1)}{2a}}v$ and $\varphi_0 - \vdagger = \left( T^{\prime \ast}_{\vdagger} T^{\prime}_{\vdagger} \right)^{\frac{s}{2a}}v$. That is, if $\varphi^\prime_0 - \vdagger^\prime$ has regularity $\beta$, then $\varphi_0 - \vdagger$ has regularity $\beta + (a+1)^{-1}$ with respect to the operator $A_{\vdagger}$. Moreover, Assumption \ref{ass:linkcond} also implies $\varphi_0 - \vdagger = \left( T^{\prime \ast}_{\vdagger} T^{\prime}_{\vdagger} \right)^{\frac{s}{2a}}v = D^{-s}v$, which implies differentiability of $\varphi_0 - \vdagger$ up to the $s^{th}$ order. Assumption \ref{ass:sourcecond} could be extended to encompass a logarithmic source condition as in \citet{dunker2018}. This is a generalization that we do not pursue here.

The conditions imposed so far are not sufficient to derive an upper bound in the $L^2$ norm for our estimator. We also need to restrict further the local behavior of the Fr\'echet derivative and its adjoint. For any pair of functions $\varphi,\tilde{\varphi} \in \mathcal{E}$, and $\psi \in L^2_{U\times W}$, we can write
\begin{align*}
\left( \hat{T}^\prime_\varphi \tilde{\varphi}\right)(u,w) =& \int \hat{t}(\varphi (x) + u,x,w) \tilde{\varphi} (x) dx \\
\left( \hat{A}^{\ast}_\varphi \psi\right)(x) =& \int \int \int \hat{t}^\ast(\varphi (\xi,x_{-1}) + u,x,w) \left(\psi (u,w) - \frac{1}{n} \sum_{i = 1}^n \psi (u,W_i) \right) d\xi du dw,
\end{align*}
where we take
\begin{align*}
\hat{t}(\varphi (x) + u,x,w) =& \hat{f}_{Y,X\vert W}(\varphi (x) + u ,x\vert w) - \hat{f}_{Y,X}(\varphi (x) + u,x)\\
\hat{t}^\ast(\varphi (\xi,x_{-1}) + u,x,w) =& \frac{1}{\hat{f}_X (x)} \un(\xi \geq x_1) \hat{f}_{Y,X,W} (\varphi (\xi,x_{-1}) + u, \xi,x_{-1},w) \hat{f}_{Y - \varphi(X)}(u),
\end{align*}
to be the kernels of $\hat{T}^\prime_\varphi$ and $\hat{A}^{\ast}_\varphi$, respectively. Since $W$ is a discrete variable, integrals can be replaced by sums, where appropriate, and we use the integral notation without loss of generality. 

The marginal density of $W$ is estimated using a discrete kernel $L(\cdot)$ with bandwidth $h_w$ such that
\[
\hat{f}_W(w) = \frac{1}{n} \sum_{i = 1}^n L_{h_w}(w - W_i).
\]
The properties of this function are given in Assumption \ref{assapp3b} \citep[see][]{aitch1976,liracine2007}. We only note here that when the bandwidth $h_w = 0$, $L(\cdot)$ reduces to the product of indicator functions, and $\hat{f}_W(w)$ is the usual frequency estimator. Then we let
\begin{align*}
\hat{f}_X(x) =& \frac{1}{n h_x^p} \sum_{i = 1}^n C_{h_x}\left(x-X_i,x\right)\\
\hat{f}_U(u) =& \frac{1}{n h_u} \sum_{i = 1}^n C_{h_u}\left(u-U_i\right)\\
\hat{f}_{Y,X}(y ,x) =& \frac{1}{n h_x^{p}h_u} \sum_{i = 1}^n C_{h_u}\left(y-Y_i\right)C_{X,h_x}\left(x-X_i,x\right) \\
\hat{f}_{Y,X,W} (y, x,w) =& \frac{1}{n h_x^{p}h_u} \sum_{i = 1}^n C_{h_u}\left(y-Y_i\right)C_{X,h_x}\left(x-X_i,x\right) L_{h_w}(w - W_i)\\
\hat{f}_{Y,X \vert W} (y,x\vert w) =& \frac{\hat{f}_{Y,X,W} (y, x,w)}{\hat{f}_W(w)},
\end{align*}

We choose the points $u$ in expanding sets of the form $\{ u: \vert u \vert \leq \ell_n\}$, where $\ell_n$ is a sequence that is either bounded or diverging slowly to infinity so that we can allow for error distributions with unbounded support \citep[see][]{hansen2008}. Finally, we make additional assumptions about the estimated operators and their convergence, and we appropriately restrict the behavior of the tuning parameters.

\begin{assumption}\label{ass:frediff}
$\hat{T}$ is Fr\'echet differentiable for all $\varphi \in \mathcal{E}^\prime$.
\end{assumption}

\begin{assumption} \label{ass:rateconv}
For $\vdagger \in \mathcal{E}^\prime$, with $h_w = O(n^{-1})$, we have that
\begin{align*}
E \Vert \hat{T}(\vdagger) - T(\vdagger) \Vert^2 =& O\left( \frac{1}{n} + h_u^{2\rho} \right) = O(\delta^2_n) \\
E \Vert \hat{A}^{\ast}_{\vdagger} - A^{\ast}_{\vdagger} \Vert^2  =& O\left( \frac{1}{nh^p_x} + h_x^{2\rho} \right)  = O(\gamma^2_n),
\end{align*}
with $\delta_n,\gamma_n \rightarrow 0$, as $n \rightarrow \infty$. 
\end{assumption}

\begin{assumption}[Tuning parameters]~ \label{ass:restrvarth}
Let $\epsilon \in (0,1/2)$, and $$\kappa_n = \inf_{\{ u: \vert u \vert \leq \ell_n\}} f_U(u).$$ The tuning parameters satisfy the following restrictions
\begin{itemize}
\item[(i)] $\delta_n N = O(1)$ and $\gamma_n N^{\frac{\epsilon + 1}{2}} = o(1)$.
\item[(ii)] 
\begin{equation*}
\frac{(\delta_n \vee \gamma_n) \sqrt{N}}{h_u^2 \kappa_n } = o(1).
\end{equation*}
\item[(iii)]
\begin{equation*}
\frac{N^{-\epsilon/2}}{h^2_u \kappa_n} = O(1).
\end{equation*}
\end{itemize}
\end{assumption}

Assumption \ref{ass:frediff} guarantees that we can locally control the behavior of the estimated operators. Assumption \ref{ass:rateconv} gives the rate of convergence of the operators in Mean Integrated Squared Error (MISE) \citep{li2008,darolles2011,florens2012}. As all conditioning IVs are discrete, and following \citet{li2008}, the bias term only depends on $h_w$, and it can be made asymptotically negligible by choosing $h_w = O(n^{-1})$. This result does not depend on the dimension of $W$. Similarly, one can choose $h_u = O(n^{-1/2\rho})$ so that $\hat{T}(\varphi)$ has parametric rates of convergence for all $\varphi \in \mathcal{E}^\prime$. $\hat{A}^{\ast}_\varphi$ is the partial integral of a conditional expectation operator, and its rate depends on the dimension of the conditioning variable. Its estimation also depends on $\hat{f}_{U}$, the density of the error term at $\varphi$. The latter is negligible as long as $h_u = o(1)$, and $nh^p_x \rightarrow \infty$. This result can be proven using a standard argument for U-statistics, and it is omitted here for brevity. In \citet{darolles2011}, with $W$ continuous, the variance term depends on $h_w^q$. However, with $W$ discrete, the variance term does not depend on $h_w$, and the bias generated by smoothing with respect to the discrete instruments is negligible when $h_w = O(n^{-1})$. We specialize this example in Section \ref{sec:mcsim} to the case of a binary instrument $W$. Assumption \ref{ass:restrvarth} imposes restrictions on the tuning parameters. Proposing a data-driven procedure for the choice of these parameters is an essential step to be pursued in future research. 

\begin{proposition}\label{prop:koper}
Let Assumptions \ref{ass:identification}-\ref{ass:frediff} and \ref{assapp1}-\ref{assapp5} hold. Then we have the following:
\begin{enumerate}
\item[(a)] There exists a positive constant $\hat{M} < \infty$, such that $$\Vert \hat{T}(\hat\varphi_j) - \hat{T}(\vdagger) - \hat{T}^\prime_{\vdagger}\left( \hat\varphi_j - \vdagger \right) \Vert \leq \hat{M} \Vert \hat\varphi_j - \vdagger \Vert^2$$
almost surely.
\item[(b)] Let $L^2_{\hat{U}_j \times W}(x)$, be the space of square-integrable functions with respect to the joint distribution of $(\hat{U}_j,W)$ given $X = x$. There exists an operator $K_{\vdagger,\hat\varphi_j}: L^2_{\hat{U}_j \times W}(x) \rightarrow L^2_{U \times W}(x)$ defined as 
\[
\left( K_{\vdagger,\hat\varphi_j} \psi \right)(v,w) = \frac{\psi(v - \hat\varphi_j(x) + \vdagger(x),w) \hat{f}_{\hat{U}_j}(v - \hat\varphi_j(x) + \vdagger(x))}{\hat{f}_U(v)},
\]
where $\psi \in L^2_{\hat{U}_j\times W}(x)$, such that
\[
\left( \hat{A}^{\ast}_{\hat\varphi_j} \psi\right)(x) = \left( \hat{A}^{\ast}_{\vdagger} K_{\vdagger,\hat\varphi_j} \psi\right)(x),
\]
and
\begin{align*}
\Vert & K_{\vdagger,\hat\varphi_j} - I\Vert \leq M h^{-2}_u \kappa^{-1}_n \Vert \hat\varphi_j - \vdagger \Vert,
\end{align*}
\end{enumerate}
where $I$ is the identity operator, and $M = M_2M_3$, as defined in Assumption \ref{ass:scalab}.
\end{proposition}

Proof of this Proposition is in Section \ref{annexA3} of the Appendix. This Proposition restricts the local behavior of the Fr\'echet derivatives to ensure convergence of the LF algorithm. It also shows that we have to pay a price to approximate the density of the unknown error term by the density of the regression residuals. This is tantamount to the usual estimation error with nonparametric generated regressors \citep{mammen2012}. When the density of the error term is uniformly bounded away from $0$ over its support, $\kappa_n$ is a constant. However, as we want to allow for distributions with unbounded support, we let $\kappa_n$ converge to $0$ as $n$ approaches $\infty$.

The following Theorem contains the main result of this Section.

\begin{theorem}\label{thm:mainconv}
Let Assumptions \ref{ass:identification}-\ref{ass:restrvarth}, and \ref{assapp1}-\ref{assapp6} hold for $\beta_\epsilon = \beta \wedge (1 - \epsilon)$, with $\epsilon > 0$, as defined in Assumption \ref{ass:restrvarth}(iii). Then,
\begin{align*}
\left( E\Vert \hat\varphi^\prime_N - \vdagger^\prime \Vert^2\right)^{1/2} =& O\left( \delta_n \sqrt{N} + \gamma_n N^{\frac{1-\beta}{2}} \ln(N)^{\un(\beta = 2)} N^{\left( \frac{\beta}{2}-1 \right)\vee 0} \right.\\
& \quad \left. +N^{-\beta_\epsilon/2} + \left( E || \hat\varphi_0 - \varphi_0 ||^2 \right)^{1/2} \right),\\
\left( E\Vert \hat\varphi_N - \vdagger \Vert^2\right)^{1/2} =& O\left( \delta_n N^{\frac{a}{2(a + 1)}} + \gamma_n N^{\frac{a}{2(a + 1)} -\frac{\beta}{2}} \ln(N)^{\un(\beta = 2)} N^{\left( \frac{\beta}{2}-1 \right)\vee 0} \right.\\
& \quad \left. + N^{-\beta_\epsilon/2 - 1/(2(a + 1))} + N^{-\frac{1}{2(a + 1)}}\left( E || \hat\varphi_0 - \varphi_0 ||^2 \right)^{1/2} \right).
\end{align*}
\end{theorem}

The result of this Theorem gives an upper bound on the Mean Integrated Squared Error (MISE) of our estimator. Loosely speaking, the MISE depends on a regularization bias, $N^{-\beta_\epsilon/2}$, which decreases with the number of iterations, $N$, and varies with the regularity of the function, $\beta$; and a \textit{variance} term, which is a function of the estimation error, and increases as $N$ diverges. The restriction $\beta \leq 1$ for LF regularization in nonlinear ill-posed inverse problems is not original to us. It is well-established in the mathematical literature that the regularization bias at iteration $j+1$ is of order $(j+1)^{-\beta/2}$. This regularization bias accumulates across $N$ iterations, and, for $\beta$ sufficiently large, it would converge to a non-zero constant as $N\rightarrow \infty$. This could be considered a \textit{saturation} effect for LF regularization in the context of nonlinear ill-posed inverse problems \citep[see][for a discussion of the saturation effect in linear inverse problems with Tikhonov regularization]{carrasco2007h}. Moreover, there is an additional saturation effect on $\beta$ in our case, as we need the regularization bias to go to zero fast enough so that the additional error from the nonparametric estimation of the unknown density of the residuals is negligible. Therefore, we need $\beta \leq 1-\epsilon$, where $\epsilon > 0$ satisfies Assumption \ref{ass:restrvarth}(iii). For $\beta > 1 - \epsilon$, we simply cannot take full advantage of the source condition in Assumption \ref{ass:sourcecond}, and the bias converges at a lower speed. Under this restriction on $\beta$, the upper bound is the same one we would get in the context of a linear ill-posed inverse problem with H\"older source condition \citep{florens2012}. Finally, the estimation of the first derivative allows us to extend our result to more regular functions, as the order of convergence of $\hat{\varphi}_N$ to $\vdagger$ is improved by integration. This result could be further extended by considering a general approach to LF regularization in Hilbert scales \citep[][p. 30]{kalten2008}. 

Finally, the rate of convergence depends on the approximation properties of the initial condition. One could choose $\hat\varphi_0$ in a way that it converges to $\varphi_0$ at a rate faster than $\delta_n \sqrt{N} \vee \gamma_n N^{\frac{1-\beta}{2}}$. This is satisfied by parametric regression models and nonparametric regressions.

There are several important differences between our result and the one in \citet{dunker2018}. First of all, the result in \citet{dunker2018} is of more general scope than the one provided here and obtained under high-level regularity conditions. The Newton method used by \citet{dunker2018} does not suffer from the saturation effect of LF regularization, although its implementation is more cumbersome as it involves outer and inner iterations of the regularization scheme. Moreover, to the best of our understanding, \citet{dunker2018} directly treats $\hat{T}^{\prime}_{\hat\varphi_j}$ and $\hat{T}^{\prime}_{\vdagger}$ as operators with the same range and does not consider the estimation error that arises from the use of the regression residuals instead of the true error. In his running example of quantile IV regression, this approach seems appropriate. However, in our case, it is important to explicitly consider the nonlinearity in the approximation to provide guidance for the choice of tuning parameters.

Finally, \citet{dunker2018} provides conditions under which its rate is optimal, i.e.,
\[
\left( E\Vert \hat\varphi^\prime_N - \vdagger^\prime \Vert^2\right)^{1/2} = O\left( \delta^{\frac{\beta}{\beta+1}}_n\right).
\]
It is unclear whether our rates are optimal. When $\beta < 1-\epsilon$, the dominating variance term depends on $\beta$, and the relationship between $\delta_n$ and $\gamma_n$. When $\beta \geq 1-\epsilon$, there is a saturation effect that does not allow us to reach the optimal rate. We have the following. 

\begin{corollary}
Let
\begin{equation} \label{eq:gammadeltan}
\gamma_n N^{\frac{1-\beta}{2}} \ln(N)^{\un(\beta = 2)} N^{\left( \frac{\beta}{2}-1 \right)\vee 0} =o \left( \delta_n \sqrt{N} \right).
\end{equation}
Then, $N \asymp \delta_n^{-\frac{2}{\beta_\epsilon + 1}}$, and
\begin{align*}
\left( E\Vert \hat\varphi^\prime_N - \vdagger^\prime \Vert^2\right)^{1/2} =& O\left( \delta_n^{\frac{\beta_\epsilon}{\beta_\epsilon + 1}} \right)\\
\left( E\Vert \hat\varphi_N - \vdagger \Vert^2\right)^{1/2} =& O\left( \delta_n^{\frac{\beta_\epsilon (a + 1 ) + 1}{( \beta_\epsilon + 1)(a + 1)}} \right). 
\end{align*}
\end{corollary}
Provided the variance term $\delta_n \sqrt{N}$ dominates, the result of this Corollary is obtained by balancing the variance and the squared bias terms in Theorem \ref{thm:mainconv}. E.g., if $\beta= 1$, condition \ref{eq:gammadeltan} is satisfied if $\gamma_n =o \left( \delta_n \sqrt{N} \right)$. When $\beta \leq 1 - \epsilon$, i.e. $\beta_\epsilon = \beta$, we can reach the optimal rate. In particular, with $\beta = (s-1)/(a+1)$, we have
\begin{align*}
\left( E\Vert \hat\varphi^\prime_N - \vdagger^\prime \Vert^2\right)^{1/2} =& O\left( \delta_n^{\frac{s-1}{s + a}} \right)\\
\left( E\Vert \hat\varphi_N - \vdagger \Vert^2\right)^{1/2} =& O\left( \delta_n^{\frac{s}{s + a}} \right).
\end{align*}
However we cannot reach the optimal rate for any $\beta > 1 - \epsilon$. Whether LF regularization in Hilbert scales can reach the optimal rate for nonlinear ill-posed problems is an interesting question to be pursued in future research.

\section{Estimation with a binary instrument. Simulations and example} \label{sec:mcsim}

We consider in this Section a particular example where $X\in \mathbbm{R}$ is a continuous variable and $W$ is a binary instrument. That is, $W\in \{0,1\}$.

Recall that, for simplicity, we take $E(Y) = E(\vdagger) = 0$. Under mean independence, the model is not identified. The restriction $E(U\vert W) =0$ reduces to two conditions
\[
\int \vdagger (x) f_{X\vert W} (x\vert w=0) dx = \int \vdagger (x) f_{X\vert W} (x\vert w=1) dx =0
\]
which cannot imply $\vdagger=0$, except when $X$ is also binary or when $\vdagger$ is a two-parameter function in $X$. 

When the completeness condition fails, a smooth regularized estimator of $\varphi$ obtained from the sample counterpart of $E(Y\vert W) = E(\vdagger(X)\vert W)$ has a well-defined probability limit, which is the pseudo-true value
\[
\vdagger^{P}(x) = \lambda_0 \frac{f_{X\vert W} (x\vert w=0)}{f_X(x)} + \lambda_1 \frac{f_{X\vert W} (x\vert w=1)}{f_X(x)},
\]
defined as the projection of $\vdagger$ on the orthogonal complement of the null space of the conditional expectation operator. $\lambda_0$ and $\lambda_1$ are the unique solutions of the system
\[
\int \vdagger(x) f_{X\vert W} (x\vert w=l)dx = \sum_{j=0,1} \lambda_j\int \frac{f_{X\vert W} (x\vert w=l) f_{X\vert W} (x\vert w=j)}{f_X (x)} dx, \text{ for } l = \{0,1 \}.
\]

We provide proof of this result in Section \ref{annexA4} of the Appendix and refer to \citet{babii2017a} for a more general framework. Then to estimate $\vdagger$ using $W$ as an instrument, we require additional assumptions, and the assumption $U\upmodels W$ with $E(U) = 0$ is convenient for this purpose. Nonetheless, we need some auxiliary completeness conditions to obtain identification.

Let us consider first the local condition in Assumption \ref{ass:condcomp}. In this case, it is equivalent to
\[
E(\vdagger (X)\vert U=u, W=1) \overset{a.s.}{=} E(\vdagger(X) \vert U=u, W=0) \Rightarrow \vdagger \overset{a.s.}{=} 0.
\]
We can, therefore, write the following decomposition
\[
\frac{f_{U,X\vert W} (u,x\vert j)}{f_{X\vert W} (x\vert j) f_U (u)} = \sum_{k=1}^{\infty}\lambda^{(j)}_{k} \chi_{k,j} (x) \psi_{k,j} (u) = \sum_{k=1}^{\infty}\lambda^{(j)}_{k} \chi_{k} (x) \psi_{k} (u),
\]
where the functions $\chi$ and $\psi$ form an orthonormal basis in $L^2_{U\times X}$, and they do not depend on $j$ (this is for the sake of exposition and slightly generalizes our Example \ref{ex:example1}, where the densities of $(X,U) \vert W = 0$ and $(X,U) \vert W = 1$ have the same eigenfunctions).

Assumption \ref{ass:condcomp} in this context can be interpreted as stating that if the singular value decomposition of the conditional expectation operators is the same when $W$ is equal to $0$ or to $1$, then the instrument does not have any identifying power. This is true, whenever $\vert \lambda^{(0)}_{k}\vert=\vert\lambda^{(1)}_{k}\vert$, for all $k \geq 0$. On the contrary, whenever $\vert \lambda^{(0)}_{k} \vert \neq \vert \lambda^{(1)}_{k}\vert$, for all $k \geq 1$, then the condition in Assumption \ref{ass:condcomp} is true only if $\vdagger\overset{a.s.}{=}0$. Hence, the model is locally identified if the dependence structure between $X$ and $U$, as captured by the sequence $\lambda^{(j)}_{k}$, is different for $W=1$ and $W=0$.

We now consider estimation in the particular case of binary instruments. First, consider a parametric approach with
\[
\varphi(x) = b(x)^\prime \theta,
\]
where $\theta \in \IR^k$ and $b$ is a given vector of $k$ basis functions (polynomials, splines, etc.). This model is parametric if $k$ is fixed but may be viewed as a sieve estimation if $k$ is allowed to increase with $n$, although we do not pursue this interpretation further. The parametric estimation is based on the exogeneity restriction
\[
P(U\leq u\vert W=1) = P(U\leq u\vert W=0).
\]
Let us define
\begin{equation}
U_i (\theta) =Y_i - b(X_i) \theta - \left(\frac{1}{n} \sum^n_{j=1} (Y_i - b(X_i) \theta \right)
\end{equation}
and we estimate $P(U\leq u\vert W=j)$ by
\begin{equation} \label{eq:estcondcdf}
\hat{F}_{\theta} (u\vert j) = \frac{1}{n_j} \sum^n_{i=1} \bar{C}_{h_u} \left(u-U_i (\theta)\right) \un (W_i = j)
\end{equation}
where $\bar{C}$ is the cdf of a kernel function of order $\rho \geq 2$, $\un$ is the indicator function, $h_u$ is a bandwidth, and $n_j = \sum^n_{i=1} \un (W_i =j)$.

We minimize the Cram\'er-von Mises distance between the two conditional cdfs with respect to $\theta$
\begin{equation}
\sum_l (\hat{F}_\theta (u_l\vert 1) - \hat{F}_{\theta} (u_l\vert 0))^2
\end{equation}
for a suitable grid of points $u_l$. An example of the application of this method is provided below.

A more flexible method (not dependent on a parametric functional form) follows from the application of the LF algorithm described in Section \ref{sec:estimation}. One detail specific to the implementation with a scalar binary instrument is that we do not smooth the conditional cdf with respect to $W$, but we sort the data for $W=0$ and $W=1$. The Fr\'echet differentiability of the operator $T$ holds as long as the joint pdf of $(Y,X)$ given $W = w$ satisfies Assumption \ref{ass:differentiability} \citep[see also][]{dunker2014,dunker2018}.

Upon an appropriate choice of the bandwidth parameter, i.e., $h_u \asymp n^{-1/2\rho}$, we can take $\delta_n \asymp n^{-1/2}$ in Assumption \ref{ass:rateconv}. Similarly, for $A^\ast_\varphi$, we write
\[
\left( \hat{A}^\ast_{\vdagger} \psi \right)(x) = \int_{-\infty}^x \int \frac{\left[\psi(u,1) \hat{f}_{Y,X,W}(\vdagger(x) + u,\xi,1) - \psi(u,0) \hat{f}_{Y,X,W}(\vdagger(x) + u,\xi,0) \right]\hat{f}_{U}(u)}{\hat{f}_{X}(x)} dud \xi,
\]
with $\psi \in L^2_{U\times W}$, $\hat{f}_{Y,X,\cdot}(\vdagger(x) + u,x,\cdot)$, $\hat{f}_{X}(x)$ nonparametric estimators of the densities of observable components of the model, as described in Section \ref{sec:asymprop}, with bandwidth parameters $(h_x,h_u)$; and $\hat{f}_{U}(u)$, nonparametric estimator of the density of the error term, with bandwidth $h_u$. Using a second order kernel for the estimation of $\hat{f}_{Y,X,\cdot}(\vdagger(x) + u,x,\cdot)$ and $\hat{f}_{X}(x)$, and using Assumption \ref{ass:rateconv}
\[
E \Vert \hat{A}^\ast_{\vdagger} - A^\ast_{\vdagger} \Vert^2 = O\left( \frac{1}{nh_x} + h_x^{4}\right),
\]
so that, upon an appropriate choice of the bandwidth parameter, i.e., $h_x \asymp n^{-1/5}$, we can take $\gamma_n = n^{-2/5}$ in Assumption \ref{ass:rateconv}. Both Silverman's rule-of-thumb and leave-one-out cross-validation satisfy these requirements. We use the former in our simulation study and empirical application to speed up computations.

To satisfy the conditions of Assumption \ref{ass:restrvarth}(ii), we then require
\[
\frac{n^{-\frac{2}{5}  + \frac{1}{\rho}}\sqrt{N}}{\kappa_n} = o(1),
\]
because $\delta_n \vee \gamma_n = \gamma_n$. Let $\kappa_n \asymp n^{-\nu}$, with $\nu > 0$, we can pick $N$ such that
\begin{equation}\label{eq:restrvar}
N \ln(N) \asymp n^{\frac{2\rho(2 - 5\nu) - 10}{5\rho}}.
\end{equation}

A sufficient restriction for this condition to hold is
\[
\rho > \frac{5}{2 - 5\nu},
\]
with $\nu \in (0,2/5)$, which implies it may be advisable to take higher-order kernels for the estimation of the cdf and pdf of $\hat{U}$. 

The condition on the bias term given in Assumption \ref{ass:restrvarth}(iii) has a more cumbersome interpretation, as it depends on the unknown parameters $\epsilon$ and $\kappa_n$. However we would like to choose $h_u$ that goes to $0$ as slowly as possible. In this respect, using higher-order kernel should also help reduce the bias of our estimator. Moreover, the order of the bias depends nonlinearly on $\beta$, and it is, therefore, more challenging to obtain an optimal order for $N$ based on the usual squared bias-variance trade-off. We thus propose to select the maximum number of iterations, $N_{max}$, according to equation \eqref{eq:restrvar}. This rule applies up to a constant, which is inversely proportional to the value of $c$ chosen (the smaller $c$, the higher the number of iterations we need for convergence), and we increase it with the variation in the dependent variable $Y$. In practice, we take this constant to be equal to $C_N = \alpha \left( Y_{max} - Y_{min} \right)/c$, with $\alpha>0$. We further take $\rho = 8$. As $\nu$, the decay at the tails of the distribution of $U$, is unknown in practice, we take it to be equal to $0$ (which implicitly assumes compact support), and then floor the maximum number of iterations $N_{max}$. That is, we take
\[
N_{max} \ln(N_{max}) = \lfloor C_N n^{\frac{4 \rho - 10}{5\rho}} \rfloor = \lfloor C_N n^{\frac{11}{20}} \rfloor,
\]
where, for a scalar $a$, $\lfloor a \rfloor$ denotes the integer smaller than or equal to $a$. This choice of $N_{max}$ avoids overfitting whenever the error term does not have bounded support. For every $N \leq N_{max}$, we proceed as described in Section \ref{sec:estimation}: we compute the Euclidean norm of $\hat{T}(\hat\varphi_{j})$ at every iteration $j$ and take $N_0$ as the argmin of $\Vert \hat{T}(\hat\varphi_{j})\Vert^2$, for $j \in [1, N_{max}]$.

\subsection{Finite sample properties} The finite sample properties are illustrated via the following simulations. The instrument $W_i\in \{0,1\}$ with probabilities $\left(\frac{1}{3}, \frac{2}{3}\right)$ and $U_i \sim N(0,1)$. The endogenous variable is given by
\[
X_i = 1 + 0.5 U_i - 0.1 U_i^2 + (2 + 0.5 U_i - 0.1 U_i^2) W_i + \varepsilon_i,
\]
where $\varepsilon_i \sim N(0,1)$. The dependent variable is defined as follows
\begin{align*}
Y_i=& -1.5 X_i+0.3X_i^2+U_i \tag{$DGP_1$},\\
Y_i =& 1.5\sin(0.5\pi X_i) + U_i \tag{$DGP_2$}. 
\end{align*}

In both DGPs, the noise-to-signal ratio is about equal to one. We run $1000$ simulations for each DGP. The initial condition, $\hat\varphi^\prime_0$, is selected in three different ways.
\begin{enumerate}
\item[1.] $\hat\varphi^\prime_{0,1}$: first derivative of the nonparametric local linear regression of $Y$ on $X$.
\item[2.] $\hat\varphi^\prime_{0,2}$: first derivative of the nonparametric control function estimator obtained under the assumption that $X = g_2(W,V)$, with $V \upmodels W$. The control function is estimated as the empirical cdf of $X$ conditional of $W$, and $\hat\varphi^\prime_{0,2}$ is obtained by one-step backfitting. In the first step, one estimates $E\left[ U \vert V \right]$ by marginal integration of a nonparametric local linear regression of $Y$ on $(X,\hat{V})$. In the second step, one estimates $\hat\varphi^\prime_{0,2}$ using a local linear regression of $Y - \hat{E}\left[ U \vert \hat{V} \right]$ on $X$.
\item[3.] $\hat\varphi^\prime_{0,3}$: Two-stage least squares (TSLS) estimator.
\end{enumerate}

All starting values are inconsistent estimators of $\vdagger$. The control function assumption is not satisfied in our simulations, as there is no monotone relation between $X$ and $U$. TSLS is the furthest from the true regression function, and it serves as a benchmark to assess the performance of our estimator with a poorly chosen starting value.

\subsection{Monte-Carlo with $n = 1000$}

\begin{knitrout}\scriptsize
\definecolor{shadecolor}{rgb}{0.969, 0.969, 0.969}\color{fgcolor}\begin{figure}[!h]

{\centering \includegraphics[width=0.48\textwidth,height=7cm,]{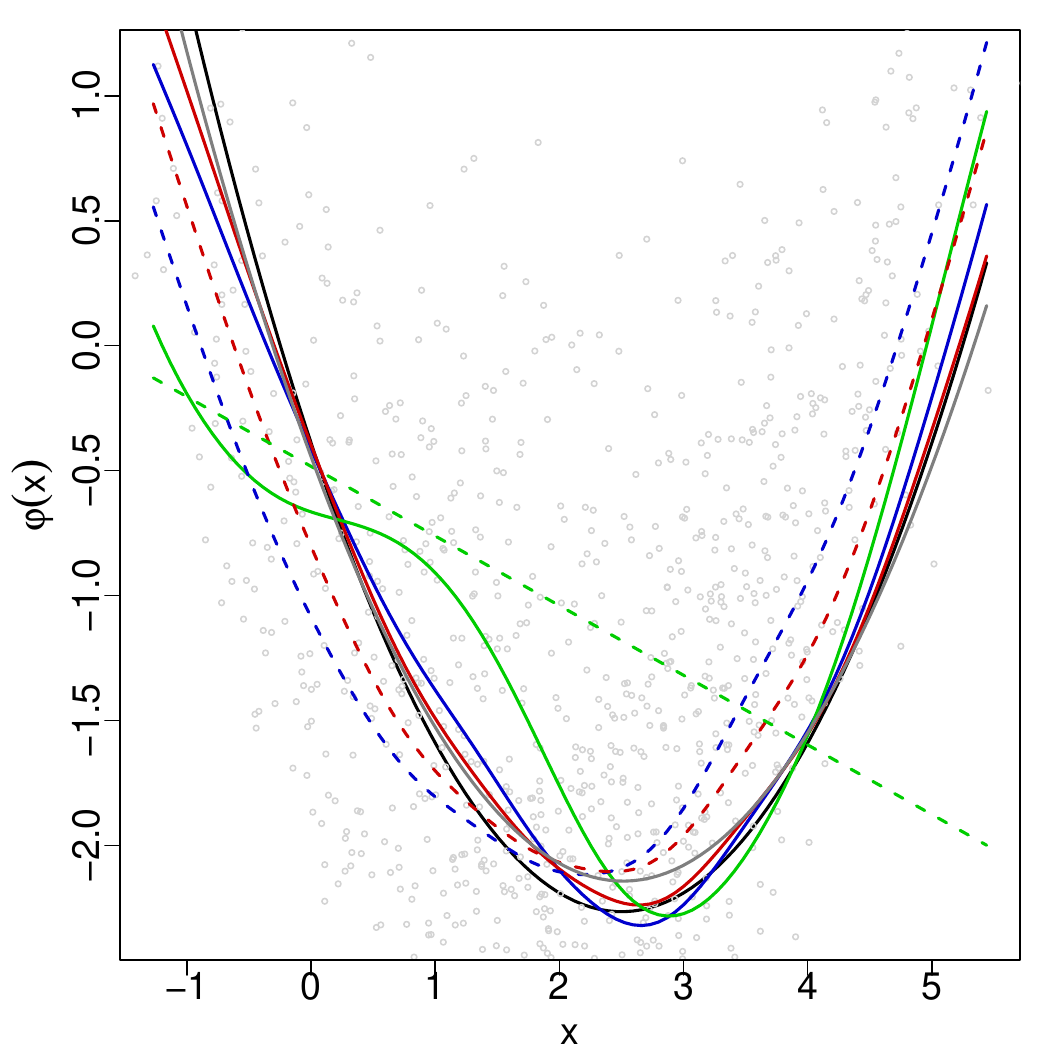} 
\includegraphics[width=0.48\textwidth,height=7cm,]{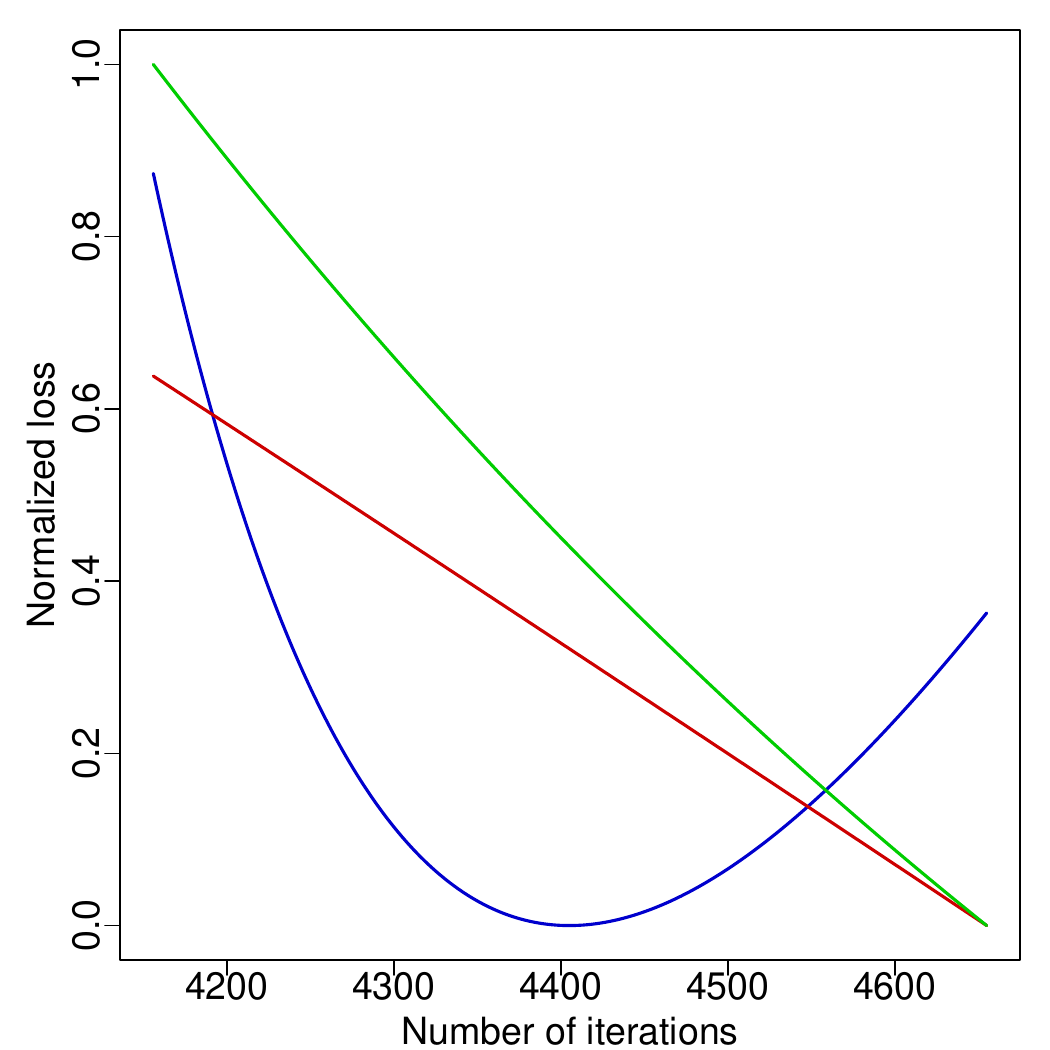} 
\includegraphics[width=0.72\textwidth,height=1cm,trim=0cm 8cm 0cm 8cm,clip=true]{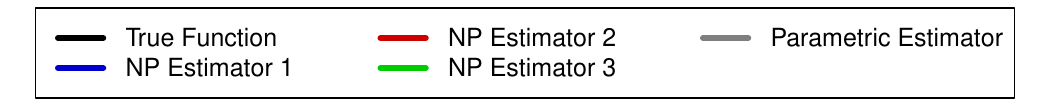} 

}

\caption[Results for a randomly chosen sample of $n=1000$ observations satisfying $DGP_1$]{Results for a randomly chosen sample of $n=1000$ observations satisfying $DGP_1$.}\label{fig:fct_sim1}
\end{figure}

\end{knitrout}

We start by presenting the results for $DGP_1$. The left panel of Figure \ref{fig:fct_sim1} shows the sample, the true curve, and the parametric and nonparametric estimators for one sample of simulated data. The dotted lines represent the starting values of our iterative procedure. The solid lines are the LF estimators obtained using each of the initial conditions, and the gray line is the correctly specified parametric estimator. Our nonparametric estimator behaves reasonably well compared to the fully parametric estimator. On the right panel, we show the decline of the Euclidean norm of $\hat{T}(\varphi)$ normalized in the interval $[0</->,1]$ for the last $500$ iterations, using different starting values, with $N_{max} = 4655$. For starting value 1, the stopping rule reaches its minimum at $N_0=1465$. Meanwhile, for starting values 2 and 3, the stopping rule is binding. Increasing the number of iterations would have further increased the variance of our estimator without necessarily improving the bias, as discussed above.

\begin{knitrout}\scriptsize
\definecolor{shadecolor}{rgb}{0.969, 0.969, 0.969}\color{fgcolor}\begin{figure}[!h]

{\centering \includegraphics[width=0.48\textwidth,height=7cm,]{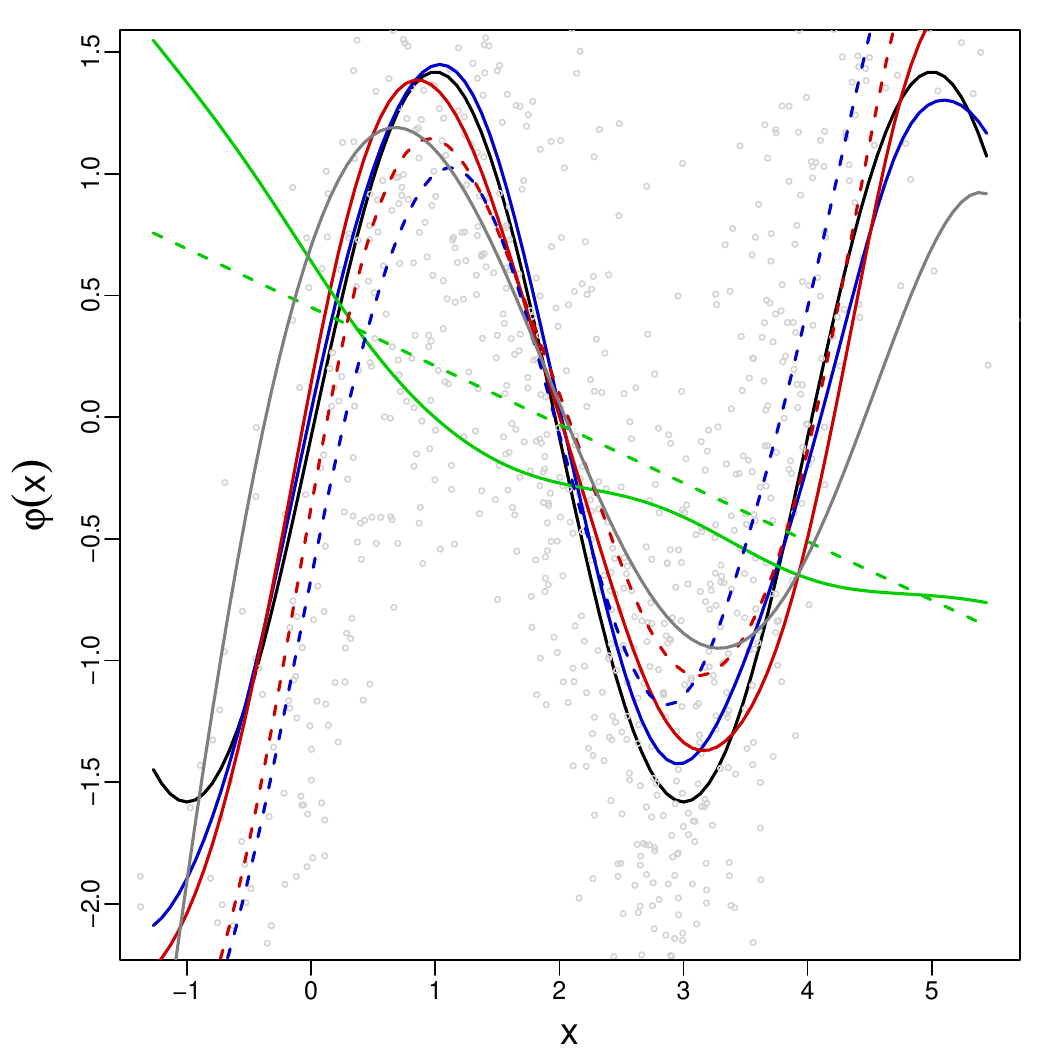} 
\includegraphics[width=0.48\textwidth,height=7cm,]{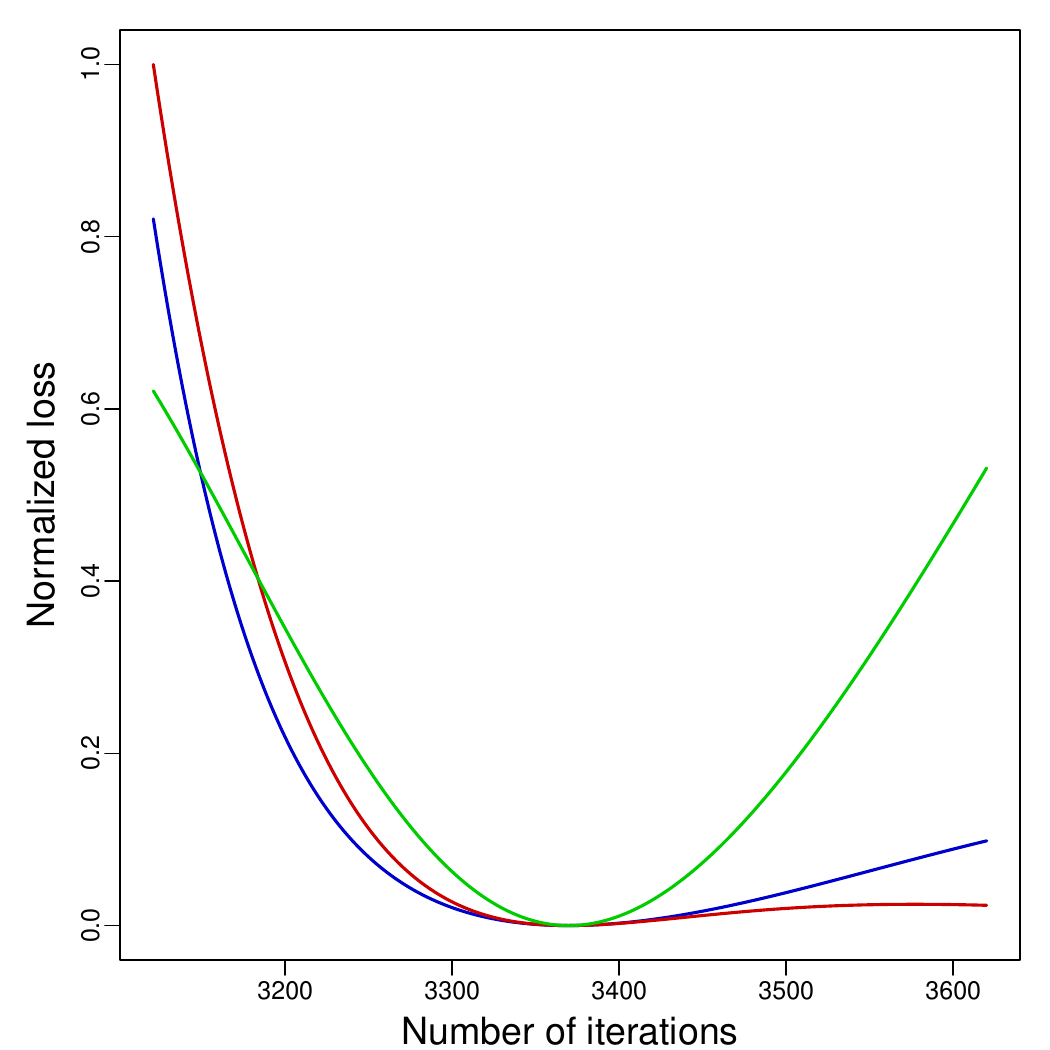} 
\includegraphics[width=0.72\textwidth,height=1cm,trim=0cm 8cm 0cm 8cm,clip=true]{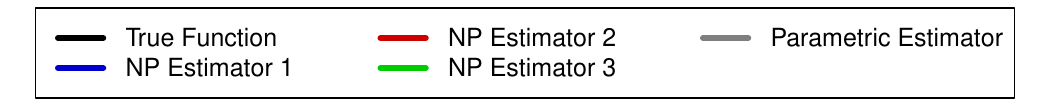} 

}

\caption[Results for a randomly chosen sample of $n=1000$ observations satisfying $DGP_2$]{Results for a randomly chosen sample of $n=1000$ observations satisfying $DGP_2$.}\label{fig:fct_sim2}
\end{figure}

\end{knitrout}

For $DGP_2$, we have similar results for starting values 1 and 2, but they substantially differ when starting from 3. The left panel of Figure \ref{fig:fct_sim2} shows the sample, the true curve, and the nonparametric estimator for one sample of simulated data (where the dotted lines are the starting values of our iterative procedure). Because of the approximation error, the parametric estimator is farther from the true value and has a substantial bias. The nonparametric procedure appears to behave better in this case. On the right panel, we show the behavior of the empirical square norm of $\hat{T}(\varphi)$, which reaches its minimum at $N_0 = 944$ for starting value 1, and at $N_0= 443$ for starting value $2$. For starting value $3$, we do not obtain a consistent estimator of $\vdagger$. $N_0 = 263$, and the final estimator is a noisy version of the starting value, and the norm of $\hat{T}(\hat\varphi_j)$ diverges as $j$ grows. This last case exemplifies the scenario in which a \textit{bad} initial condition is chosen, and the LF algorithm does not converge.

\begin{knitrout}\scriptsize
\definecolor{shadecolor}{rgb}{0.969, 0.969, 0.969}\color{fgcolor}\begin{figure}[!h]

{\centering \includegraphics[width=0.48\textwidth,height=7cm,]{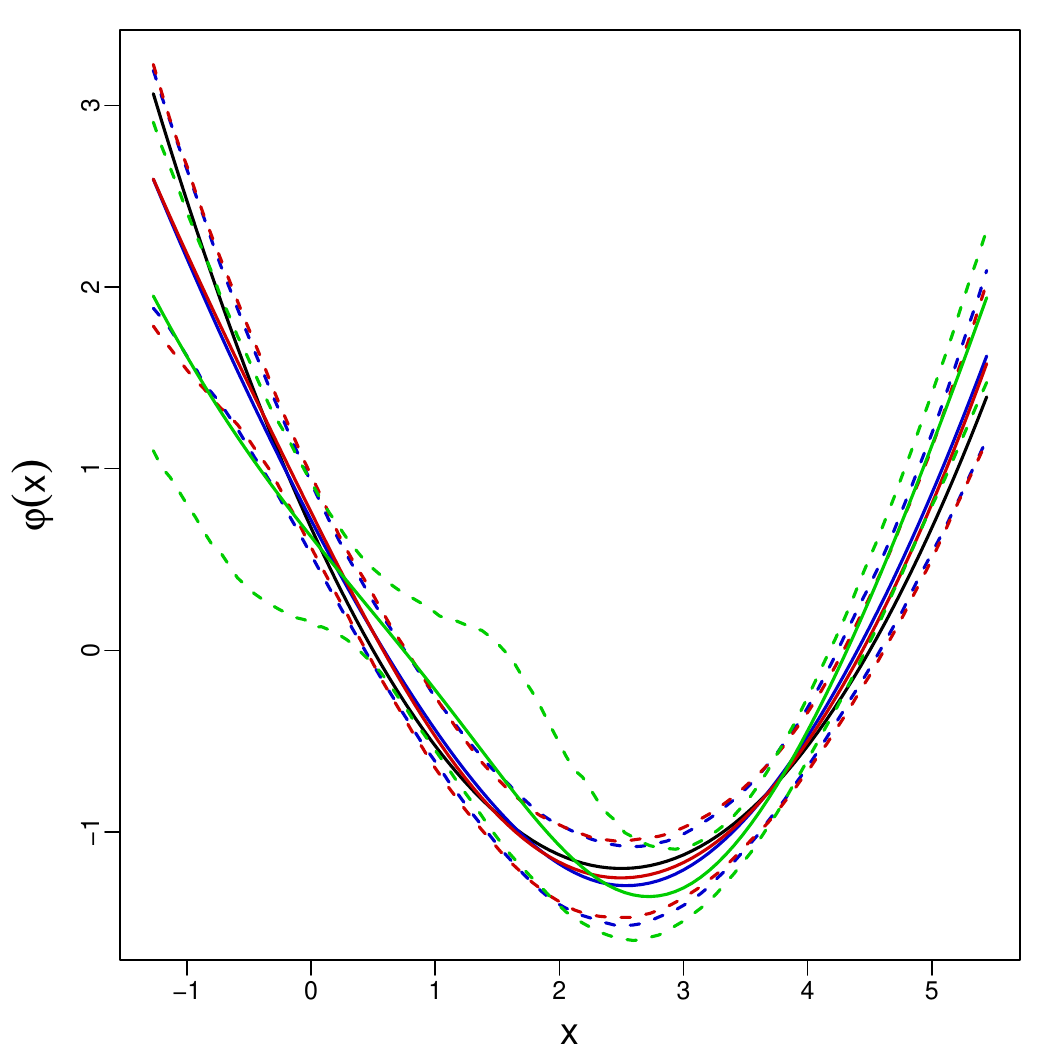} 
\includegraphics[width=0.48\textwidth,height=7cm,]{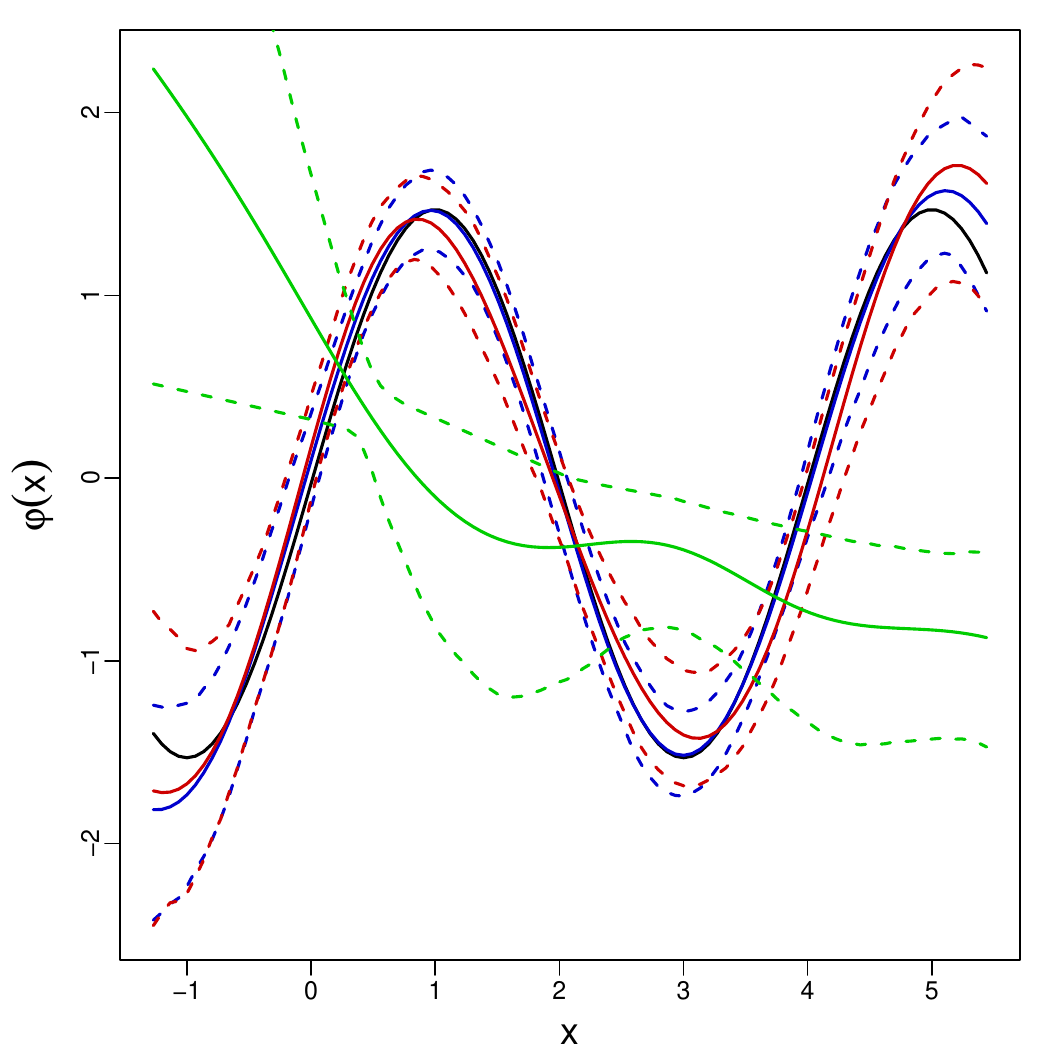} 
\includegraphics[width=0.72\textwidth,height=1cm,trim=0cm 8cm 0cm 8cm,clip=true]{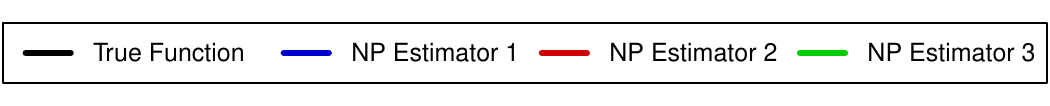} 

}

\caption[LF Monte-Carlo means and pointwise $2.5\%$ and $97.5\%$ quantiles for $DGP_1$ (left panel) and $DGP_2$ (right panel)]{LF Monte-Carlo means and pointwise $2.5\%$ and $97.5\%$ quantiles for $DGP_1$ (left panel) and $DGP_2$ (right panel).}\label{fig:fct_sim3}
\end{figure}

\end{knitrout}

The sampling distributions of both our nonparametric estimates are presented in Figure \ref{fig:fct_sim3}, where the dotted lines are the pointwise $2.5\%$ and $97.5\%$ quantiles of the values of the estimates over $1000$ simulation draws.

\subsection{Monte-Carlo with varying sample size} We also provide evidence of our estimator performance as the sample size increases. We take the same setting and number of simulations as above, with $n = \lbrace 500,1000,2000 \rbrace$. For each DGP and initial condition, we compute the Mean Integrated Squared Error (MISE). Results are reported in Table \ref{table:mc0}.

\begin{table}[!h]
\centering
\begin{adjustbox}{width=1\textwidth}
\begin{tabular}{l | c c c c c c | c c c c c c } \hline \hline
~ & \multicolumn{6}{c |}{$DGP_1$} & \multicolumn{6}{c}{$DGP_2$}\\ \hline
~ & $\hat\varphi_{0,1}$ & $med(N_{0,1})$ & $\hat\varphi_{0,2}$ & $med(N_{0,2})$ & $\hat\varphi_{0,3}$ & $med(N_{0,3})$ & $\hat\varphi_{0,1}$ & $med(N_{0,1})$ & $\hat\varphi_{0,2}$ & $med(N_{0,2})$ & $\hat\varphi_{0,3}$ & $med(N_{0,3})$ \\ \hline
  \hline
500 & 0.671 & 2260 & 0.556 & 2115 & 1.795 & 2520 & 1.080 & 1027 & 1.496 & 918 & 18.443 & 520 \\ 
  1000 & 0.246 & 3782 & 0.208 & 3718 & 1.160 & 4175 & 0.215 & 1276 & 0.458 & 974 & 20.891 & 482 \\ 
  2000 & 0.156 & 6280 & 0.145 & 6275 & 0.989 & 6715 & 0.179 & 5252 & 0.339 & 1422 & 21.228 & 378 \\ 
   \hline

\hline \hline 
\end{tabular}
\end{adjustbox}
\caption{\textit{MISE of the LF estimator and the median number of iterations for varying sample sizes.}}
\label{table:mc0}
\end{table}

The estimator behaves as expected with the MISE decreasing and the median optimal number of iterations increasing with the sample size. The only exception is when we initialize $DGP_2$ using the TSLS estimator. In this case, the estimator is not consistent, the MISE increases, and $N$ decreases with $n$.

\begin{table}[!h]
\centering
\begin{adjustbox}{width=1\textwidth}
\begin{tabular}{l | c c c c | c c c c } \hline \hline
~ & \multicolumn{4}{c |}{$DGP_1$} & \multicolumn{4}{c}{$DGP_2$}\\ \hline
~ & $h_{u,1}$ & $h_{u,2}$ & $h_{u,3}$ & $h_{x}$ &  $h_{u,1}$ & $h_{u,2}$ & $h_{u,3}$ & $h_{x}$ \\ \hline
  \hline
500 & 0.690 & 0.694 & 0.699 & 0.494 & 0.703 & 0.723 & 1.504 & 0.494 \\ 
   & [0.618 , 0.774] & [0.629 , 0.773] & [0.634 , 0.923] & [0.468 , 0.523] & [0.624 , 0.831] & [0.643 , 0.891] & [1.118 , 2.311] & [0.468 , 0.523] \\ 
  1000 & 0.663 & 0.667 & 0.665 & 0.430 & 0.673 & 0.691 & 1.667 & 0.430 \\ 
   & [0.612 , 0.719] & [0.621 , 0.722] & [0.620 , 0.798] & [0.413 , 0.447] & [0.617 , 0.747] & [0.633 , 0.790] & [1.100 , 2.868] & [0.413 , 0.447] \\ 
  2000 & 0.639 & 0.642 & 0.638 & 0.374 & 0.644 & 0.658 & 2.060 & 0.374 \\ 
   & [0.604 , 0.674] & [0.609 , 0.678] & [0.606 , 0.693] & [0.363 , 0.384] & [0.609 , 0.694] & [0.618 , 0.735] & [1.074 , 3.553] & [0.363 , 0.384] \\ 
   \hline

\hline \hline 
\end{tabular}
\end{adjustbox}
\caption{\textit{Monte-Carlo median and $95\%$ range of the bandwidth parameters.}}
\label{table:mc1}
\end{table}

In Table \ref{table:mc1}, we also report the median value and $95\%$ range of the chosen bandwidth parameters. We do not smooth with respect to $W$, and we use indicator functions instead. That is, $h_w = 0$ for all DGPs and sample sizes. We use a kernel of order $8$ for the residuals, which, in line with our theoretical results, leads to a rate for $h_u$ proportional to $n^{-1/16}$. The bandwidth $h_x$ is instead chosen to be proportional to $n^{-1/5}$. As $h_u$ is chosen according to the distribution of the residuals, its value changes depending on the initial condition taken. Moreover, for $DGP_2$, when starting from $\hat{\varphi}^\prime_{0,3}$, the LF algorithm does not converge. In this case, the median value of $h_u$ diverges instead of converging to $0$ at the chosen rate.

\section{Estimation of returns to education}\label{sec:empiricalapp}

We illustrate our methodology for estimating returns to education using data from the 1979 National
Longitudinal Survey of Youth \citep[NLSY, see][]{card1995,torgovitsky2017}. The model is the one used in the main body of the paper
\[
Y_{i} = \vdagger(X_{i}) + U_{i}, \text{ with } i = 1,\dots,n,
\]
where $Y$ denotes the logarithm of the wage; $X$ represents years of education, and $E\left[ U_{i} \vert X_{i}\right] \neq 0$. To address the concern about years of schooling being discrete, we add an idiosyncratic $Unif[-1,1]$ noise to each observation. As an instrument, $W$, we use a dummy variable equal to $1$ if the individual grew up near an accredited four-year college and $0$ otherwise. The sample is restricted to the $n = 939$ individuals older than 29 and who have completed at least $8$ years of education. As the instrument is binary, we follow the nonparametric approach outlined in Section \ref{sec:mcsim}. Summary statistics for these variables are provided in Table \ref{tab:sumstat}.

\begin{table}[ht]
\centering
\begin{tabular}{lcccc}
  \hline
\hline
 & Mean & St. Dev. & Min & Max \\ 
  \hline
Log of Wage & 6.43 & 0.45 & 4.61 & 7.78 \\ 
  Education & 13.36 & 2.81 & 7.10 & 18.96 \\ 
  Near 4-year college & 0.71 & 0.45 & 0.00 & 1.00 \\ 
   \hline
\end{tabular}
\caption{Summary statistics} 
\label{tab:sumstat}
\end{table}

The initial condition is taken as the first derivative of a local linear nonparametric regression ($\hat\varphi^\prime_{0,1}$), a control function approach, where the control function is the conditional cdf of $X \mid W$ ($\hat\varphi^\prime_{0,2}$), and a flexible parametric model (third order polynomial in $X$), which uses the assumption of independence between $U$ and $W$ (see Section \ref{sec:mcsim}, $\hat\varphi^\prime_{0,3}$). Tuning parameters (bandwidths and regularization parameters) are selected as explained in Section \ref{sec:mcsim}.

We report in Figure \ref{fig:fct_card_1} the results of our empirical exercise. In the figure, the solid blue, red, and green lines are our LF estimators, $\hat\varphi_{N,1}$, $\hat\varphi_{N,2}$, and $\hat\varphi_{N,3}$, respectively. The dashed blue, red, and green lines are the initial conditions, while the black solid line is the TSLS estimator. The dotted lines show the $95\%$ pointwise confidence intervals for $\hat\varphi_{N,1}$, $\hat\varphi_{N,2}$, $\hat\varphi_{N,3}$ obtained by nonparametric bootstrap with $99$ replications. The confidence intervals for the nonparametric estimator are presented only to illustrate its variability but should not be considered valid.

\begin{knitrout}\scriptsize
\definecolor{shadecolor}{rgb}{0.969, 0.969, 0.969}\color{fgcolor}\begin{figure}[!ht]

{\centering \includegraphics[width=0.48\textwidth,height=7cm,]{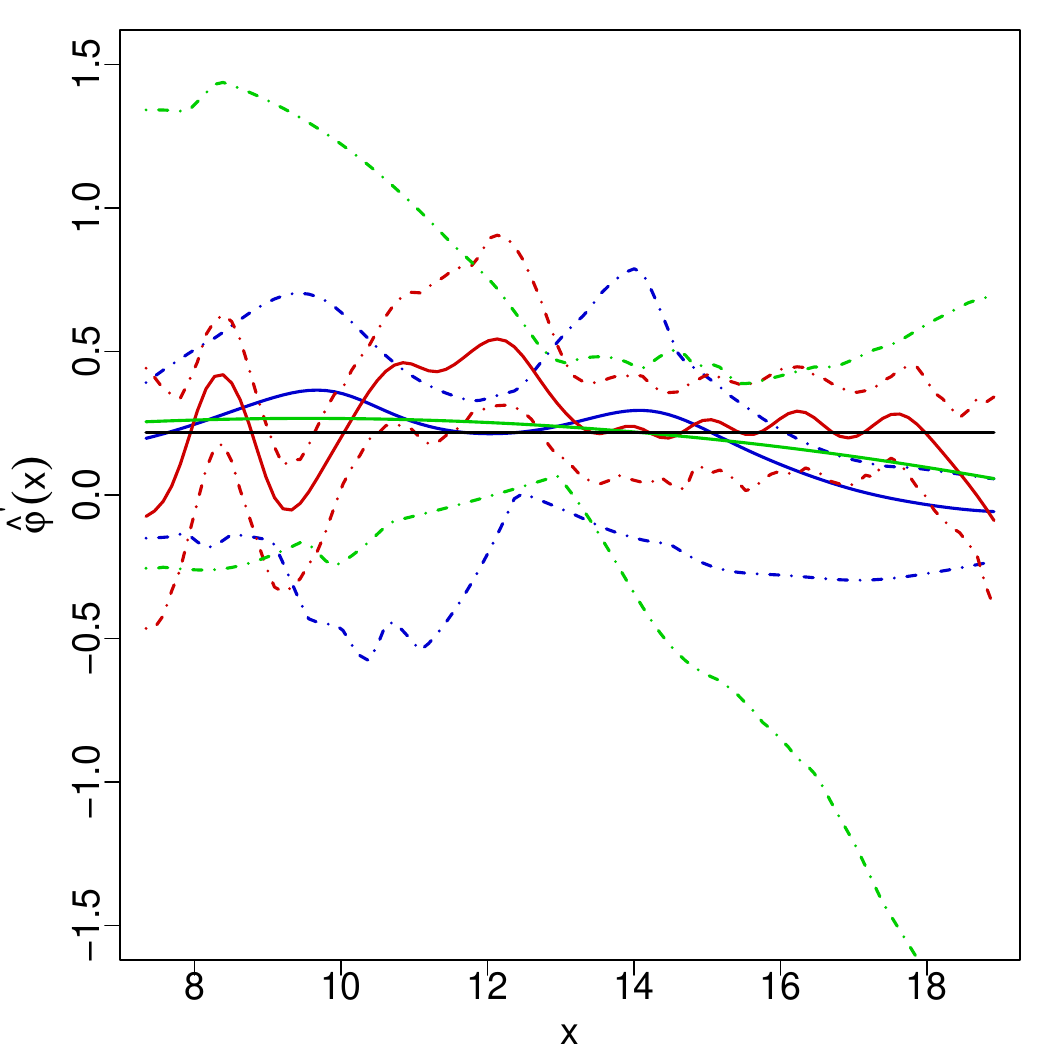} 
\includegraphics[width=0.48\textwidth,height=7cm,]{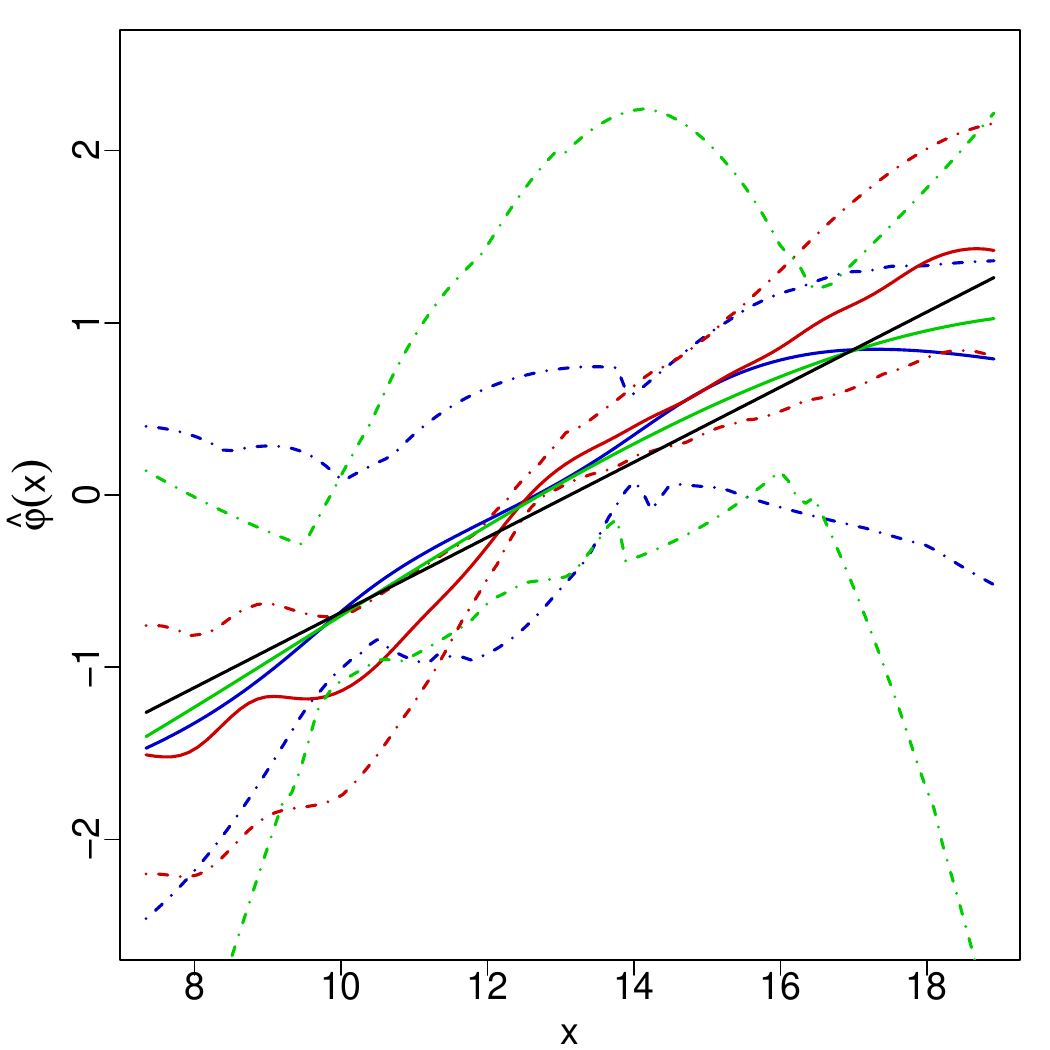} 
\includegraphics[width=0.72\textwidth,height=1cm,trim=0cm 8cm 0cm 8cm,clip=true]{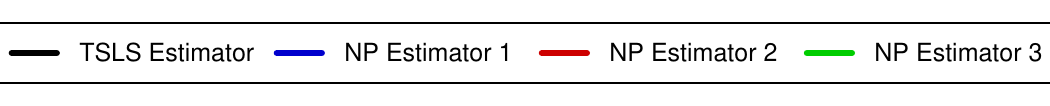} 

}

\caption[Estimation of the returns to education using \citeauthor{card1995}'s \citeyearpar{card1995} data]{Estimation of the returns to education using \citeauthor{card1995}'s \citeyearpar{card1995} data.}\label{fig:fct_card_1}
\end{figure}

\end{knitrout}

When starting from $\hat\varphi^\prime_{0,1}$ (blue line), we stop the LF algorithm at $N_0 = 40$ iterations. However, when starting from $\hat\varphi^\prime_{0,2}$ and $\hat\varphi^\prime_{0,3}$ (red and green lines, respectively), the LF algorithm remains at the initial condition. All nonparametric estimators confirm the log-linear relationship between education and wages, though some variation exists depending on the initial condition used. Marginal effects remain positive and are comparable to the constant effect estimated by TSLS. The wide confidence intervals (dashed-dotted lines) indicate higher uncertainty in the estimates. Assuming the instrumental variable satisfies our relevance assumptions, the exogeneity of the instrument can be violated, for instance, because of omitted variables.

\section{Conclusions}\label{sec:conclusions}

In this paper, we propose a nonparametric IV estimator with continuous endogenous variables and discrete instruments. We take the IVs and the structural error term to be independent, and we discuss conditions for identification and present an estimator of the regression function and its first derivative based on a smooth iterative regularization procedure. We derive an upper bound for the Mean Integrated Squared Error and present the finite sample properties in a simulation with one continuous endogenous regressor and a binary instrument. An empirical application to estimating the returns to schooling demonstrates its features compared with a linear instrumental variable estimator. Future work should focus on providing valid inference procedures and carefully studying the case with additional exogenous controls. 

\bibliographystyle{agsm}
\bibliography{instrumental}

\newpage

\setcounter{section}{0}

\renewcommand{\thesection}{\Alph{section}}
\renewcommand{\theequation}{\thesection.\arabic{equation}}

\section{Appendix}

\subsection{Additional Assumptions} \label{annexA1}

To obtain uniform consistency of the nonparametric estimators, we must impose some additional assumptions. These are listed below. Without loss of generality, we use the word density irrespective of $W$ being discrete. In this case, the probability density function can be defined as
\[
f_W(w) = \sum_{k} p_W(w_k) \Delta(w - w_k),
\]
where $p_W(w_k)$ is the probability mass function at $w_k$, and $\Delta$ is the Dirac's delta function.

\begin{assumption}~\label{assapp1}
\begin{itemize}
\item[(i)] The conditional probability density function $f_{Y,X \vert W}(y,x \vert w)$ and the density function $f_{Y,X}(y,x)$ are $d$ times continuously differentiable and uniformly bounded away from $\infty$.
\item[(ii)] The densities $f_{X,W} (x,w)$, $f_{X} (x)$ and $f_W(w)$ are uniformly bounded away from $0$ and $\infty$.
\end{itemize}
\end{assumption}

\begin{assumption} \label{assapp2}
The density of the error term $f_U(u)$ is absolutely continuous with respect to the Lebesgue measure, and $d$ times continuously differentiable.
\end{assumption}

\begin{assumption} \label{assapp3a}
The multivariate kernel $C_{\cdot,h}$ is a product kernel generated by the univariate generalized kernel functions $C_h$ satisfying the following properties: 
\begin{itemize}
\item[(i)] $C_h (\cdot, \cdot)$ is of order $l \geq 2$.
\item[(ii)] For each $t \in [0,1]$, the function $C_h(h\cdot,t)$ is supported on $[(t-1)/h,t/h] \cap \mathcal{C}$, where $\mathcal{C}$ is a compact interval that does not depend on t and
\[
\sup_{h>0,t\in [0, 1], u \in \mathcal{C} } \vert C_h(hu,t) \vert < \infty
\]
\item[(iii)] $C_h(\cdot,1) = C_h(\cdot)$ is a symmetric Lipschitz continuous kernel function with compact support.
\end{itemize}
\end{assumption}

\begin{assumption}[\citealp{aitch1976,liracine2007}, p. 131] \label{assapp3b}
The function $L_{h_w}(\cdot)$ is a discrete product kernel generated by the univariate generalized kernel functions
\[
l_{h_{w,j}}(w_j - w_{jk}) = \begin{cases} 1 & w_{jk} = w_j\\ h_{w,j} & \text{ otherwise } \end{cases},
\]
where $j = 1,\dots,q$, and $h_{w,j} \in [0,1]$. That is
\[
L_{h_w}(w - w_k) = \prod_{j= 1}^q l_{h_{w,j}}(w_j - w_{jk})= \prod_{j= 1}^q h_{w,j}^{1 - \un(w_{jk} = w_j)}.
\]
\end{assumption}

\begin{assumption} \label{assapp4}
Let $\ell_n$ a real sequence that is either bounded or diverges slowly to $\infty$ with $n$. The density of the error term, $U$, satisfies
\[
\kappa_n = \inf_{\vert u \vert \leq \ell_n} f_U (u) > 0, 
\]
with $\kappa_n \rightarrow 0$, as $n \rightarrow \infty$. 
\end{assumption}

\begin{assumption} \label{assapp5}
The smoothing parameters satisfy $h_u,h_w,h_x \rightarrow 0$, and $(nh_x^p h_u^3)^{-1} \ln n \rightarrow 0$.
\end{assumption}

Assumption \ref{assapp1} is a standard regularity condition of conditional and unconditional densities. Part $(ii)$ is not restrictive as long as we maintain that the joint support of $(X,W)$ is compact. Assumption \ref{assapp2} restricts the density of the error term to be continuous and differentiable. Assumption \ref{assapp3a} defines generalized kernel functions as in \citet{muller1991} and imposes a Lipschitz continuity condition on the kernel function to obtain uniform convergence results for the nonparametric density estimation when a random variable has unbounded support \citep{hansen2008}. Assumption \ref{assapp3b} introduces a discrete kernel to smooth with respect to the discrete instrument. When $h_w = 0$, the function $L_{h_w}$ is the product of indicator functions \citep[][pp. 126-131]{liracine2007}. Finally, Assumptions \ref{assapp4} and \ref{assapp5} are usual conditions imposed on the bandwidth parameter to achieve uniform consistency of the nonparametric density estimator.

\subsection{Proofs Section \ref{sec:identification}} \label{annexA2}

\subsubsection{Proof of Proposition \ref{prop:onetoone}}

The proof of this Proposition follows the proof of Theorem 2 in \citet{chenlee2013}. We prove by contradiction. Let us assume that there exists a $\varphi^\ast \in \mathcal{B}_\dagger$, such that $\varphi^\ast \neq \vdagger$ and $T(\varphi^\ast) = 0$. By condition $(i)$, the Fr\'echet derivative of $T$ exists. We have two cases, 
\begin{enumerate}
\item[1)] $T^\prime_{\vdagger} (\varphi^\ast - \vdagger)= 0$, and by the injectivity condition in $(ii)$, $\varphi^\ast = \vdagger$.
\item[2)] $T^\prime_{\vdagger} (\varphi^\ast - \vdagger) \neq 0$. In this case, $\Vert T^\prime_{\vdagger} (\varphi^\ast - \vdagger) \Vert > 0$, and
\begin{align*}
1 =& \frac{\Vert T^\prime_{\vdagger} (\varphi^\ast - \vdagger) \Vert }{\Vert T^\prime_{\vdagger} (\varphi^\ast - \vdagger) \Vert} = \frac{\Vert T(\varphi^\ast) - T(\vdagger) - T^\prime_{\vdagger} (\varphi^\ast - \vdagger) \Vert }{\Vert T^\prime_{\vdagger} (\varphi^\ast - \vdagger) \Vert} \\
\leq& \frac{M \Vert \varphi^\ast - \vdagger \Vert^2}{\Vert T^\prime_{\vdagger} (\varphi^\ast - \vdagger) \Vert} \\
<& \frac{M \Vert \varphi^\ast - \vdagger \Vert^2}{M \Vert \varphi^\ast - \vdagger \Vert^2} = 1,
\end{align*}
where the second line follows from condition $(iii)$, and the last line from the restriction on the set $\mathcal{B}_M$. 
\end{enumerate}
Therefore, we must have that either $\varphi^\ast = \vdagger$ or $T(\varphi^\ast) \neq 0$, for all $\varphi^\ast \in \mathcal{B}_\dagger$. This concludes the proof.

\subsubsection{Proof that Assumption \ref{ass:scalab} implies \ref{prop:onetoone}(iii)} We prove the statement in several steps. We first show that 
\[
\left\vert \frac{\partial}{\partial y} f_{Y \vert X }(y \vert x) \right\vert \leq M_2 M_3.
\]
We have
\begin{align*}
f_{Y \vert X }(y \vert x) =& \int f_{Y \vert X,W }(y \vert x,w) f_{W \vert X} (w \vert x) dw = \int f_{Y \vert X,W }(y \vert x,w) \frac{f_{X \vert W}(x \vert w)}{f_X(x)}f_{W} (w) dw.
\end{align*}
Therefore,
\begin{equation} \label{prop:onetooneres1}
\begin{aligned}
\left\vert \frac{\partial f_{Y \vert X }(y \vert x)}{\partial y} \right\vert =& \left\vert \int \frac{\partial f_{Y \vert X,W }(y \vert x,w)}{\partial y} \frac{f_{X \vert W}(x \vert w)}{f_X(x)}f_{W} (w)d w\right\vert \\
\leq & \int \left\vert \frac{\partial f_{Y \vert X,W }(y \vert x,w)}{\partial y} \frac{f_{X \vert W}(x \vert w)}{f_X(x)}\right\vert f_{W} (w) d w \leq M_2 M_3 \int f_{W} (w) dw = M_2 M_3.
\end{aligned}
\end{equation}

The result in \eqref{prop:onetooneres1} implies that 
\begin{equation} \label{prop:onetooneres2}
\left\vert \frac{\partial}{\partial y} f_{Y \vert X,W }(y \vert x,w) \frac{f_{X\vert W}(x \vert x)}{f_X(x)} - \frac{\partial}{\partial y} f_{Y \vert X }(y \vert x) \right\vert \leq 2 M_2 M_3,
\end{equation}
which follows from the triangle inequality, and Assumption \ref{ass:scalab}.

Finally, let $\varphi_t = t\varphi + (1-t)\vdagger$, for $t \in [0,1]$. By an application of the mean-value theorem for functional derivatives, we have
\[
T(\varphi) - T(\vdagger) = \int_{0}^1 T^\prime_{\varphi_t} dt (\varphi - \vdagger),
\]
so that 
\[
\left\Vert T(\varphi) - T(\vdagger) - T^\prime_{\vdagger} (\varphi - \vdagger) \right\Vert = \left\Vert \int_{0}^1 \left( T^\prime_{\varphi_t} - T^\prime_{\vdagger} \right) dt (\varphi - \vdagger) \right\Vert.
\]
Hence,
\begin{align*}
& \left\Vert \int_{0}^1 \left( T^\prime_{\varphi_t} - T^\prime_{\vdagger}\right) dt \left(\varphi - \vdagger\right) \right\Vert^2 \\
=& \int \int \biggl\lbrace \int_{0}^1 \int \Bigl[ f_{Y,X\vert W} (\varphi_t(x) + u,x \vert w) - f_{Y,X\vert W} (\vdagger(x) + u,x \vert w) \\
& \quad + f_{Y,X} (\vdagger(x) + u,x) - f_{Y,X} (\varphi_t(x) + u,x) \Bigr] \left( \varphi - \vdagger \right)(x) dx dt \biggl\rbrace^2 f_U(u) f_W(w) du dw \\
=& \int \int \left\lbrace \int_{0}^1 t \int \left[ \frac{\partial f_{Y,\vert X, W} (\bar{\varphi}_t(x) + u\vert x, w)}{\partial y} \frac{f_{X\vert W}(x \vert w)}{f_X(x)} \right. \right. \\
& \quad \left. \left. - \frac{\partial f_{Y\vert X} (\bar{\varphi}_t(x) + u \vert x)}{\partial y} \right]\left( \varphi - \vdagger \right)^2(x) f_X(x) dx dt \right\rbrace^2 f_U(u) f_W(w) du dw \\
\leq & 4 M^2_2 M^2_3 \left\lbrace \int_{0}^1 t dt \Vert \varphi - \vdagger \Vert^2 \right\rbrace^2 = M^2_2 M^2_3 \Vert \varphi - \vdagger \Vert^4,
\end{align*}
where the second equality follows from the mean-value theorem and the fact that $\varphi_t - \vdagger = t(\varphi - \vdagger)$, and the last inequality follows from \eqref{prop:onetooneres2}.
Hence,
\[
\Vert T(\varphi) - T(\vdagger) - T^\prime_{\vdagger} (\varphi - \vdagger) \Vert \leq M \Vert \varphi - \vdagger \Vert^2,
\]
with $M = M_2 M_3$, which is condition (iii) in Proposition \ref{prop:onetoone}.

Moreover, we show that the conditions in Assumption \ref{ass:scalab} imply that $T^\prime_{\varphi}$ is a Hilbert-Schmidt operator. That is, we prove that the Hilbert-Schmidt norm of $T^\prime_{\varphi}$ is finite under the maintained assumptions.
\begin{align*}
\Vert T^\prime_{\varphi} \Vert^2_{HS} =& \int_w \int_u \int_x \left[ \frac{f_{Y,X \vert W}(\varphi(x) + u,x \vert w)}{f_X(x)} - \frac{f_{Y,X}(\varphi(x) + u,x)}{f_X(x)}\right]^2 f_X(x) f_W(w) f_U(u) dx dw du \\
=& \int_w \int_u \int_x \left[ f_{Y \vert X, W}(\varphi(x) + u \vert x, w)\frac{f_{X\vert W}(x\vert w)}{f_X(x)} - f_{Y \vert X}(\varphi(x) + u \vert x)\right]^2 f_X(x) f_W(w) f_U(u) dx dw du \\
\leq & 2 \int_w \int_u \int_x \left[ f^2_{Y \vert X, W}(\varphi(x) + u \vert x, w)\frac{f^2_{X\vert W}(x\vert w)}{f^2_X(x)} + f^2_{Y \vert X}(\varphi(x) + u \vert x)\right] f_X(x) f_W(w) f_U(u) dx dw du\\
\leq & 2 \int_w \int_u \int_x \left[ M^2_3 f^2_{Y \vert X, W}(\varphi(x) + u \vert x, w) + f^2_{Y \vert X}(\varphi(x) + u \vert x)\right] f_X(x) f_W(w) f_U(u) dx dw du \\
\leq & 2 M^2_1\left( M^2_3 + 1\right) < \infty
\end{align*}
where the third line follows from Young's inequality, the fourth line follows from Assumption \ref{ass:scalab}(ii), and the last line from Assumption \ref{ass:scalab}(i), and the monotone convergence theorem.

\subsubsection{Proof of Proposition \ref{prop:condcomp}}

We follow the same steps as in the proof of Proposition \ref{prop:onetoone}. Let $\varphi^\ast \in \mathcal{B}_\dagger$ such that $T(\varphi^\ast) = 0$. 

If $\varphi^\ast - \vdagger$ is in the null space of the operator $T^\prime_{\vdagger}$, then for $\tilde{\varphi} = \vdagger - \varphi^\ast$, we have that
\begin{eqnarray*}
E(\tilde{\varphi} (X)\vert U=u,W=w) &=&\int \tilde{\varphi} (x) \frac{f_{Y,X\vert W}(\vdagger (x) +u,x\vert w)}{f_{U \vert W }(u\vert w)} dx\\
E(\tilde{\varphi} (X)\vert U=u) &=& \int \tilde{\varphi} (x) \frac{f_{Y,X}(\vdagger (x) + u, x)}{f_U(u)}dx
\end{eqnarray*}
As $f_{U\vert W}(u\vert w) = f_U(u)$, $T^\prime_{\vdagger} \tilde{\varphi}=0$ is equivalent to $E(\tilde{\varphi}\vert U,W)= E(\tilde{\varphi}\vert U)$, which, by Assumption \ref{ass:condcomp} implies $\tilde{\varphi} = 0$, and therefore $\varphi^\ast = \vdagger$. 

If $\varphi^\ast - \vdagger$ is not in the null space of the operator $T^\prime_{\vdagger}$, then $\varphi^\ast$ can be written as
\[
\varphi^\ast - \vdagger = \sum_{j = 1}^\infty b_j \chi_j,
\]
where $\Vert \varphi^\ast - \vdagger \Vert^2 = \sum_{j = 1}^\infty b_j^2$, and $\Vert T^\prime_{\vdagger} \left(\varphi^\ast - \vdagger \right)\Vert^2 = \sum_{j = 1}^\infty t^2_j b_j^2$. By Assumption \ref{ass:scalab}, and since $T(\vdagger) = T(\varphi^\ast) = 0$
\begin{equation} \label{prop:condcompres1}
\Vert T^\prime_{\vdagger} \left(\varphi^\ast - \vdagger \right)\Vert \leq M \Vert \varphi^\ast - \vdagger \Vert^2.
\end{equation}

However, by Definition \ref{ass:sourcecondID} and the Cauchy-Schwartz inequality
\begin{equation} \label{prop:condcompres2}
\begin{aligned}
\Vert \varphi^\ast - \vdagger \Vert^2 = &\sum_{j = 1}^\infty b_j^2 = \sum_{j = 1}^\infty \left( \frac{b_j}{t_j} \right) \left(t_j b_j \right) \leq \left( \sum_{j = 1}^\infty \frac{b^2_j}{t^2_j} \right)^{1/2}\left( \sum_{j = 1}^\infty t^2_j b^2_j \right)^{1/2} < \frac{1}{M} \Vert T^\prime_{\vdagger} \left(\varphi^\ast - \vdagger \right)\Vert.
\end{aligned}
\end{equation}

Equations \eqref{prop:condcompres1} and \eqref{prop:condcompres2} imply that
\[
M \Vert \varphi^\ast - \vdagger \Vert^2 < \Vert T^\prime_{\vdagger} \left(\varphi^\ast - \vdagger \right)\Vert \leq M \Vert \varphi^\ast - \vdagger \Vert^2,
\]
which is a contradiction. Hence, we must have that $T(\varphi^\ast) \neq 0$. This concludes the proof.

\subsection{Additional proofs for Section \ref{sec:asymprop}} \label{annexA3}

\subsubsection{Proof of Proposition \ref{prop:convlf}}

We wish to prove that, under the maintained regularity conditions
\begin{equation} 
\Vert \varphi^\prime_{j+1} - \vdagger^\prime \Vert^2 - \Vert \varphi^\prime_{j} - \vdagger^\prime \Vert^2 \leq 0,
\end{equation} 
with equality only when $\varphi^\prime_{j+1} = \varphi^\prime_{j} = \vdagger^\prime$. This implies that $\varphi_j \in \mathcal{B}_\dagger$ for all $j = 1,2,3,\dots$, and that the LF algorithm converges to $\vdagger^\prime$ and therefore that $\varphi_j = D^{-1}\varphi_j^\prime$ converges to $\vdagger$. 

Let $\varphi_{j} \in \mathcal{B}_\dagger$ and write
\begin{align*}
\Vert \varphi^\prime_{j+1} - \vdagger^\prime \Vert^2 -& \Vert \varphi^\prime_{j} - \vdagger^\prime \Vert^2 = \langle cA^\ast_{\varphi_j} T(\varphi_j), cA^\ast_{\varphi_j} T(\varphi_j)\rangle - 2 \langle \varphi^\prime_j - \vdagger^\prime, c A^\ast_{\varphi_j} T(\varphi_j) \rangle \\
=& \langle cA_{\varphi_j} A^\ast_{\varphi_j} T(\varphi_j), c T(\varphi_j)\rangle - 2 \langle A_{\varphi_j} \left( \varphi^\prime_j - \vdagger^\prime \right), c T(\varphi_j) \rangle \\
=& \langle cA_{\varphi_j} A^\ast_{\varphi_j} T(\varphi_j), c T(\varphi_j)\rangle - 2 \langle T^\prime_{\varphi_j} \left( \varphi_j - \vdagger \right), c T(\varphi_j) \rangle \\
=& - \langle \left(I - c A_{\varphi_j} A^\ast_{\varphi_j} \right) T(\varphi_j), c T(\varphi_j)\rangle - c\Vert T^{\prime}_{\varphi_j} \left( \varphi_j - \vdagger \right) \Vert^2 + c \Vert T(\varphi_j) - T^{\prime}_{\varphi_j} \left( \varphi_j - \vdagger \right) \Vert^2\\
\leq& -c\Vert \left(I - c A_{\varphi_j} A^\ast_{\varphi_j} \right)^{\frac{1}{2}} T(\varphi_j) \Vert^2 - c\Vert T^{\prime}_{\varphi_j} \left( \varphi_j - \vdagger \right) \Vert^2 + cM \Vert \varphi_j - \vdagger \Vert^4 \\
\leq & -c\Vert \left(I - c A_{\varphi_j} A^\ast_{\varphi_j} \right)^{\frac{1}{2}} T(\varphi_j) \Vert^2 - c\Vert T^{\prime}_{\varphi_j} \left( \varphi_j - \vdagger \right) \Vert^2 + c \Vert T^{\prime}_{\vdagger} \left( \varphi_j - \vdagger \right) \Vert^2 \\
\leq& -c\Vert \left(I - c A_{\varphi_j} A^\ast_{\varphi_j} \right)^{\frac{1}{2}} T(\varphi_j) \Vert^2 - c\Vert T^{\prime}_{\varphi_j} \left( \varphi_j - \vdagger \right) \Vert^2 \\
& \quad + c \Vert \left( T^{\prime\ast}_{\vdagger} T^{\prime}_{\vdagger} \right)^{1/2} \left( T^{\prime\ast}_{\varphi_j} T^{\prime}_{\varphi_j} \right)^{-1/2} \left( T^{\prime\ast}_{\varphi_j} T^{\prime}_{\varphi_j} \right)^{1/2} \left( \varphi_j - \vdagger \right) \Vert^2\\
\leq& -c\Vert \left(I - c A_{\varphi_j} A^\ast_{\varphi_j} \right)^{\frac{1}{2}} T(\varphi_j) \Vert^2 - c\Vert T^{\prime}_{\varphi_j} \left( \varphi_j - \vdagger \right) \Vert^2 + c \Vert \left( T^{\prime\ast}_{\vdagger} T^{\prime}_{\vdagger} \right) \left( T^{\prime\ast}_{\varphi_j} T^{\prime}_{\varphi_j} \right)^{-1} \Vert \Vert T^{\prime}_{\varphi_j} \left( \varphi_j - \vdagger \right) \Vert^2\\
\leq&  -c\Vert \left(I - cA_{\varphi_j} A^\ast_{\varphi_j} \right)^{\frac{1}{2}} T(\varphi_j) \Vert^2 - c \left( 1 - \frac{1}{M_4}\right)\Vert T^{\prime}_{\varphi_j} \left( \varphi_j - \vdagger \right) \Vert^2 \leq 0,
\end{align*}
where the third line follows from $A_{\varphi_j} = T^\prime_{\varphi_j} D^{-1}$, the fourth line from the assumptions on the continuity of the Fr\' echet derivative, the fifth line from the fact that $\varphi_j \in \mathcal{B}_\dagger$, and the last line from the Assumption in equation \eqref{ass:lfconv}. In line fifth, we keep the equality to allow for the limiting case in which $\varphi_j = \vdagger$.  As $\varphi_0 \in \mathcal{B}_\dagger$ by Assumption \ref{ass:initcond}, then this result implies that $\varphi_j \in \mathcal{B}_\dagger$ for all $j > 0$. Also, when $\varphi_j = \vdagger$, the upper bound is exactly equal to $0$, as $T(\varphi_j) = T(\vdagger) = 0$. 

Suppose there exists a sequence $\varphi^\prime_j = \vdagger^\prime + v_j$, where $\Vert v_j \Vert^2 \rightarrow 1$, as $j \rightarrow \infty$. Then
\begin{align*}
\Vert \varphi^\prime_{j+1} - \vdagger^\prime \Vert^2 -& \Vert \varphi^\prime_{j} - \vdagger^\prime \Vert^2 = \Vert v_{j+1} \Vert^2 - \Vert v_j \Vert^2\\
& \quad \leq -c\Vert \left(I - cA_{\varphi_j} A^\ast_{\varphi_j} \right)^{\frac{1}{2}} T( \vdagger + D^{-1} v_j) \Vert^2 - c \left( 1 - \frac{1}{M_4}\right)\Vert T^{\prime}_{\varphi_j} D^{-1} v_j \Vert^2.
\end{align*}

As $j \rightarrow \infty$, the lhs of the inequality would converge to zero, but the rhs remains negative, as $T( \vdagger + D^{-1} v_j) \neq 0$ almost surely on $\mathcal{B}_\dagger$ from Proposition \ref{prop:condcomp}, unless $v_j = 0$. This leads to a contradiction. Hence, the LF algorithm converges to $\vdagger$ as $j \rightarrow \infty$. The result of the Proposition follows.

\subsubsection{Proof of Proposition \ref{prop:koper}}

For part a), Assumption \ref{ass:frediff} requires the Fr\'echet differentiability of $\hat{T}$. Therefore, for $$\hat\varphi_{jt} = t \hat\varphi_j + (1-t)\vdagger,$$ we have that
\[
\hat{T} (\hat\varphi_j) - \hat{T} (\vdagger) = \int_0^1 \hat{T}^\prime_{\hat\varphi_{jt}} dt \left( \hat\varphi_j - \vdagger \right), 
\]
which implies that 
\[
\hat{T} (\hat\varphi_j) - \hat{T} (\vdagger) - \hat{T}^\prime_{\vdagger} \left( \hat\varphi_j - \vdagger \right) = \int_0^1 \left( \hat{T}^\prime_{\hat\varphi_{jt}} - \hat{T}^\prime_{\vdagger} \right) dt \left( \hat\varphi_j - \vdagger \right).
\]
Furthermore, for any function $\varphi$,
\begin{align*}
& \left(\hat{T}^\prime_\varphi (\hat\varphi_j - \vdagger)\right)(u,w)= \int \left[ \hat{f}_{Y,X \vert W } (\varphi(x) + u,x\vert w) - \hat{f}_{Y,X} (\varphi(x) + u,x) \right]  (\hat\varphi_j - \vdagger)(x) dx \\
=& \int \left[ \frac{\hat{f}_{Y, X \vert W } (\varphi(x) + u, x \vert w)}{f_X(x)} - \frac{\hat{f}_{Y , X } (\varphi(x) + u,x)}{f_X(x)} \right]  (\hat\varphi_j - \vdagger)(x) f_X(x) dx, 
\end{align*} 
which, for $\bar{\varphi}_{jt}$ a point between $\hat\varphi_{jt}$ and $\vdagger$, and the continuity of the kernel function in Assumption \ref{assapp3a}, gives
\begin{align*}
\int_0^1 & \left( \hat{T}^\prime_{\hat\varphi_{jt}} - \hat{T}^\prime_{\vdagger} \right) dt \left( \hat\varphi_j - \vdagger \right)\\
=& \int \int_0^1 t \left[ \frac{\partial \hat{f}_{Y,X \vert W } (\bar{\varphi}_{jt}(x) + u, x \vert  w)}{\partial y} \frac{1}{f_X(x)} - \frac{\partial  \hat{f}_{Y, X } (\bar{\varphi}_{jt}(x) + u, x)}{\partial y} \frac{1}{f_X(x)} \right]dt  (\hat\varphi_j - \vdagger)^2(x) f_X(x) dx.
\end{align*}
Because of Assumptions \ref{assapp1}-\ref{assapp5}, the nonparametric kernel density estimators are uniformly consistent estimators of their population counterparts and their derivatives \citep{hansen2008,li2008,darolles2011}. Therefore, reasoning as in \citet[][Lemma 4, p. 6168]{dunker2018}, for a sufficiently large constant $M_n$, we have that 
\begin{align*}
\left\vert \frac{\partial \hat{f}_{Y,X \vert W}(y, x \vert w)}{\partial y } \right\vert \leq & \left\vert \frac{\partial f_{Y,X \vert W}(y, x \vert w)}{\partial y } \right\vert  + M_n \leq M_2 M_3 f_X(x) + M_n \\
\left\vert \frac{\partial \hat{f}_{Y,X}(y, x)}{\partial y } \right\vert \leq & \left\vert \frac{\partial f_{Y,X }(y, x )}{\partial y } \right\vert  + M_n \leq M_2 M_3 f_X(x) + M_n,
\end{align*}
where the last inequality follows from Assumption \ref{ass:scalab}. Therefore,
\begin{align*}
& \left\vert \frac{\partial \hat{f}_{Y,X \vert W } (\bar{\varphi}_{jt}(x) + u, x \vert  w)}{\partial y} \frac{1}{f_X(x)} - \frac{\partial  \hat{f}_{Y, X } (\bar{\varphi}_{jt}(x) + u, x)}{\partial y} \frac{1}{f_X(x)} \right\vert \\
& \qquad \leq 2M_2M_3 + 2\frac{M_n}{\inf_x f_X(x)} \leq 2M_2M_3 + 2M_n M_X,
\end{align*}
where the last bound follows from Assumption \ref{assapp1}(ii), with $\frac{1}{\inf_x f_X(x)}  \leq M_X < \infty$. Let $\hat{M} = M_2M_3 + M_n M_X$, then we conclude
\begin{align*}
\Vert & \hat{T} (\hat\varphi_j) - \hat{T} (\vdagger) - \hat{T}^\prime_{\vdagger} \left( \hat\varphi_j - \vdagger \right)  \Vert  \leq \left\Vert \int_0^1 \left( \hat{T}^\prime_{\hat\varphi_{jt}} - \hat{T}^\prime_{\vdagger} \right) dt \left( \hat\varphi_j - \vdagger \right) \right\Vert\\
& \quad \leq  \hat{M} \Vert \varphi_j - \vdagger \Vert^2.
\end{align*}

To prove part b), let $\hat{e}_j = \hat\varphi_j - \vdagger$ and $$\tilde{\psi}(u,w) = \psi(u,w) - \frac{1}{n}\sum_{i = 1}^n \psi(u,W_i).$$ Then 
\begin{align*} 
\left( \hat{A}^{\ast}_{\hat\varphi_j}\tilde{\psi} \right) (x) =& \frac{1}{\hat{f}_X (x)} \int \int \int \un(\xi \geq x_1) \hat{f}_{Y,X,W} (\hat\varphi_j (\xi,x_{-1}) + u, \xi,x_{-1},w) \hat{f}_{Y - \varphi(X)}(u) \tilde{\psi}(u,w) d\xi du dw \\
=& \frac{1}{\hat{f}_X (x)} \int \int \int \un(\xi \geq x_1) \hat{f}_{Y,X,W} (\hat{e}_j (\xi,x_{-1}) + \vdagger(\xi,x_{-1})+ u, \xi,x_{-1},w) \hat{f}_{Y - \varphi(X)}(u) \times \\
& \qquad \tilde{\psi}(u,w) \hat{f}_{X_1}(\xi) d\xi du dw \\
=& \frac{1}{\hat{f}_{X_1} (x_1)} \int \int \int  \un(\xi \geq x_1)  \hat{f}_{Y,W \vert X} (\vdagger(\xi,x_{-1}) + v, w \vert \xi,x_{-1}) \hat{f}_{\hat{U}_j} (v - \hat{e}_j (\xi,x_{-1})) \times \\
& \qquad \tilde{\psi}(v - \hat{e}_j (\xi,x_{-1}),w)\hat{f}_{X_1}(\xi) d \xi dv dw\\
=& \frac{1}{\hat{f}_{X_1} (x_1)} \int \int \int  \un(\xi \geq x_1) \hat{f}_{Y,W \vert X} (\vdagger(\xi,x_{-1}) + v, w \vert x) \hat{f}_{U} (v) \times \\
& \qquad \left\lbrace \frac{\hat{f}_{\hat{U}_j} (v - \hat{e}_j (\xi,x_{-1}) )}{\hat{f}_U(v)} \tilde{\psi}(v - \hat{e}_j (\xi,x_{-1}),w) \right\rbrace \hat{f}_{X_1}(\xi) d\xi dv dw\\
=& \left( \hat{A}^{\ast}_{\vdagger} K_{\hat\varphi_j,\vdagger}\tilde{\psi} \right) (x).
\end{align*}

By the continuity of the kernel function in Assumption \ref{assapp3a}, the mean-value theorem implies that
\begin{align*}
\hat{f}_{\hat{U}_j} (v - \hat{e}_j (x) ) =& \int \hat{f}_{U,X} (v - \hat{e}_j (x) + \hat{e}_j (t),t) dt \\ 
=& \hat{f}_U(v) - \int \frac{d\hat{f}_{U \vert X} (\bar{v}\vert t) }{du} \left( \hat{e}_j (x) - \hat{e}_j (t)\right) f_X(t) dt.
\end{align*}
The norm of the difference between this operator and the identity operator is thus given by
\begin{align*}
E \Vert K_{\hat\varphi_j,\vdagger} - I \Vert^2=& \int \int \left( \frac{\hat{f}_{\hat{U}_j} (v - \hat{e}_j (x) )}{\hat{f}_U(v)} - 1 \right)^2 f_{U,X}(v,x) dv dx \\
=& \int \int \left( \frac{1}{\hat{f}_U(v)} \int \frac{d\hat{f}_{U \vert X} (\bar{v}\vert t) }{du} \left( \hat{e}_j (x) - \hat{e}_j (t)\right) f_X(t) dt \right)^2 f_{U,X}(v,x) dv dx \\
\leq& M^2 h^{-4}_u \kappa_n^{-2} \int \left(\hat{e}_j (x) - E \left[ \hat{e}_j (X) \right] \right)^2 f_{X}(x) dx\\
\leq& M^2 h^{-4}_u \kappa_n^{-2} E \Vert \hat{e}_j \Vert^2,
\end{align*}
and the result of the Proposition follows. 

\subsubsection{Proof of Theorem \ref{thm:mainconv}}

As the operator $A_{\vdagger}^{\ast}A_{\vdagger}$ is compact and thus admits a singular value decomposition, we use the notation $(A_{\vdagger}^{\ast}A_{\vdagger})^{\beta/2}$ from functional analysis to signify that a function is applied to the singular values of $A_{\vdagger}^{\ast}A_{\vdagger}$.
 
 We first recall the following definition.
 \begin{definition}[Qualification]\label{qualifdef}
 A regularization procedure, $g_{N}$, is said to have qualification of order $\kappa>0$, if:
 \begin{equation} \label{qualifineq} 
 \sup_{0\leq t\leq\Vert A_{\vdagger} \Vert^{2}} \vert 1-tg_{N}(t)\vert t^{\eta}\leq M_\eta N^{-\eta},
 \end{equation}
 for $0<\eta\leq\kappa$, and a constant $M_\eta < \infty$. \hfill{}\qed
 \end{definition}

In particular, LF regularization has qualification equal to $\infty$, in the sense that for every $\eta > 0$, the inequality in equation \eqref{qualifineq} holds with
\[
1-tg_{N}(t) = (1 - ct)^N.
\]

 Moreover, we need the following.
 \begin{assumption} \label{assapp6} 
 For LF regularization, there exist two positive constants $M_\eta$ and $\eta$ such that:
 \begin{equation} \label{eq:qualifass} \sup_{0\leq t\leq\Vert
  A_{\vdagger} \Vert^{2}} t^{\beta/2- \eta} \leq M_\eta
 N^{-\beta/2} N^{\eta}.
 \end{equation} \hfill \qed 
 \end{assumption}
 
This Assumption is used repeatedly in the proof below. In the following we also use the fact that $$\hat{A}_{\varphi} \left( \hat{\varphi}^\prime_j - \vdagger^\prime \right) = \hat{T}^\prime_{\varphi} D^{-1}\left( \hat{\varphi}^\prime_j - \vdagger^\prime\right) = \hat{T}^\prime_{\varphi} \left( \hat{\varphi}_j - \vdagger\right).$$
 We have
 \begin{align*}
 \hat\varphi^\prime_N -\vdagger^\prime &=\hat\varphi^\prime_{N-1} -\vdagger^\prime - c \hat{A}^\ast_{\hat\varphi_{N-1}} \left( \hat{T} (\hat\varphi_{N-1})\right) \\
 &=\hat\varphi^\prime_{N-1} -\vdagger^\prime - c \hat{A}^\ast_{\hat\varphi_{N-1}} \left( \hat{T} (\hat\varphi_{N-1}) - \hat{T} (\vdagger)\right) - c \hat{A}^\ast_{\hat\varphi_{N-1}} \left( \hat{T} (\vdagger) - T(\vdagger) \right) \\
 &= \hat\varphi^\prime_{N-1} -\vdagger^\prime - c \hat{A}^\ast_{\vdagger} \left( \hat{T} (\hat\varphi_{N-1}) - \hat{T} (\vdagger)\right) \\
 &\quad- c \left( \hat{A}^\ast_{\hat\varphi_{N-1}} - \hat{A}^\ast_{\vdagger} \right) \left( \hat{T} (\hat\varphi_{N-1}) - T(\vdagger) \right) \\
 &\quad-c\hat{A}^\ast_{\vdagger}\left( \hat{T} (\vdagger) - T(\vdagger) \right)\\
 &= \hat\varphi^\prime_{N-1} -\vdagger^\prime - c \hat{A}^\ast_{\vdagger} \hat{A}_{\vdagger} \left( \hat\varphi^\prime_{N-1} - \vdagger^\prime \right) \\
 &\quad- c \hat{A}^\ast_{\vdagger} \left( \hat{T} (\hat\varphi_{N-1}) - \hat{T} (\vdagger) - \hat{T}^{\prime}_{\vdagger} (\hat\varphi_{N-1} - \vdagger) \right) \\
  &\quad- c \left( \hat{A}^\ast_{\hat\varphi_{N-1}} - \hat{A}^\ast_{\vdagger} \right) \left( \hat{T} (\hat\varphi_{N-1}) - T(\vdagger) \right) \\
 &\quad-c\hat{A}^\ast_{\vdagger}\left( \hat{T} (\vdagger) - T(\vdagger) \right),
 \end{align*}
 where the second line follows from $T(\vdagger) = 0$. By replacing iteratively $\hat{\varphi}_j$, for all $j = 0,\dots,N-2$, and letting $\hat{e}_j = \hat\varphi_j - \vdagger$, and $\hat{e}^\prime_j = \hat\varphi^\prime_j - \vdagger^\prime$, for all $j = 0,1,2,\dots$, we finally obtain
 \begin{align*}
 \hat{e}^\prime_N &=\left( I - c \hat{A}^\ast_{\vdagger} \hat{A}_{\vdagger} \right)^N \left( \hat\varphi^\prime_0 - \vdagger^\prime\right) \\
 &\quad-c\sum_{j = 0}^{N-1} \left( I - c \hat{A}^\ast_{\vdagger} \hat{A}_{\vdagger} \right)^{j} \hat{A}^\ast_{\vdagger}\left( \hat{T} (\vdagger) - T(\vdagger) \right)\\
 &\quad-c\sum_{j = 0}^{N-1} \left( I - c \hat{A}^\ast_{\vdagger} \hat{A}_{\vdagger} \right)^{N-j-1} \hat{A}^\ast_{\vdagger} \left( \hat{T} (\hat\varphi_j) - \hat{T} (\vdagger) - \hat{T}^{\prime}_{\vdagger} \hat{e}_j \right) \\
 &\quad-c\sum_{j = 0}^{N-1} \left( I - c \hat{A}^\ast_{\vdagger} \hat{A}^{\prime}_{\vdagger} \right)^{N-j-1} \left( \hat{A}^\ast_{\hat\varphi_j} - \hat{A}^\ast_{\vdagger} \right) \left( \hat{T} (\hat\varphi_j) - T(\vdagger) \right)\\
 &=I+II+III+IV.
 \end{align*}
The first two terms are similar as in the asymptotic expansion of LF regularization for linear inverse problems, and their bounds are rather standard in this literature \citep[see][for a comparison]{florens2012}. By contrast, the terms in $III$ and $IV$ come from the nonlinearity of the inverse problem in our framework. These latter terms are identically zero when the ill-posed inverse problem is linear. To control these terms, we use the main result provided in Proposition \ref{prop:koper}. 

Let $\delta_n$ and $\gamma_n$ to be defined as in Assumption \ref{ass:rateconv}. We start by considering the term in $I$. Because of the source condition in Assumption \ref{ass:sourcecond}, we have that 
 \begin{align*}
 E|| I ||^2 &\leq 2 E||\left( I - c \hat{A}^\ast_{\vdagger} \hat{A}_{\vdagger} \right)^N \left( \hat\varphi_0 - \varphi_0 \right) ||^2 \\
 &\quad+ 2 E||\left( I - c \hat{A}^\ast_{\vdagger} \hat{A}_{\vdagger} \right)^N \left( \varphi_0 - \vdagger \right) ||^2 \\
&= 2 E||\left( I - c \hat{A}^\ast_{\vdagger} \hat{A}_{\vdagger} \right)^N \left( \hat\varphi_0 - \varphi_0 \right) ||^2 \\
 &\quad+ 4 E||\left[ \left( I - c \hat{A}^\ast_{\vdagger} \hat{A}_{\vdagger} \right)^N - \left( I - c A^\ast_{\vdagger} A_{\vdagger} \right)^N \right](A_{\vdagger}^{\ast} A_{\vdagger})^{\beta/2} v ||^2 \\
 &\quad+ 4 E||\left( I - c A^\ast_{\vdagger} A_{\vdagger} \right)^N (A_{\vdagger}^{\ast} A_{\vdagger})^{\beta/2} v ||^2\\
 &= 2 E|| I_a ||^2 + 4E|| I_b ||^2 + 4E|| I_c ||^2,
 \end{align*}
 
where the last step follows from Young's inequality and Assumption \ref{ass:sourcecond}. Using Assumptions \ref{ass:sourcecond} and \ref{assapp6}, we directly have $E|| I_c ||^2= O\left(N^{-\beta}\right)$. Moreover, by telescoping the difference of two operators, $A^N - B^N = \sum_{j =0}^{N-1} A^{N-j-1}(A - B) B^{j}$, we obtain 
 \begin{align*}
 \left( E|| I_b ||^2 \right)^{1/2} &= \left( E \left\Vert c \sum_{j =0}^{N-1} \left( I - c \hat{A}^\ast_{\vdagger} \hat{A}_{\vdagger} \right)^{N-j-1} \left( \hat{A}^\ast_{\vdagger} \hat{A}_{\vdagger} - A^\ast_{\vdagger} A_{\vdagger} \right) \left( I - c A^\ast_{\vdagger} A_{\vdagger} \right)^{j} (A_{\vdagger}^{\ast} A_{\vdagger})^{\beta/2} v \right\Vert^2 \right)^{1/2}\\
 &\leq \left( E \left\Vert c \sum_{j =0}^{N-1} \left( I - c \hat{A}^\ast_{\vdagger} \hat{A}_{\vdagger} \right)^{N-j-1} \hat{A}^\ast_{\vdagger} \left( \hat{A}_{\vdagger} - A_{\vdagger} \right) \left( I - c A^\ast_{\vdagger} A_{\vdagger} \right)^{j} (A_{\vdagger}^\ast A_{\vdagger})^{\beta/2} v \right\Vert^2 \right)^{1/2}\\
 &\quad+ \left( E \left\Vert c \sum_{j =0}^{N-1} \left( I - c \hat{A}^\ast_{\vdagger} \hat{A}_{\vdagger} \right)^{N-j-1} \left( \hat{A}^\ast_{\vdagger} - A^\ast_{\vdagger} \right) A_{\vdagger} \left( I - c A^\ast_{\vdagger} A_{\vdagger} \right)^{j} (A^{\ast} A_{\vdagger})^{\beta/2} v \right\Vert^2 \right)^{1/2}\\
 &\leq \sum_{j =0}^{N-1} \left( E \left\Vert \left( I - c \hat{A}^\ast_{\vdagger} \hat{A}_{\vdagger} \right)^{N-j-1} \hat{A}^\ast_{\vdagger} \right\Vert^2 || \hat{A}_{\vdagger} - A_{\vdagger} ||^2 \left\Vert c\left( I - c A^\ast_{\vdagger} A_{\vdagger} \right)^{j} (A_{\vdagger}^\ast A_{\vdagger})^{\beta/2} v \right\Vert^2 \right)^{1/2}\\
 &\quad+ \sum_{j =0}^{N-1} \left( E \left\Vert \left( I - c \hat{A}^\ast_{\vdagger} \hat{A}_{\vdagger} \right)^{N-j-1} \right\Vert^2 || \hat{A}^\ast_{\vdagger} - A^\ast_{\vdagger} ||^2 \left\Vert c A_{\vdagger} \left( I - c A^\ast_{\vdagger} A_{\vdagger} \right)^{j} (A_{\vdagger}^\ast A_{\vdagger})^{\beta/2} v \right\Vert^2 \right)^{1/2}\\
 &\leq \sum_{j =0}^{N-1} \left( E \left\Vert \left( I - c \hat{A}^\ast_{\vdagger} \hat{A}_{\vdagger} \right)^{N-j-1} \hat{A}^\ast_{\vdagger} \right\Vert^4 E || \hat{A}_{\vdagger} - A_{\vdagger} ||^4 \right)^{1/4} \left\Vert c\left( I - c A^\ast_{\vdagger} A_{\vdagger} \right)^{j} (A_{\vdagger}^\ast A_{\vdagger})^{\beta/2} v \right\Vert\\
 &\quad+ \sum_{j =0}^{N-1} \left( E \left\Vert \left( I - c \hat{A}^\ast_{\vdagger} \hat{A}_{\vdagger} \right)^{N-j-1} \right\Vert^4 E|| \hat{A}^\ast_{\vdagger} - A^\ast_{\vdagger} ||^4 \right)^{1/4} \left\Vert c A_{\vdagger} \left( I - c A^\ast_{\vdagger} A_{\vdagger} \right)^{j} (A_{\vdagger}^\ast A_{\vdagger})^{\beta/2} v \right\Vert,
 \end{align*}
where the second and third steps follow from repeated applications of the Minkowski's inequality. From Lemma \ref{lemapp3}, we can bound
 \begin{align*}
 E || \left( I - c \hat{A}^\ast_{\vdagger} \hat{A}_{\vdagger} \right)^{N-j-1} ||^4 =& O(1)\\
 E || \left( I - c \hat{A}^\ast_{\vdagger} \hat{A}_{\vdagger} \right)^{N-j-1}\hat{A}^\ast_{\vdagger} ||^4 =& O((N-j)^{-2}),
 \end{align*} 
 which implies that
 \begin{align*}
 \left( E|| I_b ||^2 \right)^{1/2} &\leq c \left( E || \hat{A}_{\vdagger} - A_{\vdagger} ||^2 \right)^{1/2} \sum_{j = 0}^{N - 1} \left[ (N - j)^{-1/2} \sup_{0< t \leq \Vert A_{\vdagger}\Vert^2} (1 - ct)^j t^{\beta/2} \right] \\
 &\quad+\left( E || \hat{A}^\ast_{\vdagger} - A^\ast_{\vdagger} ||^2 \right)^{1/2} \left[ \sup_{0< t \leq \Vert A_{\vdagger}\Vert^2} \sum_{j = 0}^{N - 1} \left( 1 - c t \right)^{j} (ct) t^{(\beta-1)/2} \right] \\
 &\leq c \left( E ||\hat{A}_{\vdagger} - A_{\vdagger} ||^2 \right)^{1/2} \sum_{j = 0}^{N - 1} (N - j)^{-1/2} (j+1)^{-\beta/2} \\
 &\quad+ \left( E || \hat{A}^\ast_{\vdagger} - A^\ast_{\vdagger} ||^2 \right)^{1/2} \left[ \sup_{0< t \leq \Vert A_{\vdagger}\Vert^2} \left( 1 - (1 - ct)^N \right) t^{(\beta-1)/2} \right]  \\
 &= O\left( \gamma_n N^{-\beta/2}\sqrt{N}\ln(N)^{\un(\beta = 2)} N^{\left( \frac{\beta}{2}-1 \right)\vee 0}\right),
 \end{align*}
where the last line follows from Definition \ref{qualifdef}, Assumption \ref{assapp6} and Lemma \ref{lemapp2}, with
 \[
 \sum_{j = 0}^{N - 1} (N - j)^{-1/2} (j + 1)^{-\beta/2} =O\left(N^{-\beta/2}\sqrt{N} \begin{cases} 1 & \text{ if } \beta < 2 \\ \ln(N) & \text{ if } \beta = 2\\ N^{\beta/2 - 1} & \text{ if } \beta > 2  \end{cases}\right).
 \]

 An equivalent line of proof can be used to show that 
 \[
 \left( E|| I_a ||^2 \right)^{1/2} \leq \left( E ||\left( I - c \hat{A}^\ast_{\vdagger} \hat{A}_{\vdagger} \right)^N ||^4 E || \hat\varphi_0 - \varphi_0 ||^4 \right)^{1/4} = O(1) \left( E || \hat\varphi_0 - \varphi_0 ||^4 \right)^{1/4}.
 \]
 Similarly, for $II$, we have 
 \begin{align*}
 E|| II ||^2 &\leq \left( E || c\sum_{j = 0}^{N-1} \left( I - c \hat{A}^\ast_{\vdagger} \hat{A}_{\vdagger} \right)^{j} \hat{A}^\ast_{\vdagger} ||^4 \right)^{1/2} \left( E || \hat{T} (\vdagger) - T(\vdagger) ||^4 \right)^{1/2}\\
 &= O\left( N \delta^2_n\right),
 \end{align*}
 where the last bound follows from Lemma \ref{lemapp3} and Assumption \ref{ass:rateconv}. 
 
 We now control the nonlinear terms. Let
 \begin{align*}
 \left( E|| III ||^2 \right)^{1/2} & \leq c \sum_{j = 0}^{N-1} \left( E\left\Vert \left( I - c \hat{A}^\ast_{\vdagger} \hat{A}_{\vdagger} \right)^{N-j-1} \hat{A}^\ast_{\vdagger} \right\Vert^2 || \hat{T} (\hat\varphi_{j}) - \hat{T} (\vdagger) - \hat{T}^{\prime}_{\vdagger} \hat{e}_{j} ||^2 \right)^{1/2} \\
& \leq c \hat{M} \sum_{j = 0}^{N-1} \left( E\left\Vert \left( I - c \hat{A}^\ast_{\vdagger} \hat{A}_{\vdagger} \right)^{N-j-1} \hat{A}^\ast_{\vdagger} \right\Vert^2 || \hat{e}_{j} ||^4 \right)^{1/2} \\
 &\leq \hat{M} \sum_{j = 0}^{N-1} \left( N-j \right)^{-1/2} \left( E || \hat{e}_{j} ||^8 \right)^{1/4},
 \end{align*}
where the second line follows from Proposition \ref{prop:koper}(a), the third line from Cauchy-Schwarz inequality, and the bound from Lemma \ref{lemapp3}. Similarly,
\begin{align*}
\left( E|| IV ||^2 \right)^{1/2} &= \left(E \left\Vert c\sum_{j = 0}^{N-1} \left( I - c \hat{A}^\ast_{\vdagger} \hat{A}_{\vdagger} \right)^{N-j-1} \hat{A}^\ast_{\vdagger} \left( K_{\hat\varphi_{j},\vdagger} - I \right) \left( \hat{T} (\hat\varphi_{j}) - T(\vdagger) \right) \right\Vert^2 \right)^{1/2} \\
&\leq c \sum_{j = 0}^{N-1}\left( E \left\Vert \left( I - c \hat{A}^\ast_{\vdagger} \hat{A}_{\vdagger} \right)^{N-j-1} \hat{A}^\ast_{\vdagger} \right\Vert^2 \left\Vert K_{\hat\varphi_{j},\vdagger} - I \right\Vert^2 \left\Vert \hat{T} (\hat\varphi_j) - \hat{T}(\vdagger) - \hat{T}^\prime_{\vdagger} \hat{e}_j \right\Vert^2 \right)^{1/2}\\
& \quad + c\sum_{j = 0}^{N-1}\left( E \left\Vert \left( I - c \hat{A}^\ast_{\vdagger} \hat{A}_{\vdagger} \right)^{N-j-1} \hat{A}^\ast_{\vdagger} \right\Vert^2 \left\Vert K_{\hat\varphi_{j},\vdagger} - I \right\Vert^2 \left\Vert \hat{T} (\vdagger) - T(\vdagger) \right\Vert^2 \right)^{1/2}\\
& \quad + c\sum_{j = 0}^{N-1}\left( E \left\Vert \left( I - c \hat{A}^\ast_{\vdagger} \hat{A}_{\vdagger} \right)^{N-j-1} \hat{A}^\ast_{\vdagger} \right\Vert^2 \left\Vert K_{\hat\varphi_{j},\vdagger} - I \right\Vert^2 \left\Vert\hat{T}^\prime_{\vdagger} \hat{e}_j \right\Vert^2 \right)^{1/2}\\
 &\leq c \sum_{j = 0}^{N-1} \left( N-j \right)^{-1/2} \left(E\left\Vert K_{\hat\varphi_{j},\vdagger} - I\right\Vert^4 \left\Vert \hat{T} (\hat\varphi_j) - \hat{T}(\vdagger) - \hat{T}^\prime_{\vdagger} \hat{e}_j \right\Vert^4 \right)^{1/4}\\
& \quad + c\sum_{j = 0}^{N-1}\left( N-j \right)^{-1/2} \left(E \left\Vert K_{\hat\varphi_{j},\vdagger} - I \right\Vert^4 \left\Vert \hat{T} (\vdagger) - T(\vdagger) \right\Vert^4 \right)^{1/4}\\
& \quad + c\sum_{j = 0}^{N-1} \left( N-j \right)^{-1/2} \left(E \left\Vert K_{\hat\varphi_{j},\vdagger} - I \right\Vert^4 \left\Vert\hat{T}^\prime_{\vdagger} \hat{e}_j \right\Vert^4 \right)^{1/4}\\
&\leq c \hat{M}^2 (h^2_u \kappa_n)^{-1} \sum_{j = 0}^{N-1} \left( N-j \right)^{-1/2} \left(E\left\Vert \hat{e}_j \right\Vert^{12} \right)^{1/4}\\
& \quad + c M (h^2_u \kappa_n)^{-1} \delta_n \sum_{j = 0}^{N-1}\left( N-j \right)^{-1/2} \left(E \left\Vert \hat{e}_j \right\Vert^8 \right)^{1/8}\\
& \quad + c M (h^2_u \kappa_n)^{-1} \sum_{j = 0}^{N-1} \left( N-j \right)^{-1/2} \left(E \left\Vert \hat{e}_j \right\Vert^8 E\left\Vert\hat{T}^\prime_{\vdagger} \hat{e}_j \right\Vert^8 \right)^{1/8}\\
=& \left( E \Vert IV_a \Vert^2 \right)^{1/2} + \left( E \Vert IV_b \Vert^2 \right)^{1/2} + \left( E \Vert IV_c \Vert^2 \right)^{1/2},
\end{align*}
where the second inequality follows from Cauchy-Schwarz and Lemma \ref{lemapp3}, and the third inequality follows from Proposition \ref{prop:koper}(b). Finally, the last line is simply for the convenience of the reader, as it assigns a label to each one of the terms in the bound.

We prove the result of the Theorem by induction. To simplify notations, let $$\tilde{\beta} = \frac{1}{a}\left( (a + 1)\beta + 1\right) = \frac{s}{a}.$$ We further let 
\begin{align*}
E \left\Vert \hat{e}_j \right\Vert^2 \leq& 2 E \left\Vert \hat\varphi_j - \varphi_j \right\Vert^2 + 2 E \left\Vert e_j \right\Vert^2 < 2 E \left\Vert \hat\varphi_j - \varphi_j \right\Vert^2 + \frac{2}{M} E \left\Vert T^\prime_{\vdagger} e_j \right\Vert \\
=& O \left( (j+1) \delta^2_n + (j+1)^{1-\tilde{\beta}}\gamma^2_n + (j+1)^{-\frac{\tilde{\beta}+1}{2}} \right),
\end{align*}
where the last inequality on the first line follows from Assumption \ref{ass:identification}, using the restriction imposed by Definition \ref{ass:sourcecondID} about the regularity of the functions in $\mathcal{B}_\dagger$, and
\begin{align*}
E \left\Vert \hat{T}^\prime_{\vdagger}\hat{e}_j \right\Vert^2 =& O \left( \delta^2_n + (j+1)^{-\tilde{\beta}}\gamma^2_n + (j+1)^{-(\tilde{\beta} + 1)}\right),
\end{align*}
and for $l = \lbrace 1,2,3,4,5,6 \rbrace$,
\begin{align*}
\left( E \left\Vert \hat{e}_j \right\Vert^{2l} \right)^{1/2l} =& O \left( E \left\Vert \hat{e}_j \right\Vert^{2}\right)^{1/2}\\
\left(E \left\Vert \hat{T}^{\prime}_{\vdagger}\hat{e}_j \right\Vert^{2l} \right)^{1/2l} =& O \left( E \left\Vert \hat{T}^{\prime}_{\vdagger}\hat{e}_j \right\Vert^2 \right)^{1/2}.
\end{align*}

The final result is established by controlling the remaining terms. We have

\begin{align*}
\left( E|| III ||^2 \right)^{1/2} & \leq c \hat{M} \left( \delta^2_n \sum_{j = 0}^{N-1} \left( N-j \right)^{-1/2} \left( j + 1 \right) \right. \\
&\quad + \left.\gamma^2_n \sum_{j = 0}^{N-1} \left( N-j \right)^{-1/2} (j+1)^{1-\tilde{\beta}} + \frac{1}{M} \sum_{j = 0}^{N-1} \left( N-j \right)^{-1/2} (j+1)^{-\frac{\tilde{\beta}+1}{2}} \right),
\end{align*}
and we analyze these terms one by one. 

First
\begin{align*}
\delta^2_n \sum_{j = 0}^{N-1} \left( N-j \right)^{-1/2} (j+1) &= O\left( \delta^2_n N \sqrt{N}\right),
\end{align*}
as
\begin{align*}
\sum_{j = 0}^{N-1} & \left( N-j \right)^{-1/2} (j+1) \leq (N+1)^{3/2}\int_{\frac{1}{N+1}}^{\frac{N}{N+1}} \frac{v}{\sqrt{1-v}} dv \\
=& 2 (N+1)^{3/2}\left[ \frac{1}{N+1}\sqrt{\frac{N}{N+1}} - \frac{N}{N+1}\sqrt{\frac{1}{N+1}} + \frac{2}{3}\frac{N}{(N+1)^{3/2}} - \frac{2}{3}\left( \frac{1}{N+1} \right)^{3/2} \right]\\
=& 2 \left[ \sqrt{N} - N + \frac{2}{3} N^{3/2} - \frac{2}{3} \right] = O\left(N \sqrt{N}\right).
\end{align*}

Similarly, for $\tilde{\beta} \leq 1$,
\[
\gamma^2_n \sum_{j = 0}^{N-1} \left( N-j \right)^{-1/2} (j+1)^{1-\tilde{\beta}} = O\left( \gamma^2_n N^{3/2 - \tilde{\beta}} \right). 
\]
as,
\begin{align*}
\sum_{j = 0}^{N-1} & \left( N-j \right)^{-1/2} (j+1)^{1-\tilde{\beta}} \leq (N+1)^{3/2-\tilde{\beta}}\int_{\frac{1}{N+1}}^{\frac{N}{N+1}} \frac{v^{1-\tilde{\beta}}}{\sqrt{1-v}} dv \\
\leq& (N+1)^{3/2-\tilde{\beta}}\int_{\frac{1}{N+1}}^{\frac{N}{N+1}} \frac{1}{\sqrt{1-v}} dv\\
=& 2 (N+1)^{3/2-\tilde{\beta}} \left[ \sqrt{\frac{N}{N+1}} - \sqrt{\frac{1}{N+1}} \right] = O\left(N^{3/2-\tilde{\beta}} \right).
\end{align*}

Otherwise, if $\tilde{\beta} > 1$,
\[
\gamma^2_n \sum_{j = 0}^{N-1} \left( N-j \right)^{-1/2} (j+1)^{1-\tilde{\beta}} = O\left( \gamma^2_n N^{3/2-\tilde{\beta}} \begin{cases} 1 & 1 < \tilde{\beta} < 2 \\ \ln(N) & \tilde{\beta}=2 \\ N^{\tilde{\beta} - 2} & \tilde{\beta} > 2 \end{cases}\right), 
\]
which finally gives
\[
\gamma^2_n \sum_{j = 0}^{N-1} \left( N-j \right)^{-1/2} (j+1)^{1-\tilde{\beta}} = O\left( \gamma^2_n N^{3/2-\tilde{\beta}} \begin{cases} 1 &  \tilde{\beta}< 2\\ \ln(N) & \tilde{\beta}=2 \\ N^{\tilde{\beta} - 2} & \tilde{\beta}>2 \end{cases}\right), 
\]

For the bias component, we have instead
\[
\frac{1}{M}\sum_{j = 0}^{N-1} \left( N-j \right)^{-1/2} (j+1)^{-\frac{\tilde{\beta}+1}{2}} \leq \begin{cases} N^{-\tilde{\beta}/2} & \tilde{\beta} < 1 \\ N^{-1/2} \ln(N) & \tilde{\beta}=1  \\ N^{-1/2} & \tilde{\beta} > 1 \end{cases}.
\]
Thus,
\begin{align*}
\left( E|| III ||^2 \right)^{1/2} =& O\left(\delta^2_n N \sqrt{N} + \gamma^2_n N^{1-\tilde{\beta}} \sqrt{N} \ln(N)^{\un(\tilde{\beta} = 2)} N^{(\tilde{\beta} -2)\vee 0} + N^{-(\tilde{\beta} \wedge 1)/2}\ln(N)^{\un\left(\tilde{\beta} = 1\right)} \right)\\
=& o \left(\delta_n \sqrt{N} +  \gamma_n N^{\frac{1-\beta}{2}} \ln(N)^{\un (\beta = 2)} N^{\frac{\beta}{2} - 1 \vee 0} + N^{-(\beta \wedge 1-\epsilon)/2}\right),
\end{align*}
where the last line follows from Assumption \ref{ass:restrvarth}.

Reasoning as above, we have that
\begin{align*}
\sum_{j = 0}^{N-1} & \left( N-j \right)^{-1/2} \left( j + 1 \right)^{3/2} = O\left( N^2\right),\\
\sum_{j = 0}^{N-1} & \left( N-j \right)^{-1/2} \left( j + 1 \right)^{1/2} = O\left( N \right),\\
\sum_{j = 0}^{N-1} & \left( N-j \right)^{-1/2} (j+1)^{\frac{3}{2}\left( 1-\tilde{\beta}\right)} = O\left( N^{2 - \frac{3}{2}\tilde{\beta}} \begin{cases} 1 & \tilde{\beta} < \frac{5}{3} \\ \ln(N) & \tilde{\beta}=\frac{5}{3} \\ N^{\frac{3}{2}\tilde{\beta} - \frac{5}{2}} & \tilde{\beta}>\frac{5}{3}  \end{cases} \right),\\
\sum_{j = 0}^{N-1} & \left( N-j \right)^{-1/2} (j+1)^{\frac{1}{2}\left( 1-\tilde{\beta} \right)} = O\left( N^{1 - \frac{\tilde{\beta}}{2}} \begin{cases} 1 & \tilde{\beta} < 3 \\ \ln(N) & \tilde{\beta}=3  \\ N^{\frac{\tilde{\beta}}{2} - \frac{3}{2}} & \tilde{\beta}>3  \end{cases}\right),\\
\sum_{j = 0}^{N-1} &\left( N-j \right)^{-1/2} (j+1)^{-\frac{3}{4}\left( \tilde{\beta}+1 \right)}= O\left( N^{-\frac{3\tilde{\beta}+1}{4}} \begin{cases} 1 & \tilde{\beta} < \frac{1}{3} \\ \ln(N) & \tilde{\beta}=\frac{1}{3}  \\ N^{\frac{3\tilde{\beta}-1}{4}} & \tilde{\beta}>\frac{1}{3}   \end{cases}\right),\\
\sum_{j = 0}^{N-1} &\left( N-j \right)^{-1/2} (j+1)^{-\frac{1}{4}\left( \tilde{\beta}+1 \right)}= O\left( N^{-\frac{\tilde{\beta}-1}{4}} \begin{cases} 1 & \tilde{\beta} < 3 \\ \ln(N) & \tilde{\beta}=3  \\ N^{\frac{\tilde{\beta}-3}{4}} & \tilde{\beta}>3  \end{cases}\right).
\end{align*}
This implies
\begin{align*}
\left( E|| IV_a ||^2 \right)^{1/2} & \leq c \hat{M}^2 (h^2_u \kappa_n)^{-1} \left( \delta^3_n \sum_{j = 0}^{N-1} \left( N-j \right)^{-1/2} \left( j + 1 \right)^{3/2} \right. \\
&\quad + \left.\gamma^3_n \sum_{j = 0}^{N-1} \left( N-j \right)^{-1/2} (j+1)^{\frac{3}{2}\left( 1-\tilde{\beta} \right)} + \frac{1}{M} \sum_{j = 0}^{N-1} \left( N-j \right)^{-1/2} (j+1)^{-\frac{3}{4}\left( \tilde{\beta}+1 \right)} \right)\\
=& O\left( (h^2_u \kappa_n)^{-1} \left( \delta_n^3 N^2 + \gamma^3_n N^{2 - \frac{3}{2}\tilde{\beta}}\ln(N)^{\un\left(\tilde{\beta} = \frac{5}{3}\right)}N^{\left(\frac{3}{2}\tilde{\beta} - \frac{5}{2} \right)\vee 0}+ N^{-\frac{3\tilde{\beta} + 1}{4}}\ln(N)^{\un\left(\tilde{\beta} = \frac{1}{3}\right)} N^{\frac{3\tilde{\beta} - 1}{4}\vee 0}\right) \right)\\
=& o\left( \delta_n \sqrt{N} + \gamma_n N^{\frac{1-\beta}{2}} \ln(N)^{\un (\beta = 2)} N^{\frac{\beta}{2} - 1 \vee 0}\right) + O \left( N^{-(\beta \wedge  (1-\epsilon))/2}\right),\\
\left( E|| IV_b ||^2 \right)^{1/2} & \leq c M (h^2_u \kappa_n)^{-1} \delta_n\left( \delta_n \sum_{j = 0}^{N-1}\left( N-j \right)^{-1/2} \left(j+1\right)^{1/2} \right. \\
&\quad + \left. \gamma_n \sum_{j = 0}^{N-1}\left( N-j \right)^{-1/2} \left(j+1\right)^{(1-\tilde{\beta})/2}+  \sum_{j = 0}^{N-1}\left( N-j \right)^{-1/2} \left(j+1\right)^{-(\tilde{\beta}+1)/4}\right)\\
=& O\left( \frac{\delta_n \sqrt{N}}{h^2_u \kappa_n} \left( \delta_n \sqrt{N}  + \gamma_n N^{\frac{1 - \tilde{\beta}}{2}}\ln(N)^{\un\left(\tilde{\beta} = 3\right)}N^{\left(\frac{\tilde{\beta}-3}{2} \right)\vee 0}+ N^{-\frac{\tilde{\beta} + 1}{4}}\ln(N)^{\un\left(\tilde{\beta} =3 \right)} N^{\frac{\tilde{\beta} - 3}{4}\vee 0}\right) \right)\\
=& o\left( \delta_n \sqrt{N} + \gamma_n N^{\frac{1-\beta}{2}} \ln(N)^{\un (\beta = 2)} N^{\frac{\beta}{2} - 1 \vee 0} + N^{-(\beta \wedge  (1-\epsilon))/2}\right),\\
\left( E|| IV_c ||^2 \right)^{1/2} & \leq c M (h^2_u \kappa_n)^{-1} \left( \delta_n^2 \sum_{j = 0}^{N-1} \left( N-j \right)^{-1/2} \left(j+1\right)^{1/2} \right. \\
&\quad + \left. \gamma^2_n \sum_{j = 0}^{N-1} \left( N-j \right)^{-1/2} \left(j+1\right)^{1/2 - \tilde{\beta}} +   \sum_{j = 0}^{N-1} \left( N-j \right)^{-1/2} \left(j+1\right)^{-\frac{3}{4}\left(\tilde{\beta}+1 \right)}\right)\\
=& O\left(  (h^2_u \kappa_n)^{-1} \left( \delta^2_n N  + \gamma^2_n N^{1 - \tilde{\beta}}\ln(N)^{\un\left(\tilde{\beta} = \frac{3}{2}\right)}N^{\left(\tilde{\beta}-\frac{3}{2} \right)\vee 0}+ N^{-\frac{3\tilde{\beta} + 1}{4}}\ln(N)^{\un\left(\tilde{\beta} = \frac{1}{3}\right)} N^{\frac{3\tilde{\beta} - 1}{4}\vee 0}\right)\right)\\
=& o\left( \delta_n \sqrt{N} + \gamma_n N^{\frac{1-\beta}{2}} \ln(N)^{\un (\beta = 2)} N^{\frac{\beta}{2} - 1 \vee 0}\right) + O \left( N^{-(\beta \wedge (1-\epsilon))/2}\right).
\end{align*}

By the restrictions imposed in Assumption \ref{ass:restrvarth}, we have that the nonlinear terms are therefore of the same order as or of smaller order than the linear ones. The proof for $\hat{\varphi}_N$ follows similar arguments. Assumption \ref{ass:linkcond} implies 
\begin{align*}
D^{-1} \sim \left( T^{\prime \ast}_{\vdagger}T^{\prime}_{\vdagger}\right)^{\frac{1}{2a}} \sim \left( A^{\ast}_{\vdagger}A_{\vdagger}\right)^{\frac{1}{2(a+1)}},
\end{align*}
and 
\begin{align*}
E\Vert & \left( A^{\ast}_{\vdagger}A_{\vdagger}\right)^{\frac{1}{2(a+1)}} \left(I - c \hat{A}^{\ast}_{\vdagger}\hat{A}_{\vdagger} \right)^N  \Vert^2 =  O\left( N^{-\frac{1}{a+1}} \right) \\
E\Vert & \left( A^{\ast}_{\vdagger}A_{\vdagger}\right)^{\frac{1}{2(a+1)}} \left(I - c \hat{A}^{\ast}_{\vdagger}\hat{A}_{\vdagger} \right)^N  \hat{A}^{\ast}_{\vdagger} \Vert^2 =  O\left( N^{-\frac{a+2}{a+1}} \right),
\end{align*}
by Lemma \ref{lemapp3}. Thus, 
\begin{align*}
\hat{e}_N &=\left( A^{\ast}_{\vdagger}A_{\vdagger}\right)^{\frac{1}{2(a+1)}}\left( I - c \hat{A}^\ast_{\vdagger} \hat{A}_{\vdagger} \right)^N \left( \hat\varphi^\prime_0 - \vdagger^\prime\right) \\
&\quad-c\sum_{j = 0}^{N-1} \left( A^{\ast}_{\vdagger}A_{\vdagger}\right)^{\frac{1}{2(a+1)}} \left( I - c \hat{A}^\ast_{\vdagger} \hat{A}_{\vdagger} \right)^{j} \hat{A}^\ast_{\vdagger}\left( \hat{T} (\vdagger) - T(\vdagger) \right)\\
&\quad-c\sum_{j = 0}^{N-1} \left( A^{\ast}_{\vdagger}A_{\vdagger}\right)^{\frac{1}{2(a+1)}}\left( I - c \hat{A}^\ast_{\vdagger} \hat{A}_{\vdagger} \right)^{N-j-1} \hat{A}^\ast_{\vdagger} \left( \hat{T} (\hat\varphi_j) - \hat{T} (\vdagger) - \hat{T}^{\prime}_{\vdagger} \hat{e}_j \right) \\
&\quad-c\sum_{j = 0}^{N-1} \left( A^{\ast}_{\vdagger}A_{\vdagger}\right)^{\frac{1}{2(a+1)}}\left( I - c \hat{A}^\ast_{\vdagger} \hat{A}^{\prime}_{\vdagger} \right)^{N-j-1} \left( \hat{A}^\ast_{\hat\varphi_j} - \hat{A}^\ast_{\vdagger} \right) \left( \hat{T} (\hat\varphi_j) - T(\vdagger) \right).
\end{align*}
Therefore, we have that
\begin{align*}
\left( E \Vert \hat{e}_N \Vert^2 \right)^{1/2} =& O\left( N^{-\frac{1}{2(a + 1)}}\left(E \Vert  \hat\varphi^\prime_0 - \vdagger^\prime \Vert^2\right)^{1/2} + \delta_n N^{\frac{a}{2(a + 1)}} + \gamma_n N^{\frac{a}{2(a + 1)} -\frac{\beta}{2}} \ln(N)^{\un(\beta = 2)} N^{\left( \frac{\beta}{2}-1 \right)\vee 0} \right. \\
\quad & + \left. N^{- \frac{\beta \wedge (1- \epsilon)}{2} - \frac{1}{2(a + 1)}} \right).
\end{align*}

The result of the Theorem follows. 


\subsection{Proofs for Section \ref{sec:mcsim}} \label{annexA4}

We provide here a proof of the partial identification result for the nonparametric instrumental variable estimator under mean independence. We use the example of a continuous endogenous variable and a binary instrument. The proof can be easily extended to any discrete instrument with a finite number of support points. When the number of support point is not finite, the proof is different, and we provide a simple example of the lack of identification below. Recall that $K$ is the conditional expectation operator such that
\[
K\varphi=E(\varphi(X)\vert W),
\]
and let $\mathcal{N} (K)$ be the null space of $K$. That is,
\[
\mathcal{N} (K) = \lbrace \psi \in L^2_X : K\psi = 0 \rbrace. 
\]

\begin{proposition}
Let $X \in \IR$ and $W \in \lbrace 0,1\rbrace$, such that 
\[
\eta_0(x) = \frac{f_{X\vert W}(x \vert w = 0)}{f_X(x)}, \text{ and } \eta_1(x) = \frac{f_{X\vert W}(x \vert w = 1)}{f_X(x)}
\]
are linearly independent. Then, the pseudo-true solution to the integral equation 
\[
K\varphi = r,
\]
with $r = E(Y \vert W)$, is equal to
\[
\tilde{\varphi} = \sum_{j = \lbrace 0,1 \rbrace} \lambda_j \eta_j,
\]
where the $\lambda_j$'s are the unique solutions to the system of equations
\[
r(l)  = \sum_{j = \lbrace 0,1 \rbrace} \lambda_j \int\eta_j(x) f_{X\vert W}(x \vert w = l) dx,
\]
for $l = \{ 0,1 \}$. 
\begin{proof}
Take $\psi \in \mathcal{N}(K)$, and let $\eta_j$, for $j = \lbrace 0,1\rbrace$, as defined in the Proposition. Then, we have that
\begin{align*}
(K\psi)(j) =& \int \psi(x) f_{X\vert W} (x \vert w = j) dx \\
=& \int \psi(x) \eta_j(x) f_{X}(x ) dx = E(\psi \eta_j) = 0.
\end{align*}
This implies that $\eta_j \in \mathcal{N}(K)^{\perp}$, the orthogonal space to $\mathcal{N}(K)$. Furthermore, as they are linearly independent they form a basis for $\mathcal{N}(K)^{\perp}$. That is, every element of $\mathcal{N}(K)^{\perp}$ that is orthogonal to $\eta_0$ and $\eta_1$ is in $\mathcal{N}(K)$, and it is therefore equal to $0$. We conclude that the dimension of $\mathcal{N}(K)^{\perp}$ is $2$ and that every element of $\mathcal{N}(K)^{\perp}$ can be written as a linear combination of $\eta_0$ and $\eta_1$. 

Thus, there must exist constants $\lambda_0$ and $\lambda_1$ such that, for a $\tilde{\varphi} \in \mathcal{N}(K)^{\perp}$
\[
\tilde{\varphi} = \sum_{j = \lbrace 0,1 \rbrace} \lambda_j \eta_j. 
\]

Moreover, these $\lambda_j$'s must necessarily be unique as, for $\psi = \varphi - \tilde{\varphi} \in \mathcal{N}(K)^{\perp}$, either $K \psi = 0$, and $\psi \in \mathcal{N}(K)$, which implies $\varphi = \tilde{\varphi}$; or $K \psi \neq 0$, and thus $\varphi \neq \tilde{\varphi}$. This concludes the proof. 
\end{proof}
\end{proposition}

\begin{example}[Identification with $W$ discrete with infinite support] \label{ex:infsuppins}
We let $(Y,X,W) \in \IR^3$, $Y = \vdagger(X) + U$, with $E( U \vert W) = 0$ and $X \vert W = w \sim \mathcal{U} \left[ 2\pi w, 2 \pi(w + 1)\right]$, with $w = \lbrace 0,1,2,\dots \rbrace$. The conditional expectation operator
\[
(K \vdagger)(w) = \frac{1}{2\pi} \int_{2\pi w}^{2\pi (w + 1)} \vdagger(x) dx. 
\]
For the completeness condition to hold, we need to have that $K \varphi = 0$, implies $\varphi= 0$, for all $\varphi \in L^2_X$. Let $\psi$ be a $2\pi$-periodic function, i.e., $\psi(x + 2\pi) = \psi(x)$, such that 
\[
\int_0^{2\pi} \psi(x)dx = 0,
\]
i.e., $\psi \in \mathcal{N}(K)$ and $\tilde{\varphi} = \vdagger + \psi$. E.g., $\psi(x) = \sin(x)$. Then 
\[
(K (\tilde{\varphi} - \vdagger))(w) = \frac{1}{2\pi} \int_{2\pi w}^{2\pi (w + 1)} \psi(x) dx = 0,
\]
where the last line follows from the fact that $w$ is an integer and the $2\pi$ periodicity of $\psi$. The last equality implies that the function $\vdagger$ is not uniquely identified in this example.
\end{example}

\subsection{Useful lemmas} \label{annexA5}

We collect here some lemmas useful for the proofs above. Results are given without proof, and we refer the reader to the corresponding source. 

\begin{lemma}[\citeauthor{kalten2008} (\citeyear{kalten2008}, Lemma 2.9, p. 17)] \label{lemapp2}
Let $a$ and $b$ be non-negative. Then there is a positive constant $M(a,b)$ independent of $N$ so that
\begin{equation*}
\sum_{j = 0}^{N - 1} (N- j)^{-a} (j+1)^{-b} \leq M(a,b) N^{1-a-b}\begin{cases} 1 & a \vee b < 1 \\ \ln(N) & a \vee b =1 \\ N^{a \vee b - 1} & a \vee b > 1\end{cases}.
\end{equation*}
\end{lemma}

\begin{lemma}[\citeauthor{kalten2008} (\citeyear{kalten2008}, Lemma 2.10, p. 18)] \label{lemapp1}
Let $T$ be a compact operator such that $c\Vert T \Vert^2 \leq 1$, with $T^\ast$ be its adjoint. Further let $a \in [0,1]$, and $N\geq 0$, an integer. Then the following estimates hold
\begin{align*}
||\left( I - cT^\ast T\right)^N \left( T^\ast T \right)^a || =& O\left( N^{-a}\right), \\
\left\Vert c \sum_{j = 0}^{N-1}\left( I - cT^\ast T\right)^j \left( T^\ast T \right)^a\right\Vert =& O\left( N^{1-a}\right).
\end{align*}
\end{lemma}

\begin{lemma} \label{lemapp3}
Let $T$ be a compact operator and $T^\ast$ its adjoint, such that $c\Vert T^\ast T \Vert < 1$, for $c > 0$. Further let $\hat{T}$ and $\hat{T}^\ast$ be nonparametric estimators of $T$ and $T^\ast$, respectively, such that for every positive integer $l$
\[
E || \hat{T}^\ast - T^\ast ||^{2l} =E || \hat{T} - T ||^{2l} = O\left( \gamma^{2l}_n\right),
\]
with $\gamma_n$ satisfying Assumption \ref{ass:restrvarth}(i)-(ii). Further let $N\geq 1$, an integer, and $a \in [0,0.5)$. Then
\begin{align}
\left( E || \left(T^\ast T\right)^a \left( I - c\hat{T}^\ast \hat{T}\right)^N ||^{2l} \right)^{1/2l}=& O\left( N^{-a}\right), \label{lemmaA3eq1}\\
\left( E || \left(T^\ast T\right)^a \left( I - c\hat{T}^\ast \hat{T}\right)^N \hat{T}^\ast||^{2l} \right)^{1/2l}=& O\left( N^{-1/2 - a}\right). \label{lemmaA3eq2}
\end{align}
\begin{proof}
To simplify notations, we let 
\begin{align*}
\hat{Q} =& I - c\hat{T}^\ast \hat{T}\\
Q =& I - c T^\ast T\\
\hat{P}=& - \left( \hat{T}^\ast \hat{T} - T^\ast T \right) = \hat{P}_1 + \hat{P}_2 + \hat{P}_3 \\
=& -T^\ast \left( \hat{T} - T \right) - \left( \hat{T}^\ast - T^\ast \right) T - \left( \hat{T}^\ast - T^\ast \right) \left( \hat{T} - T \right).
\end{align*}
To prove \ref{lemmaA3eq1}, we write
\begin{align*}
\left( E || \left(T^\ast T\right)^a \hat{Q}^N ||^{2l} \right)^{1/2l} \leq& \left( E || \left(T^\ast T\right)^a  \left[ \left( Q + c\hat{P}\right)^N - Q^N \right] ||^{2l} \right)^{1/2l} + || \left(T^\ast T\right)^a Q^N ||.
\end{align*}
The second term is $O(N^{-a})$ from Lemma \ref{lemapp1}. Further, for indexes $j_1,j_2 \geq 0$, and using the result in Lemma \ref{lemapp1}, we have that 
\begin{equation} \label{lemmaA3b1}
\begin{aligned}
\left( E ||  \left(T^\ast T\right)^a Q^{j_1} \hat{P} Q^{j_2} ||^{2l} \right)^{1/2l} =& \left( E || \left(T^\ast T\right)^a Q^{j_1} \hat{P}_1  Q^{j_2} ||^{2l} \right)^{1/2l} + \left( E || \left(T^\ast T\right)^a  Q^{j_1}  \hat{P}_2 Q^{j_2} ||^{2l} \right)^{1/2l} \\
& \quad + \left( E || \left(T^\ast T\right)^a  Q^{j_1} \hat{P}_3  Q^{j_2} ||^{2l}\right)^{1/2l} \\
=& \left( E || \left(T^\ast T\right)^a Q^{j_1} \hat{P}_1 ||^{2l} \right)^{1/2l} + \left( E || \left(T^\ast T\right)^a  Q^{j_1}  \hat{P}_2 Q^{j_2} ||^{2l} \right)^{1/2l} \\
& \quad + \left( E || \left(T^\ast T\right)^a  Q^{j_1} \hat{P}_3 ||^{2l}\right)^{1/2l} \\
=& O\left( (j_1 + 1)^{-1/2 - a}\gamma_n +  (j_1 + 1)^{-a} (j_2 + 1)^{-1/2}\gamma_n +  (j_1 + 1)^{-a} \gamma_n^2 \right).
\end{aligned}
\end{equation}
We wish to show that
\[
 \left( E || \left(T^\ast T\right)^a  \left[ \left( Q + c\hat{P}\right)^N - Q^N \right]||^{2l} \right)^{1/2l} = c\left( E \left\Vert \sum_{j = 0}^{N -1} \left(T^\ast T\right)^a Q^j \hat{P} Q^{N - j - 1} \right\Vert^{2l} \right)^{1/2l} (1 + o(1)),
\]
where the leading term is the first-order directional derivative of $Q^N$ at $\hat{P}$. We have
\begin{align*}
& \left( E \left\Vert \sum_{j = 0}^{N -1} \left(T^\ast T\right)^a Q^j \hat{P} Q^{N - j - 1} \right\Vert^{2l} \right)^{1/2l} \\
& \quad \leq \left( E \left\Vert \sum_{j = 0}^{N -1} \left(T^\ast T\right)^a Q^j \hat{P}_1 Q^{N - j - 1} \right\Vert^{2l} \right)^{1/2l} \\
& \quad + \left( E \left\Vert \sum_{j = 0}^{N -1} \left(T^\ast T\right)^a Q^j \hat{P}_2 Q^{N - j - 1} \right\Vert^{2l} \right)^{1/2l} + \left( E \left\Vert \sum_{j = 0}^{N -1} \left(T^\ast T\right)^a Q^j \hat{P}_3 Q^{N - j - 1} \right\Vert^{2l} \right)^{1/2l} \\
& \quad \leq \sum_{j = 0}^{N -1} \left[ \left( E \left\Vert \left(T^\ast T\right)^a Q^j \hat{P}_1 Q^{N - j - 1} \right\Vert^{2l} \right)^{1/2l} \right.\\
& \quad + \left. \left( E \left\Vert \left(T^\ast T\right)^a Q^j \hat{P}_2 Q^{N - j - 1} \right\Vert^{2l} \right)^{1/2l} + \left( E \left\Vert \left(T^\ast T\right)^a Q^j \hat{P}_3 Q^{N - j - 1} \right\Vert^{2l} \right)^{1/2l}\right] \\
& \quad= O \left( \sum_{j = 0}^{N -1} (j + 1)^{-1/2 - a} \gamma_n + (N - j)^{-1/2} (j + 1)^{-a}\gamma_n + (j + 1)^{-a} \gamma_n^2 \right) \\
& \quad = O \left( N^{\frac{1}{2}-a} \gamma_n + N^{1-a} \gamma^2_n \right) = O \left( N^{\frac{1}{2}-a} \gamma_n (1 + \sqrt{N} \gamma_n) \right) = O \left( N^{\frac{1}{2}-a} \gamma_n \right),
\end{align*}
where the fourth line follows directly from the bound in \ref{lemmaA3b1}, and the last line from Lemma \ref{lemapp2}, and Assumption \ref{ass:restrvarth}(i). To conclude the proof, we show that higher order terms in the Taylor approximation are negligible under the maintained assumptions. The second directional derivative can be written as
{\tiny
\begin{align*}
& \left( E \left\Vert \sum_{j = 0}^{N - 2} \sum_{k = 0}^{N - 2 - j}\left(T^\ast T\right)^a Q^j \hat{P} Q^k \hat{P} Q^{N - j - k - 2} \right\Vert^{2l} \right)^{1/2l} \\
& \quad \leq \left( E \left\Vert \sum_{j = 0}^{N - 2} \sum_{k = 0}^{N - 2 - j} \left(T^\ast T\right)^a Q^j \hat{P}_1 Q^k \hat{P}_1 Q^{N - j - k - 2} \right\Vert^{2l} \right)^{1/2l} + \left( E \left\Vert \sum_{j = 0}^{N - 2} \sum_{k = 0}^{N - 2 - j}\left(T^\ast T\right)^a Q^j \hat{P}_1 Q^k \hat{P}_2 Q^{N - j - k - 2} \right\Vert^{2l} \right)^{1/2l} \\
& \quad + \left( E \left\Vert \sum_{j = 0}^{N - 2} \sum_{k = 0}^{N - 2 - j} \left(T^\ast T\right)^a Q^j \hat{P}_1 Q^k \hat{P}_3 Q^{N - j - k - 2} \right\Vert^{2l} \right)^{1/2l} \\
& \quad + \left( E \left\Vert \sum_{j = 0}^{N - 2} \sum_{k = 0}^{N - 2 - j} \left(T^\ast T\right)^a Q^j \hat{P}_2 Q^k \hat{P}_1 Q^{N - j - k - 2} \right\Vert^{2l} \right)^{1/2l} + \left( E \left\Vert \sum_{j = 0}^{N - 2} \sum_{k = 0}^{N - 2 - j} \left(T^\ast T\right)^a Q^j \hat{P}_2 Q^k \hat{P}_2 Q^{N - j - k - 2} \right\Vert^{2l}\right)^{1/2l} \\
& \quad + \left( E \left\Vert \sum_{j = 0}^{N - 2} \sum_{k = 0}^{N - 2 - j} \left(T^\ast T\right)^a Q^j \hat{P}_2 Q^k \hat{P}_3 Q^{N - j - k - 2} \right\Vert^{2l} \right)^{1/2l} \\
& \quad + \left( E \left\Vert \sum_{j = 0}^{N - 2} \sum_{k = 0}^{N - 2 - j} \left(T^\ast T\right)^a Q^j \hat{P}_3 Q^k \hat{P}_1 Q^{N - j - k - 2} \right\Vert^{2l} \right)^{1/2l} + \left( E \left\Vert \sum_{j = 0}^{N - 2} \sum_{k = 0}^{N - 2 - j} \left(T^\ast T\right)^a Q^j \hat{P}_3 Q^k \hat{P}_2 Q^{N - j - k - 2} \right\Vert^{2l} \right)^{1/2l} \\
& \quad + \left( E \left\Vert \sum_{j = 0}^{N - 2} \sum_{k = 0}^{N - 2 - j} \left(T^\ast T\right)^a Q^j \hat{P}_3 Q^k \hat{P}_3 Q^{N - j - k - 2} \right\Vert^{2l} \right)^{1/2l} \\
\leq & \sum_{j = 0}^{N - 2} \sum_{k = 0}^{N - 2 - j} \left[ \left( E \left\Vert \left(T^\ast T\right)^a Q^j \hat{P}_1 Q^k \hat{P}_1 Q^{N - j - k - 2} \right\Vert^{2l} \right)^{1/2l} + \left( E \left\Vert \left(T^\ast T\right)^a Q^j \hat{P}_1 Q^k \hat{P}_2 Q^{N - j - k - 2} \right\Vert^{2l} \right)^{1/2l} + \left( E \left\Vert \left(T^\ast T\right)^a Q^j \hat{P}_1 Q^k \hat{P}_3 Q^{N - j - k - 2} \right\Vert^{2l} \right)^{1/2l}\right. \\
& \quad + \left( E \left\Vert \left(T^\ast T\right)^a Q^j \hat{P}_2 Q^k \hat{P}_1 Q^{N - j - k - 2} \right\Vert^{2l} \right)^{1/2l} + \left( E \left\Vert \left(T^\ast T\right)^a Q^j \hat{P}_2 Q^k \hat{P}_2 Q^{N - j - k - 2} \right\Vert^{2l} \right)^{1/2l} + \left( E \left\Vert \left(T^\ast T\right)^a Q^j \hat{P}_2 Q^k \hat{P}_3 Q^{N - j - k - 2} \right\Vert^{2l} \right)^{1/2l} \\
& \quad + \left. \left( E \left\Vert \left(T^\ast T\right)^a Q^j \hat{P}_3 Q^k \hat{P}_1 Q^{N - j - k - 2} \right\Vert^{2l} \right)^{1/2l} + \left( E \left\Vert \left(T^\ast T\right)^a Q^j \hat{P}_3 Q^k \hat{P}_2 Q^{N - j - k - 2} \right\Vert^{2l} \right)^{1/2l} + \left( E \left\Vert \left(T^\ast T\right)^a Q^j \hat{P}_3 Q^k \hat{P}_3 Q^{N - j - k - 2} \right\Vert^{2l} \right)^{1/2l} \right]\\
\leq& O \left( \sum_{j = 0}^{N - 2} \sum_{k = 0}^{N - 2 - j} (j + 1)^{-1/2-a} (k + 1)^{-1/2} \gamma^2_n + (j + 1)^{-1/2 - a} (N - 1 - j- k)^{-1/2} \gamma^2_n + (j + 1)^{-1/2 - a} \gamma_n^3 + (j+ 1)^{-a} (k+ 1)^{-1}\gamma_n^2 \right. \\
&\quad \left. + (j+ 1)^{-a}( k+ 1)^{-1/2} (N - 1 - j- k)^{-1/2} \gamma_n^2 + (j+ 1)^{-a} (k + 1)^{-1/2} \gamma_n^3 + (j+ 1)^{-a} (N - 1 - j- k)^{-1/2} \gamma_n^3 + (j+ 1)^{-a}\gamma_n^4 \right) \\
=& O \left( N^{1-a} \gamma^2_n + N^{1-a} \sqrt{N} \gamma^3_n + N^{1-a} \ln(N) \gamma^2_n + N^{2-a} \gamma^4_n\right) = O \left( N^{\frac{1}{2}-a}\gamma_n \left( \sqrt{N} \gamma_n + \sqrt{N} \ln(N) \gamma_n + N^{\frac{3}{2}} \gamma^3_n\right) \right) = o \left( N^{\frac{1}{2}-a}\gamma_n \right),
\end{align*}}
\normalsize
where the last line follows from Assumption \ref{ass:restrvarth}(i), $\gamma_n \sqrt{N} \ln(N) = o(1)$. Finally, $\gamma_n = o(N^{-(1 + \epsilon)/2})$ gives $\gamma_n N^{\frac{1}{2}-a} = o(N^{-a-\epsilon/2}) = o(N^{-a})$. The first claim of the Lemma follows.

To prove \ref{lemmaA3eq2}, we write 
\begin{align*}
\left( E || \left(T^\ast T\right)^a  \hat{Q}^N \hat{T}^\ast ||^{2l} \right)^{1/2l} \leq& \left( E || \left(T^\ast T\right)^a  \left[ \hat{Q}^N \hat{T}^\ast  - Q^N T^\ast  \right] ||^{2l}\right)^{1/2l} + || \left(T^\ast T\right)^a Q^N T^\ast|| \\
\leq& \left( E ||\left(T^\ast T\right)^a  \left[ \hat{Q}^N - Q^N \right] ||^{4l} E || \hat{T}^\ast - T^\ast ||^{4l} \right)^{1/4l} \\
& \quad + \left( E || \left(T^\ast T\right)^a  \left[ \hat{Q}^N - Q^N \right] T^\ast ||^{2l} \right)^{1/2l} \\
& \quad + ||\left(T^\ast T\right)^a  Q^N|| \left( E || \hat{T}^\ast - T^\ast ||^{2l}\right)^{1/2l} \\
& \quad + || \left(T^\ast T\right)^a Q^N T^\ast||.
\end{align*}
The last term is $O\left( N^{-1/2 - a}\right)$ directly from Lemma \ref{lemapp1}. Moreover, using the same proof as above, and Lemma \ref{lemapp1} with $a < 0.5$, one can show that 
\begin{align*}
\left( E ||\left(T^\ast T\right)^a  \left[ \hat{Q}^N - Q^N \right] ||^{4l} E || \hat{T}^\ast - T^\ast ||^{4l} \right)^{1/4l} =& O\left( \gamma^2_n N^{\frac{1}{2}-a} \right)\\
\left( E || \left(T^\ast T\right)^a  \left[ \hat{Q}^N - Q^N \right] T^\ast ||^{2l}\right)^{1/2l} =& O\left( \gamma_n N^{-a} \right) \\
|| \left(T^\ast T\right)^a  Q^N|| \left( E || \hat{T}^\ast - T^\ast ||^{2l}\right)^{1/2l} =& O\left( \gamma_n N^{-a}\right).
\end{align*}
The result follows from $\gamma_n = o(N^{-(1 + \epsilon)/2})$, which gives 
\begin{align*}
\gamma^2_n N^{\frac{1}{2}-a} =  & o(N^{-\frac{1}{2}-a-\epsilon})= o(N^{-\frac{1}{2}-a}),\\
\gamma_n N^{-a} = &  o(N^{-\frac{1}{2}-a-\epsilon/2})=  o(N^{-\frac{1}{2}-a}).
\end{align*}
\end{proof}
\end{lemma}


\clearpage

\end{document}